\documentclass[10pt]{amsart}   
\usepackage{amssymb,amsmath,amscd,amsfonts,amsthm,mathrsfs,curves}
\usepackage[active]{srcltx} 
\usepackage[all]{xy}  
\usepackage{color}

\def\thesection{\arabic{section}} 
\renewcommand{\theequation}{\arabic{section}.\arabic{equation}} 
\newtheorem{theorem}{Theorem}[section] 
\newtheorem{lemma}[theorem]{Lemma} 
\newtheorem{proposition}[theorem]{Proposition} 
\newtheorem{corollary}[theorem]{Corollary} 
\newtheorem{assumption}[theorem]{Assumption} 
\theoremstyle{definition}

\theoremstyle{remark}  
\newtheorem{rem}[theorem]{Remark}  
  
\newcommand{\Hr}{\mathscr{H}}

\newcommand{\Bc}{\mathcal{B}}  
\newcommand{\Cc}{\mathcal{C}}  
\newcommand{\Dc}{\mathcal{D}} 
\newcommand{\Fc}{\mathcal{F}}
\newcommand{\Gc}{\mathcal{G}}
\newcommand{\Hc}{\mathcal{H}}  
\newcommand{\Ic}{\mathcal{I}}  
\newcommand{\Jc}{\mathcal{J}}

\newcommand{\Sc}{\mathcal{S}} 
 
\newcommand{\Vc}{\mathcal{V}}  

\newcommand{\Sch}{\mathscr{S}}
\newcommand{\ad}{\mathrm{ad}}

\newcommand{\re}{\mathrm{Re}}   
\newcommand{\im}{\mathrm{Im}}

\newcommand{\supp}{\mathrm{supp}}  
\newcommand{\C}{\mathbb{C}}  
\newcommand{\R}{\mathbb{R}}  
\newcommand{\N}{\mathbb{N}}

\def\build#1_#2^#3{\mathrel{\mathop{\kern 0pt#1}\limits_{#2}^{#3}}} 
\def\cchi{\raisebox{.45 ex}{$\chi$}}%an upper placed \cchi
 
\newcommand{\un}{1\hskip-0.6ex {\mathsf{ I} }} 

\newcommand{\op}{{\rm Op\, }}  
\newcommand{\ph}{{\rm x}^\ast}
\newcommand{\ssi}{\Longleftrightarrow}
\newcommand{\Pfof}[1]{{\it Proof of #1.}\, }
\newcommand{\donne}{\mapsto}
\newcommand{\dans}{\longrightarrow}

\newcommand{\rL}{{\rm L}}

\begin{document}  
\title[Oscillating potentials]
{Limiting absorption principle for Schr\"odinger operators with 
oscillating potential}
\author{Thierry Jecko and Aiman Mbarek}
\address{AGM, UMR 8088 du CNRS, Universit\'e de Cergy Pontoise, Site de Saint Martin,  
2, rue Adolphe Chauvin, F-95302 Cergy-Pontoise cedex, France.}
\email{thierry.jecko@math.cnrs.fr\, ,\, aiman.mbarek@u-cergy.fr\, .}
%\subjclass[2000]{47A40,47B25, 81U99}  
\date{\today}   
\begin{abstract}
Making use of the localised Putnam theory developed in \cite{gj}, we show the
limiting absorption principle for Schr\"odinger operators with perturbed oscillating potential
on appropriate energy intervals. We focus on a certain class of oscillating potentials (larger than the one
in \cite{gj2}) that was already studied in \cite{bd,dmr,dr1,dr2,mu,ret1,ret2}. Allowing long-range and short-range
components and local singularities in the perturbation, we improve known results. A subclass of the considered potentials
actually cannot be treated by the Mourre commutator method with the generator of dilations as conjugate operator. 
Inspired by \cite{fh}, we also show, in some cases, the absence of positive eigenvalues for our 
Schr\"odinger operators. 
\end{abstract}   
\maketitle   
\tableofcontents

%%%%%%%%%%%%%%%%%%%%%%%%%%%%%%%%%%
\section{Introduction.} \label{s:intro}
\setcounter{equation}{0}
%%%%%%%%%%%%%%%%%%%%%%%%%%%%%%%%%%

In this paper, we are interested in the behaviour near the positive real axis of the resolvent of a class of continuous Schr\"odinger
operators. We shall prove a so called ``limiting absorption principle'', a very useful result to develop the scattering theory
associated to those Schr\"odinger operators. It also gives information on the nature of their essential spectrum, as a byproduct.
The main interest of our study relies on the fact that we include some oscillating contribution in the potential of our 
Schr\"odinger operators.

To set up our framework and precisely formulate our results, we need to introduce some notation.
Let $d\in\N^\ast$. We denote by $\langle \cdot ,\cdot \rangle$ and $\|\cdot\|$ the right
linear scalar product and the norm in $\rL^2(\R^d)$, the space of squared integrable, complex
functions on $\R^d$. We also denote by $\|\cdot\|$ the norm of bounded operators on $\rL^2(\R^d)$.
Writing $x=(x_1; \cdots ; x_d)$ the variable in $\R^d$, we set
\[\langle x\rangle \ :=\ \biggl(1\, +\, \sum_{j=1}^d\, x_j^2\biggr)^{1/2}\, .\]
Let $Q_j$ the multiplication operator in $\rL^2(\R^d)$
by $x_j$ and $P_j$ the self-adjoint realization of $-i\partial _{x_j}$ in $\rL^2(\R^d)$. We set $Q=(Q_1; \cdots ; Q_d)^T$
and $P=(P_1; \cdots ; P_d)^T$, where $T$ denotes the transposition. Let
\[H_0\ =\ |P|^2\ :=\ \sum_{j=1}^d\, P_j^2\ =\ P^T\cdot P\]
be the self-adjoint realization of the nonnegative Laplace operator
$-\Delta$ in $\rL^2(\R^d)$. We consider the Schr\"odinger operator $H=H_0+V(Q)$, where $V(Q)$ is the
multiplication operator by a real valued function $V$ on $\R^d$ satisfying the following
\begin{assumption}\label{a:cond-pot}
Let $\alpha, \beta\in ]0; +\infty[$. Let $\rho _{sr}, \rho _{lr}, \rho _{lr}'\in ]0; 1]$. 
Let $v\in\Cc^1(\R^d; \R^d)$ with bounded derivative. Let $\kappa\in\Cc^\infty_c(\R; \R )$ with $\kappa =1$ on 
$[-1; 1]$ and $0\leq\kappa\leq 1$. We consider functions $V_{sr}, \tilde{V}_{sr}, V_{lr}, V_c, W_{\alpha\beta}: \R^d\dans\R$ such
that $V_c$ is compactly supported and $V_c(Q)$ is $H_0$-compact, such that the functions $\langle x\rangle^{1+\rho _{sr}}V_{sr}(x)$, 
$\langle x\rangle^{1+\rho _{sr}}\tilde{V}_{sr}(x)$, $\langle x\rangle^{\rho _{lr}}V_{lr}(x)$ and the distributions  
$\langle x\rangle^{\rho _{lr}'}x\cdot\nabla V_{lr}(x)$ and $\langle x\rangle^{\rho _{sr}}(v\cdot\nabla\tilde{V}_{sr})(x)$ 
are bounded, and
\begin{equation}\label{eq:W}
W_{\alpha\beta}(x)\ =\ w\bigl(1 - \kappa (|x|)\bigr)|x|^{-\beta}\sin (k|x|^\alpha)
\end{equation}
with real $w$. Let $V=V_{sr}+v\cdot\nabla\tilde{V}_{sr}+V_{lr}+V_c+W_{\alpha\beta}$.
\end{assumption}

Under Assumption~\ref{a:cond-pot}, $V(Q)$ is $H_0$-compact. Therefore $H$ is self-adjoint on the domain $\Dc (H_0)$ 
of $H_0$, which is the Sobolev space $\Hc^2(\R^d)$ of $\rL^2(\R^d)$-functions such that their distributional derivative up to second 
order belong to $\rL^2(\R^d)$. By Weyl's theorem, the essential spectrum of $H$ is given by the spectrum of $H_0$, 
namely $[0; +\infty [$. Let $A$ be the self-adjoint realization of the operator $(P\cdot Q+Q\cdot P)/2$ in $\rL^2(\R^d)$. 
By the Mourre commutator method with $A$ as conjugate operator, one has the following Theorem , which is a consequence of 
the much more general Theorem 7.6.8 in \cite{abg}:

\begin{theorem}\label{th:tal-abg}\cite{abg}.
Consider the above operator $H$ with $w=0$ (i.e. without the oscillating part of the potential). Then the point spectrum 
of $H$ is locally finite in $]0; +\infty[$. Furthermore, for any $s>1/2$ and any compact interval $\Ic\subset ]0; +\infty[$, 
that does not intersect the point spectrum of $H$, 
\begin{equation}\label{eq:tal-A-abg}
\sup_{\Re z\in\Ic,\atop\Im z\neq 0}\bigl\|\langle A\rangle^{-s}(H-z)^{-1}\langle A\rangle^{-s}\bigr\|\ <\ +\infty
\, .
\end{equation}
\end{theorem}
\begin{rem}\label{r:tal-abg-ext}
In \cite{co,cg}, a certain class of potentials that can be written as the divergence of a short range potential 
(i.e. a potential like $V_{sr}$) were studied. Theorem~\ref{th:tal-abg} covers this case. \\
We point out that the short range conditions (on $V_{sr}$ and $\tilde{V}_{sr}$) can be relaxed to reduce to a Agmon-H\"ormander 
type condition (see Theorem 7.6.10 \cite{abg} and Theorem 2.14 in \cite{gm}). ``Strongly singular'' terms (more singular than our 
$V_c$) are also considered in Section 3 in \cite{gm}. 
\end{rem}
\begin{rem}\label{r:tal-abg}
When $w=0$, $H$ has a good enough regularity w.r.t. $A$ (see Section~\ref{s:regu} and Appendix~\ref{app:s:regu} for details) 
thus the Mourre theory based on $A$ can be applied to get Theorem~\ref{th:tal-abg}. But it actually gives more, 
not only the existence of the boundary values of the resolvent 
of $H$ (which is implied by \eqref{eq:tal-A-abg}) but also some H\"older continuity of these boundary values. It is well-known 
that all this implies that the same holds true when the weight $\langle A\rangle^{-s}$ are replaced by $\langle Q\rangle^{-s}$ 
(see Remark~\ref{r:2tal} below for a sketch).\\
Still for $w=0$, under some assumption on the form $[V_c, A]$ (roughly \eqref{eq:commut-v_c} below), it follows from 
\cite{fh,fhhh1} that $H$ has no positive eigenvalue. 
\end{rem}

Now, we turn on the oscillating part $W_{\alpha\beta}$ of the potential and ask ourselves, which result from the above ones is 
preserved. To formulate our first main result, we shall need the following 
%We shall consider compact intervals $\Jc$ included in $]0; +\infty [$. We denote by $\mathring{\Jc}$ the interior of $\Jc$.
%
\begin{assumption}\label{a:cond-alpha-beta-gene}
Let $\alpha , \beta >0$ and set $\beta _{lr}=\min (\beta ; \rho _{lr})$. Unless $|\alpha -1|+\beta >1$, we take $\alpha\geq 1$ 
and we take $\beta$ and $\rho_{lr}$ such that $\beta +\beta _{lr}>1$ or, equivalently, $\beta>1/2$ and $\rho_{lr}>1-\beta$. 
We consider a compact interval $\Ic$ such that $\Ic\subset ]0; k^2/4[$, if $\alpha =1$ and $\beta\in ]1/2; 1]$, else such 
that $\Ic\subset ]0; +\infty[$. 
\end{assumption}
\begin{rem}\label{r:W-regu}
If $\beta >1$, $W_{\alpha\beta}$ can be considered as short range potential like $V_{sr}$. If $\alpha<\beta\leq 1$, 
$W_{\alpha\beta}$ satisfies the long range condition required on $V_{lr}$. If $\alpha + \beta >2$ and $\beta\leq 1$ then, for 
$\epsilon =\alpha + \beta -2$, for some short range potentials $\hat{V}_{sr}$ $\check{V}_{sr}$ (i.e. satisfying the same requirement 
as $V_{sr}$), for some $\tilde{\kappa}\in\Cc^\infty_c(\R; \R )$ with $\tilde{\kappa }=1$ on 
$[-1/2; 1/2]$ and with support in $[-1; 1]$, and for $x\in\R^d$, 
\begin{equation}\label{eq:w=grad}
w\bigl(1 - \tilde{\kappa} (|x|)\bigr)|x|^{-1}x\cdot\nabla\check{V}_{sr}(x)\ =\ k\alpha W_{\alpha\beta}(x)\, +\, \hat{V}_{sr}(x)\, ,
\end{equation}
where $\check{V}_{sr}(x)=-(1 - \kappa (|x|))|x|^{-1-\epsilon}\cos (k|x|^{\alpha})$. 
In all cases, Theorem~\ref{th:tal-abg} applies.
\end{rem}
\begin{rem}\label{r:W-compare}
Our assumptions allow $V$ to contain the function $x\donne |x|^{-\beta}\sin (k|x|^\alpha)$ with $\beta <2+\alpha$. This 
function was considered in \cite{bd,dmr,dr1,dr2,ret1,ret2}.\\
Assumption~\ref{a:cond-alpha-beta-gene} excludes the situation where $0<\beta \leq \alpha <1$.
A reason for this is given just after Proposition~\ref{p:oscillations-energy-alpha-leq-1} in Section~\ref{s:oscillations}.\\
It turns out that our results do not change if one replaces the sinus function in $W_{\alpha\beta}$ by a cosinus function. 
\end{rem}

Let $\Pi$ be the orthogonal projection onto the pure point spectral subspace of $H$. We set $\Pi^\perp =1-\Pi$. For any complex number
$z\in\C$, we denote by $\Re z$ (resp. $\Im z$) its real (resp. imaginary) part. Our first main result is the following limiting absorption
principle (LAP). 
\begin{theorem}\label{th:main}
Suppose Assumptions~\ref{a:cond-pot} and~\ref{a:cond-alpha-beta-gene} are satisfied. For any $s>1/2$, 
\begin{equation}\label{eq:tal-Q}
\sup_{\Re z\in\Ic ,\atop \Im z\neq 0}\bigl\|\langle Q\rangle^{-s}(H-z)^{-1}\Pi^\perp\langle Q\rangle^{-s}\bigr\|\ <\ +\infty
\, .
\end{equation}
\end{theorem}
\begin{rem}\label{r:tal-proj}
In the litterature, the LAP is often proved away from the point spectrum, as in Theorem~\ref{th:tal-abg}. If $\Ic$ in 
\eqref{eq:tal-Q} does not intersect the latter, one can remove $\Pi ^\perp$ in \eqref{eq:tal-Q} and therefore get the 
usual LAP. But the LAP \eqref{eq:tal-Q} gives information on the absolutely continuous subspace of $H$ near possible 
embedded eigenvalues.\\
When $|\alpha -1|+\beta >1$ and $\Ic$ does not intersect the point spectrum of $H$, the Mourre theory gives a stronger 
result than Theorem~\ref{th:main} (cf. Theorem~\ref{th:tal-abg} and Remark~\ref{r:tal-abg}). 
\end{rem}

Historically, LAPs for Schr\"odinger operators were first obtained by pertubation, starting from the LAP for the Laplacian $H_0$.
Lavine initiated nonnegative commutator methods in \cite{l1,l2} by adapting Putnam's idea (see \cite{cfks} p. 60). 
Mourre introduced 1980 in \cite{m} a powerful, non pertubative, local commutator method, nowadays called
``Mourre commutator theory'' (see \cite{abg,gg,ggm,jmp,sa}). Nevertheless
it cannot be applied to potentials that contain some kind of oscillaroty term (cf. \cite{gj2}).
In \cite{co,cg}, the LAP was proved pertubatively for a class of oscillatory potentials. This result now follows from 
Mourre theory (cf. Remark~\ref{r:tal-abg-ext}). In \cite{bd,dmr,dr1,dr2,ret1,ret2}, the present situation with $V_c=0$ 
and a radial long range contribution $V_{lr}$ was
treated using tools of ordinary differential equations and again a pertubative argument.
Theorem~\ref{th:main} improves the results of these papers in two ways. First, we allow a long range (non radial) part in the potential.
Second, the set $\Vc$ of values of $(\alpha ;\beta)$, for which the LAP (on some interval) holds true, is here larger.
However, in the case $\alpha =1$, these old results provide a LAP also beyond $k^2/4$ in all dimension $d$, whereas we
are able to do so only in dimension $d=1$. For $\alpha =\beta =1$, the LAP at high enough energy was proved in \cite{mu}.
Another proof of this result is sketched in Remark~\ref{r:V_c} below.\\
We point out that the discrete version of the present situation is treated in \cite{man}. We also signal that the LAP
for continuous Schr\"odinger operators is studied in \cite{mar} by Mourre commutator theory but with new conjugate operators,
including the one used in \cite{n}. We also emphasize an alternative approach to the LAP based on the density of states. 
It seems however that general long range pertubations are not treated yet. We refer to \cite{ben} for details on this approach. 

In Fig.~\ref{dessin:1}, we drew the set $\Vc$ in a $(\alpha ; \beta )$-plane. It is the union of the blue and green regions. 
The papers \cite{bd,dmr,dr1,dr2,ret1,ret2}
etablished the LAP in the region above the red and black lines and, along the vertical green line, above the point $A=(1; 2/3)$. 
According to Remark~\ref{r:W-regu}, Theorem~\ref{th:tal-abg} shows the LAP in the blue region (above the red lines and the blue one). 
Both results are obtained without energy restriction. Theorem~\ref{th:main} covers the blue and green regions (the set $\Vc$), 
with a energy restriction on the vertical green line. In \cite{gj2}, the LAP with energy restriction is proved at the point 
$B=(1;1)$. In the red region (below the red lines), the LAP is still an open question.

\begin{figure}
\setlength{\unitlength}{1cm}
\begin{center}
\begin{picture}(8.5,6)(-0.5,-0.5)
\put(0,0){\vector(1,0){10}}
\put(0,0){\vector(0,1){5}}
\put(3,-0.1){\line(0,1){0.2}}
\put(-0.1,3){\line(1,0){0.2}}
\put(-0.1,2.4){\line(1,0){0.2}}
\put(-0.1,1.5){\line(1,0){0.2}}
\put(-0.1,2){\line(1,0){0.2}}
\put(-0.1,-0.5){0}
\put(5.9,-0.5){2}
\put(2.9,-0.5){1}
\put(-0.7,2.3){4/5}
\put(-0.7,1.9){2/3}
\put(-0.7,1.4){1/2}
\put(9.8,-0.5){$\alpha$}
\put(-0.4,2.9){1}
\put(-0.4,4.8){$\beta$}

\color{red}
\put(0,0){\line(1,1){2.97}}
\put(3,1.5){\line(1,0){1.5}}
\put(4.5,1.5){\line(1,-1){1.5}}
\put(2.2,1.7){red}
\put(1.5,0.8){red}
\put(4,0.8){red}
\color{green}
\put(3,1.5){\line(0,1){1.5}}
\put(3.2,1.6){green}
\color{blue}
\put(0.2,1){blue}
\put(1.4,4){blue}
\put(5.9,4){blue}
\put(5.9,1){blue}
\put(3,3){\line(1,-1){1.5}}
\color{black}
\put(0.1,2){\line(1,0){1.87}}
\put(2.9,2){\line(1,0){0.2}}
\put(3.2,1.9){A}
\put(2.6,2.9){B}
\put(3,2.4){\line(5,-2){1}}
\put(4,2){\line(1,0){6}}
\end{picture}
\end{center}
\caption{LAP. $\Vc=$ \textcolor{blue}{blue}\, $\cup$\, \textcolor{green}{green}.}\label{dessin:1}
\end{figure}

Recall that $A$ is the self-adjoint realization of the operator $(P\cdot Q+Q\cdot P)/2$ in $\rL^2(\R^d)$. 
We are able to get the following improvement of a main result in \cite{gj2}.
\begin{theorem}\label{th:tal-Wigner-gj2}
Let $\alpha =\beta =1$. Under Assumption~\ref{a:cond-pot} with $\tilde{V}_{sr}=V_c=0$, 
take a compact interval $\Ic\subset ]0;k^2/4[$. Then, for any $s>1/2$,
\begin{equation}\label{eq:tal-A}
\sup_{\Re z\in\Ic,\atop\Im z\neq 0}\bigl\|\langle A\rangle^{-s}(H-z)^{-1}\Pi^\perp\langle A\rangle^{-s}\bigr\|\ <\ +\infty
\, .
\end{equation}
\end{theorem}
\proof In \cite{gj2}, it was further assumed that, for any $\mu\in\Ic$, ${\rm Ker}(H-\mu)\subset\Dc (A)$.
Thanks to Corollary~\ref{c:regu-A}, this assumption is superfluous. \qed

\begin{rem}\label{r:V_c}
Note that Assumption~\ref{a:cond-alpha-beta-gene} is satisfied for $\alpha =\beta =1$. In dimension $d=1$, the above result is still true if
$\Ic\subset ]k^2/4 ; +\infty[$. A careful inspection of the proof in \cite{gj2} shows that Theorem~\ref{th:tal-Wigner-gj2} holds true in all
dimensions if $\Ic\subset ]a ; +\infty[$, for large enough positive $a$ (depending on $|w|$). If $|w|$ is small enough, the mentioned proof 
is even valid on any compact interval $\Ic\subset ]0 ; +\infty[$. \\
For nonzero potentials $V_c$ and $\tilde{V}_{sr}$, we believe that
one can adapt the proof in \cite{gj2} of Theorem~\ref{th:tal-Wigner-gj2}.
\end{rem}
\begin{rem}\label{r:2tal}
It is well known that \eqref{eq:tal-A} implies \eqref{eq:tal-Q}. Let us sketch this briefly. It suffices to restrict $s$ to $]1/2; 1[$.
Take $\theta\in\Cc_c^\infty(\R;\R)$ such that $\theta =1$ near $\Ic$. Then, the bound \eqref{eq:tal-Q} is valid if $(H-z)^{-1}$ is
replaced by $(1-\theta (H))(H-z)^{-1}$. The boundedness of the contribution of $\theta (H)(H-z)^{-1}$ to the l.h.s of 
\eqref{eq:tal-Q} follows from \eqref{eq:tal-A} and from the boundedness of $\langle Q\rangle^{-s}\theta (H)\langle A\rangle^s$.
To see the last property, one can write
\[\langle Q\rangle^{-s}\theta (H)\langle A\rangle^s=\langle Q\rangle^{-s}\theta (H)\langle P\rangle^s\langle Q\rangle^s\cdot
\langle Q\rangle^{-s}\langle P\rangle^{-s}\langle A\rangle^s\, .\]
The last factor is bounded by Lemma C.1 in \cite{gj2}. The boundedness of the other one is granted by the regularity of $H$ w.r.t.
$\langle Q\rangle$ (see Section~\ref{s:regu}) and the fact that $\theta (H)\langle P\rangle$ is bounded. 
\end{rem}
\begin{rem}\label{r:spec}
It is well known that \eqref{eq:tal-Q} implies the absence of singular continuous spectrum in $\Ic$ (see \cite{rs4}).
On this subject, we refer to \cite{k,rem} for more general results. 
\end{rem}

In Section~\ref{s:regu}, we show that the Mourre commutator method, with the generator $A$ of dilations as conjugate operator, 
cannot be applied to recover Theorem~\ref{th:main} in his full range of validity $\Vc$, neither the classical theory with 
$\Cc^{1,1}$ regularity (cf. \cite{abg}), nor the improved one with ``local'' $\Cc^{1+0}$ regularity (cf. \cite{sa}). Indeed 
the required regularity w.r.t. $A$ is not valid on $\Vc$. As pointed out in \cite{gj2}, Theorem~\ref{th:tal-Wigner-gj2} 
cannot be proved with these Mourre theories for the same reason. We expect that the use of known, alternative conjugate 
operators (cf. \cite{abg,n,mar}) does not cure this regularity problem. However, according to a new version of the paper 
\cite{mar}, one would be able to apply the Mourre theory in a larger region than the blue region mentioned above, this 
region still being smaller than $\Vc$ (cf. Section~\ref{s:regu}). 

The given proof of Theorem~\ref{th:tal-Wigner-gj2} relies on a kind of ``energy localised'' 
Putnam argument. This method, which is reminiscent of the works \cite{l1,l2} by Lavine, was introduced in \cite{gj} and 
improved in \cite{ge,gj2}. It was originally called ``weighted Mourre theory'' but it is closer to Putnam idea (see 
\cite{cfks} p. 60) and does not make use of differential inequalities as the Mourre theory. Note that, up to now, the latter 
gives stronger results than the former. It is indeed still unknown whether this ``localised Putnam theory'' is able to prove 
continuity properties of the boundary values of the resolvent. \\
We did not succeed in applying the ``localised Putnam theory'' formulated in 
\cite{gj2} to prove Theorem~\ref{th:main}. We believe that, again, the bad regularity of $H$ w.r.t. $A$ is the source of our 
difficulties (cf. Section~\ref{s:regu}). Instead, we follow the more complicated version presented in \cite{gj}, 
which relies on a Putnam type argument that is localised in $Q$ and $H$, and use the excellent regularity of $H$ w.r.t. 
$\langle Q\rangle$ (cf. Section~\ref{s:regu}). 

A byproduct of the proof of Theorem~\ref{th:tal-abg} is the local finitness (counting multiplicity) of the pure point 
spectrum of $H$ in $]0; +\infty[$. Thus this local finitness holds true if $|\alpha -1|+\beta >1$. We extend this result 
to the case where $|\alpha -1|+\beta \leq 1$ in the following way: the above local finitness is valid in $]0; +\infty[$, 
if $\alpha >1$, and in $]0; k^2/4[$, if $\alpha =1$ (cf. Corollary~\ref{c:finitness}). \\
In the papers \cite{fhhh2,fh}, polynomial bounds and even exponential bounds were proven on possible eigenvectors with 
positive energy. In our framework, those results fully apply when $|\alpha -1|+\beta >1$. Here we get the same polynomial 
bounds under the less restrictive Assumptions~\ref{a:cond-pot} and~\ref{a:cond-alpha-beta-gene} 
(cf. Proposition~\ref{p:decroissace-poly}). Concerning the exponential bounds, we manage to get them under 
Assumptions~\ref{a:cond-pot} and~\ref{a:cond-alpha-beta-gene}, but for $\alpha>1$ (see Proposition~\ref{p:borne-exp}). \\
In the papers \cite{fhhh2,fh} again, the absence of positive eigenvalue is proven. In our framework, this result applies when 
$\alpha <\beta$ and when $\beta >1$, provided that the form 
$[(V_c+v\cdot\nabla\tilde{V}_{sr})(Q), iA]$ is $H_0$-form-lower-bounded with relative bound $<2$
(see \eqref{eq:commut-v_c} for details). When $\alpha+\beta>2$ and $\beta\leq 1$, it applies under the same condition, provided that 
the oscillating part of the potential is small enough (i.e. if $|w|$ is small enough). Indeed, in that case, the form 
$[(V_c+v\cdot\nabla\tilde{V}_{sr}+W_{\alpha\beta})(Q), iA]$ is $H_0$-form-lower-bounded with relative bound $<2$. 
Inspired by those papers, we shall derive our second main result, namely 
\begin{theorem}\label{th:no-posivite-eigenvalue}
Under Assumptions~\ref{a:cond-pot} and~\ref{a:cond-alpha-beta-gene} with $\alpha>1$ when $|\alpha -1|+\beta\leq 1$, we 
assume further that the form $[(V_c+v\cdot\nabla\tilde{V}_{sr})(Q), iA]$ is
$H_0$-form-lower-bounded with relative bound $<2$ (see \eqref{eq:commut-v_c} for details). 
Furthermore, we require that $|w|$ is small enough if $\alpha +\beta >2$ and $\beta\leq 1/2$. 
Then $H$ has no positive eigenvalue.  
\end{theorem}
\proof The result follows from Propositions~\ref{p:borne-exp} and~\ref{p:pas-de-vp}. \qed

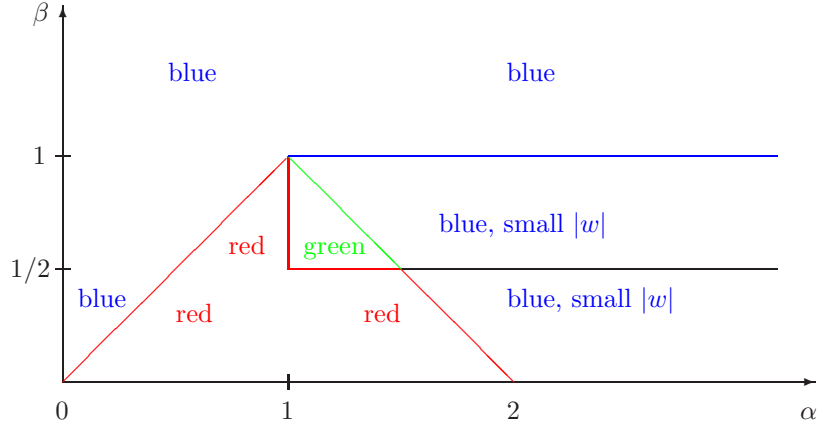
\begin{figure}
\setlength{\unitlength}{1cm}
\begin{center}
\begin{picture}(8.5,6)(-0.5,-0.5)
\put(0,0){\vector(1,0){10}}
\put(0,0){\vector(0,1){5}}
\put(3,-0.1){\line(0,1){0.2}}
\put(-0.1,3){\line(1,0){0.2}}
\put(-0.1,1.5){\line(1,0){0.2}}
\put(-0.1,-0.5){0}
\put(2.9,-0.5){1}
\put(5.9,-0.5){2}
\put(-0.7,1.4){1/2}
\put(9.8,-0.5){$\alpha$}
\put(-0.4,2.9){1}
\put(-0.4,4.8){$\beta$}
\put(4.5,1.5){\line(1,0){5}}
\color{red}
\put(0,0){\line(1,1){2.97}}
\put(3,1.5){\line(1,0){1.5}}
\put(2.2,1.7){red}
\put(1.5,0.8){red}
\put(4,0.8){red}
\put(3,1.5){\line(0,1){1.5}}
\put(4.5,1.5){\line(1,-1){1.5}}
\color{green}
\put(3.2,1.7){green}
\put(3,3){\line(1,-1){1.5}}
\color{blue}
\put(0.2,1){blue}
\put(1.4,4){blue}
\put(5.9,4){blue}
\put(5,2){blue, small $|w|$}
\put(5.9,1){blue, small $|w|$}
\put(3,3){\line(1,0){6,5}}
\end{picture}
\end{center}
\caption{No positive eigenvalue in \textcolor{blue}{blue}\, $\cup$\, \textcolor{green}{green}.}\label{dessin:2}
\end{figure}
\begin{rem}\label{r:eigenvector}
Our proof is strongly inspired by the ones in \cite{fhhh2,fh}. Actually, these proofs cover the cases $\beta >1$, 
$\alpha <\beta$, and the case where $\alpha +\beta >2$, $\beta\leq 1$, and $|w|$ is small enough. In the last case, 
namely when $\alpha>1$, $\beta>1/2$, $\rho_{lr}>1-\beta$, and $\alpha +\beta\leq 2$, 
the main new ingredient is an appropriate control on the oscillatory part of the potential. 
In particular, in the latter case, we do not need any smallness on $|w|$. 

\end{rem}
\begin{rem}\label{r:no-eigenvalue-alpha=beta=1}
In the case $\alpha =\beta=1$, assuming \eqref{eq:commut-v_c}, we can show the absence of eigenvalue at high energy. This follows from 
Remark~\ref{r:case-alpha=beta=1} and Proposition~\ref{p:pas-de-vp}. However an embedded eigenvalue does exist for an appropriate choice
of $V$ (see \cite{fh,cfks,chm}). 
\end{rem}
\begin{rem}\label{r:superfluous-pi}
Under the assumptions of Theorem~\ref{th:no-posivite-eigenvalue}, for any compact interval $\Ic\subset ]0; +\infty[$, the result of
Theorem~\ref{th:main}, namely \eqref{eq:tal-Q}, is valid with $\Pi^\perp$ replaced by the identity operator. Indeed, for any compact
interval $\Ic'\subset ]0; +\infty[$ containing $\Ic$ in its interior, $\un _{\Ic '}(H)\Pi =0$ by
Theorem~\ref{th:no-posivite-eigenvalue}. In view of Remark~\ref{r:V_c}, the LAP \eqref{eq:tal-A} is valid at high energy, when 
$\alpha =\beta=1$. Thanks to Remark~\ref{r:no-eigenvalue-alpha=beta=1}, one can also remove $\Pi^\perp$ in \eqref{eq:tal-A}. 
\end{rem}
One can find many papers on the absence of positive eigenvalue for Schr\"odinger operators: see for instance 
\cite{co,k,si,a,fhhh2,fh,ij,rs4,cfks}. They do not cover the present situation due to the oscillations in the 
potential. In Fig.~\ref{dessin:2}, we summarise results on the absence of positive eigenvalue. 
In the blue region (above the red and blue lines), the result is granted by \cite{fhhh2,fh}, with a smallness 
condition below the blue line. Theorem~\ref{th:no-posivite-eigenvalue} covers the blue and green regions 
(above the red lines), with a smallness condition below the black line. 

In Assumption~\ref{a:cond-alpha-beta-gene} with $|\alpha -1|+\beta\leq 1$, the parameter $\rho_{lr}$, that controls the 
behaviour at infinity of the long range potential $V_{lr}$, stays in a $\beta$-dependent region. One can get rid of 
this constraint if one chooses a smooth, symbol-like function as $V_{lr}$, as seen in the next 
\begin{theorem}\label{th:results-symbol}
Assume that Assumption~\ref{a:cond-pot} is satisfied with $|\alpha -1|+\beta\leq 1$ and $\beta >1/2$. Assume further that 
$V_{lr}: \R^d\dans\R$ is a smooth function such that, for some $\rho _{lr}\in ]0; 1]$, for all $\gamma\in\N^d$,
\[\sup_{x\in\R^d}\bigl|\langle x\rangle ^{\rho _{lr}+|\gamma|}(\partial_x^\gamma V_{lr})(x)\bigr|\ <\ +\infty\, .\]
Take $\alpha =1$. Then the LAP \eqref{eq:tal-Q} holds true on any compact interval $\Ic$ such that $\Ic\subset ]0; k^2/4[$, if $d\geq 2$, and
such that $\Ic\subset ]0; +\infty[\setminus\{k^2/4\}$, if $d=1$.\\
Take $\alpha >1$. Then the LAP \eqref{eq:tal-Q} holds true on any compact interval $\Ic\subset ]0; +\infty[$.
If, in addition, $[(V_c+v\cdot\nabla\tilde{V}_{sr})(Q), iA]$ is $H_0$-form-lower-bounded with relative bound $<2$
(see \eqref{eq:commut-v_c} for details), then $H$ has no positive eigenvalue. In particular, \eqref{eq:tal-Q} holds true
with $\Pi^\perp$ removed.
\end{theorem}
\begin{rem}\label{r:more-gene-oscill}
We expect that our results hold true for a larger class of oscillatory potential provided that the ``interference'' 
phenomenon exhibited in Section~\ref{s:oscillations} is preserved. In particular, we do not need that $W_{\alpha\beta}$ is radial.
\end{rem}

We point out that there still are interesting, open questions on the Schr\"odinger operators studied here. Concerning the LAP, 
for $\alpha =1$, it is expected that \eqref{eq:tal-Q} is false near $k^2/4$. Note that the Mourre estimate is false there, 
when $\beta =1$ (see \cite{gj2}). The validity of \eqref{eq:tal-Q} beyong $k^2/4$ is still open, even at high energy when 
$\beta <1$. Concerning the existence of positive eigenvalue, again for $\alpha =1$, it is known in dimension $d=1$ that 
there is at most one at $k^2/4$ if $\beta =1$ (see \cite{fh}). It is natural to expect that this is still true for 
$d\geq 2$ and $\beta =1$. We do not know what happens for $\alpha =1>\beta$.

In Section~\ref{s:oscillations}, we analyse the interaction between the oscillations in the potential $W_{\alpha\beta}$ 
and the kinetic energy operator $H_0$. In Section~\ref{s:regu}, we focus on regularity properties of $H$ w.r.t. $A$ and 
to $\langle Q\rangle$ and discuss the applicability of the Mourre theory and of the results from the papers \cite{fhhh2,fh}. 
In Section~\ref{s:mourre}, in some appropriate energy window, we show the Mourre 
estimate, which is still a crucial result. We deduce from it polynomial bounds on possible eigenvectors of $H$ in Section~\ref{s:poly-bounds}. This furnishes 
the material for the proof of Theorem~\ref{th:tal-Wigner-gj2}. In Section~\ref{s:finitness}, we show the local finitness
of the point spectrum in the mentioned energy window. In the case $\alpha >1$, we show exponential bounds on possible eigenvectors in
Section~\ref{s:exp-bounds} and prove the absence of positive eigenvalue in Section~\ref{s:strong-exp-bounds}. Independently of
Sections~\ref{s:exp-bounds} and~\ref{s:strong-exp-bounds}, we prove Theorem~\ref{th:main} in Section~\ref{s:lap}.
Section~\ref{s:symbolic-long-range} is devoted to the proof of Theorem~\ref{th:results-symbol}. Finally, we gathered 
well-known results on pseudodifferential calculus in Appendix~\ref{app:s:calcul-pseudo}, basic facts on regularity w.r.t. an
operator in Appendix~\ref{app:s:regu}, known results on commutator expansions and technical results in
Appendix~\ref{app:s:commut-expansions}, and an elementary, but lengthy argument, used in Section~\ref{s:oscillations}, in
Appendix~\ref{app:s:oscillation}.

{\bf Aknowledgement:} The first author thanks V. Georgescu, S. Gol\'enia, T. Harg\'e, I. Herbst, and P. Rejto, 
for interesting discussions on the subject. Both authors express many thanks to A. Martin, who allowed them to access 
to some result in his work in progress. Both authors are particularly grateful to the anonymous referee for his 
constructive and fruitful report.

%%%%%%%%%%%%%%%%%%%%%%%%
\section{Oscillations.} 
\label{s:oscillations}
\setcounter{equation}{0}
%%%%%%%%%%%%%%%%%%%%%%%%

In this section, we study the oscillations appearing in the considered potential $V$. It is convenient to make use
of some standard pseudodifferential calculus, that we recall in Appendix~\ref{app:s:calcul-pseudo}.
As in \cite{gj2}, our results strongly rely on the interaction of the oscillations in the potential with 
localisations in momentum (i.e. in $H_0$). This interaction is described in the following two propositions.

The oscillating part of the potential $V$ occurs in the potential $W_{\alpha\beta}$ as described in Assumption~\ref{a:cond-pot}.
By \eqref{eq:W}, for some function $\kappa\in\Cc_c^\infty (\R;\R)$ such
that $\kappa =1$ on $[-1; 1]$ and $0\leq\kappa\leq 1$, $W_{\alpha\beta}=w(2i)^{-1}(e_+^\alpha -e_-^\alpha)$, where
\begin{equation}\label{eq:def-e_pm-alpha}
e_\pm^\alpha : \R^d\dans\C\, ,\hspace{.4cm} e_\pm^\alpha (x)\ =\ \bigl(1-\kappa (|x|)\bigr)e^{\pm ik|x|^{\alpha}}\, .
\end{equation}
Let $g_0$ be the metric defined in \eqref{eq:metric}.
\begin{proposition}\label{p:oscillations-energy-alpha-leq-1}\cite{gj2}. 
Let $\alpha =1$. For any function $\theta\in \Cc_c^\infty (\R;\C)$, there exist smooth symbols $a_\pm\in\Sc(1; g_0)$, $b_\pm , c_\pm
\in\Sc(\langle x\rangle^{-1}\langle \xi\rangle^{-1}; g_0)$ such that
\begin{equation}\label{eq:oscillation-localisation}
 e_\pm ^{\alpha}\theta (H_0)\ =\ a_\pm ^w e_\pm ^{\alpha}\, +\, b_\pm ^w e_\pm ^{\alpha} \, + \, e_\pm ^{\alpha}c_\pm ^w
\end{equation}
and, near the support of $1-\kappa (|\cdot |)$, $a_\pm$ is given by
\[a_\pm (x; \xi)\ =\ \theta \Bigl(\bigl|\xi \mp \alpha k|x|^{\alpha -2}x\bigr|^2\Bigl)\ =\ 
\theta \Bigl(\bigl|\xi \mp k|x|^{-1}x\bigr|^2\Bigl)\, .\]
In particular, if $\theta $ has a small enough support in $]0; k^2/4[$, then, for
any $\epsilon\in [0; 1[$, the operator $\theta (H_0)\langle Q\rangle ^\epsilon\sin (k|Q|)\theta (H_0)$ extends to a compact
operator on $\rL^2(\R^d)$, and it is bounded if $\epsilon =1$. 
\end{proposition}  
\begin{rem}\label{r:alpha-leq-1}
In dimension $d=1$, the last result in Proposition~\ref{p:oscillations-energy-alpha-leq-1} still holds 
true if $\theta$ has small enough support in $]0; +\infty[\setminus\{k^2/4\}$ (see \cite{gj2}).
\end{rem}
\Pfof{Proposition~\ref{p:oscillations-energy-alpha-leq-1}} See Lemma 4.3 and Proposition
A.1 in \cite{gj2}. 
\qed

In any dimension $d\geq 1$, for $0<\alpha<1$, the above phenomenon is absent.
A careful inspection of the proof of \eqref{eq:oscillation-localisation} shows that it actually works if $0<\alpha<1$.
But, in constrast to the case $\alpha =1$, the principal symbol of $\theta (H_0)\langle Q\rangle ^\epsilon\sin (k|Q|)\theta (H_0)$,
which is given by
\[\R^{2d}\ni (x; \xi)\, \donne\, (2i)^{-1}\theta \bigl(|\xi |^2\bigr)(a_+-a_-)(x; \xi )\, ,\]
is not everywhere vanishing, for any choice of nonzero $\theta$ with
support in $]0; +\infty[$. The conditions ``$|\xi |^2$ in the support of $\theta$'' and ``$|\xi \mp \alpha k|x|^{\alpha -2}x|^2$
in the support of $\theta$'' are indeed compatible for large $|x|$. \\
In this setting, namely for $0<\alpha<1$ and $d\geq 1$, one can give the following, more precise picture with the help of an appropriate
pseudodifferential calculus. Take a nonzero, smooth function
$\theta$ with compact support in $]0; +\infty[$. For $\epsilon\in ]0; 1[$, on $\rL^2(\R^d)$, the operator 
\[\theta (H_0)\langle Q\rangle ^{\epsilon}\sin \bigl(k|Q|^\alpha \bigr)\theta (H_0)\hspace{.4cm} \Bigl({\rm resp.}\
\theta (H_0)\sin \bigl(k|Q|^\alpha \bigr)\theta (H_0)\Bigr)\]
is unbounded (resp. is not a compact operator). Indeed, for the function $\kappa$
given in \eqref{eq:def-e_pm-alpha}, the multiplication operator
\[\bigl(1-\kappa (|Q|)\bigr)\sin \bigl(k|Q|^\alpha \bigr)\]
is a pseudodifferential operator with symbol in $\Sc(1; g_\alpha)$ for the metric $g_\alpha$ defined in \eqref{eq:metric}.
By pseudodifferential calculus for this admissible metric $g_\alpha $, the symbol of
\[\theta (H_0)\langle Q\rangle ^{\epsilon}\bigl(1-\kappa (|Q|)\bigr)\sin \bigl(k|Q|^\alpha \bigr)
\theta (H_0)\, ,\]
namely
\[\theta \bigl(|\xi |^2\bigr)\#\langle x\rangle ^{\epsilon}\bigl(1-\kappa (|x|)\bigr)\sin \bigl(k|x|^\alpha \bigr)\#\theta
\bigl(|\xi |^2\bigr)\, ,\]
is not a bounded symbol. Thus, the operator is unbounded on $\rL^2(\R^d)$, while 
\[\theta (H_0)\langle Q\rangle ^{\epsilon}\kappa (|Q|)\sin \bigl(k|Q|^\alpha \bigr)\theta (H_0)\]
is compact
since its symbol $\theta (|\xi |^2)\#\langle x\rangle ^{\epsilon}\kappa (|x|)\sin (k|x|^\alpha )\#\theta (|\xi |^2)$ tends to $0$ at
infinity. Still for the metric $g_\alpha$, the symbol of
\[\theta (H_0)\bigl(1-\kappa (|Q|)\bigr)\sin \bigl(k|Q|^\alpha \bigr)\theta (H_0)\]
is
$\theta (|\xi |^2)\#(1-\kappa (|x|))\sin (k|x|^\alpha )\#\theta (|\xi |^2)$, that does not tend to zero at infinity. Therefore
$\theta (H_0)(1-\kappa (|Q|))\sin (k|Q|^\alpha )\theta (H_0)$ is not a compact operator, whereas so is
$\theta (H_0)\kappa (|Q|)\sin (k|Q|^\alpha )\theta (H_0)$.  \\

\begin{rem}\label{r:beta-leq-alpha-leq-1}
The difference between the cases $\alpha =1$ and $0<\alpha <1$ sketched just above
explains why we exclude the case $\beta\leq\alpha<1$ in our results. Recall that the
case $0<\alpha <\beta\leq 1$ is covered by Theorem~\ref{th:tal-abg} (cf. Remark~\ref{r:W-regu}).  
\end{rem}

In the case $\alpha >1$, one can relax the localisation to get
compactness as seen in
\begin{proposition}\label{p:oscillations-energy-alpha>1}
Let $\alpha >1$. For any real $p\geq 0$, there exist $\ell _1\geq 0$ and $\ell _2\geq 0$ such that $\langle P\rangle^{-\ell _1}
\langle Q\rangle ^p(1-\kappa (|Q|))\sin (k|Q|^\alpha)\langle P\rangle^{-\ell _2}$ extends to a compact operator on $\rL^2(\R^d)$. 
In particular, so does $\theta (H_0)\langle Q\rangle ^p(1-\kappa (|Q|))\sin (k|Q|^\alpha)\theta (H_0)$, for any $p$ and
any $\theta\in \Cc_c^\infty (\R;\C)$.
\end{proposition}  

\proof The proof is rather elementary and postponed in Appendix~\ref{app:s:oscillation}. Appropriate $\ell_1$ and $\ell _2$ depend on $p$, 
$\alpha$, and on the dimension $d$. For instance, one can choose $\ell _1$ and $\ell _2$ greater than $1$ plus the integer part of 
$(\alpha -1)^{-1}(p+d)$. \qed

\begin{rem}\label{r:alpha>1}
Take $\theta\in \Cc_c^\infty (\R;\C)$, $\tau\in\Cc_c^\infty (\R^d;\C)$ such that $\tau =1$ near zero, and $\alpha >1$. 
The smooth function
\[(x; \xi)\mapsto \bigl(1-\tau (x)\bigr)\theta \Bigl(\bigl|\xi \mp \alpha k|x|^{\alpha -2}x\bigr|^2\Bigr)\, ,\]
does not belong to $\Sc (m; g)$ for any
weight $m$ associated to the metric $g_0$. So we cannot use the proof of Proposition~\ref{p:oscillations-energy-alpha-leq-1}
in this case. \\
The proof of Proposition~\ref{p:oscillations-energy-alpha>1} shows that the oscillations manage to transform a decay in $\langle P\rangle$
in one in $\langle Q\rangle$. This is not suprising if one is aware of the following, one dimensional formula (see eq. (VII. 5; 2), p. 245,
in \cite{s}), pointed out by V. Georgescu. For any $m\in\N$, there exist $\lambda _0, \cdots , \lambda _{2m}\in\C$
such that 
\[\forall x\in\R\, , \hspace{.4cm}(1+x^2)^m\, e^{i\pi x^2}\ =\ \sum _{j=0}^{2m}\lambda _j\frac{d^j}{dx^j}e^{i\pi x^2}\, .\]
Note that the result of Proposition~\ref{p:oscillations-energy-alpha>1} is false for $\alpha\leq 1$ by
Proposition~\ref{p:oscillations-energy-alpha-leq-1} and the discussion following it. 
\end{rem}
%

%%%%%%%%%%%%%%%%%%%%%%%%
\section{Regularity issues.} 
\label{s:regu}
\setcounter{equation}{0}
%%%%%%%%%%%%%%%%%%%%%%%%

In this section, we focus on the regularity of $H$ w.r.t. the generator of dilations $A$ and also the multiplication operator 
$\langle Q\rangle$. We explain, in particular, why neither the Mourre theory with $A$ as conjugate operator nor the results 
in \cite{fhhh2,fh} on the absence of positive eigenvalue can be applied to $H$ 
in the full framework of Assumption~\ref{a:cond-alpha-beta-gene}. Fig.~\ref{dessin:3} below provides, in the plane of the 
parameters $(\alpha , \beta)$, a region where those external results apply and another where they do not. 

We denote, for $k\in\N$, by $\Hc^k(\R^d)$ or simply $\Hc^k$, the Sobolev space of $\rL^2(\R^d)$-functions such that their 
distributional derivatives up to order $k$ belong to $\rL^2(\R^d)$. Using the Fourier transform, it can be seen as the 
domain of the operator $\langle P\rangle^k$. The dual space of $\Hc^k$ can be identified with 
$\langle P\rangle^{-k}\rL^2(\R^d)$ and is denoted by $\Hc^{-k}$. Recall that $A$ is the self-adjoint realisation of 
$(P\cdot Q+Q\cdot P)/2$ in $\rL^2(\R^d)$. It is well known that the propagator $\R\ni t\donne\exp (itA)$, generated by $A$, 
acts on $\rL^2(\R^d)$ as 
\[\bigl(\exp (itA)f\bigr)(x)\ =\ e^{td/2}f\bigl(e^tx\bigr)\, .\]
It preserves all the Sobolev spaces $\Hc^k$, thus the domain $\Dc(H)=\Dc(H_0)=\Hc^2$ of $H$ and $H_0$. \\
The regularity spaces $\Cc^k(A)$, for $k\in\N^\ast\cup \{\infty\}$, are defined in
Appendix~\ref{app:s:regu}. By Theorem~\ref{th:abg2}, $H\in\Cc^1(A)$ if and only if the form $[H, A]$, defined on 
$\Dc(H)\cap\Dc(A)$, extends to a bounded form from $\Hc^2$ to $\Hc^{-2}$, that is, if and only if there exists $C>0$ such 
that, for all $f, g\in\Hc^2$, 
\begin{equation}\label{eq:bound-form-h2}
 \bigl|\langle f\, ,\, [H, A]g\rangle\bigr|\ \leq\ C\cdot \|f\|_{\Hc^2}\cdot \|g\|_{\Hc^2}\, .
\end{equation}

Before studying the regularity of $H$ w.r.t. $A$, it is convenient to first show that $H$ is very regular w.r.t. 
$\langle Q\rangle$. This latter property relies on the fact that $V(Q)$ commutes with $\langle Q\rangle$. 

\begin{lemma}\label{l:Q-regu} Assume that Assumptions~\ref{a:cond-pot} and~\ref{a:cond-alpha-beta-gene} are satisfied. 
\begin{itemize}
 \item[(1)] For $i, j\in\{1; \cdots ; d\}$, the operators $H_0$, $\langle P\rangle$, $\langle P\rangle^2$, $P_i$, 
 and $P_iP_j$ all belong to 
 $\Cc^\infty(\langle Q\rangle)$ and $\Dc(\langle Q\rangle\langle P\rangle)=\Dc(\langle P\rangle\langle Q\rangle)$.
 \item[(2)] $H\in\Cc^\infty(\langle Q\rangle)$.
 \item[(3)] For $\theta\in\Cc_c^\infty(\R; \C)$, for $i, j\in\{1; \cdots ; d\}$, the bounded operators 
 $\theta (H_0)$, $P_i\theta (H_0)$, $P_iP_j\theta (H_0)$, 
$\theta (H)$, $P_i\theta (H)$, and $P_iP_j\theta (H)$ belong to $\Cc^\infty(\langle Q\rangle)$, and we have the inclusion
$\theta (H)\Dc(\langle Q\rangle)\subset \Dc(\langle P\rangle\langle Q\rangle)\cap\Dc(H_0)$. 
\end{itemize}
\end{lemma}
\proof See Appendix~\ref{app:s:commut-expansions}. \qed

The form $[H, A]$ is defined on $\Dc(H)\cap\Dc(A)$ by $\langle f, [H, iA]g\rangle = \langle Hf, Af\rangle -\langle
Af , Hf\rangle$. Let $\cchi _c\in\Cc_c^\infty(\R^d; \R)$ such that $\cchi _c=1$ on the compact support of $V_c$. 
By statement (1) in Lemma~\ref{l:Q-regu}, the form $[H, A]$ coincides, on $\Dc(\langle P\rangle\langle Q\rangle)\cap\Dc(H_0)$, 
with the form $[H, iA]^{'}$ given by 
\begin{eqnarray}
\langle f\, , \, [H, iA]^{'}g\rangle &=& \langle f\, , \, [H_0, iA]^{'}g\rangle \, +\, \langle f\, , \, [V_{sr}(Q), iA]^{'}g\rangle 
\, +\, \langle f\, , \, [V_c(Q), iA]^{'}g\rangle \nonumber\\
&&\, +\, \langle f\, , \, [V_{lr}(Q), iA]^{'}g\rangle\, +\, \langle f\, , \, [W_{\alpha \beta}(Q), iA]^{'}g\rangle
\label{eq:commutateur}\\
&&+\, \langle f\, , \, [(v\cdot\nabla\tilde{V}_{sr})(Q), iA]^{'}g\rangle\, ,\nonumber
\end{eqnarray}
where $\langle f\, , \, [H_0, iA]^{'}g\rangle = \langle f\, ,\, 2H_0 g\rangle$, $\langle f\, , \, [V_{lr}, iA]^{'}g\rangle = 
-\langle f\, ,\, Q\cdot(\nabla V_{lr})(Q)g\rangle$, 
\begin{eqnarray}
\langle f\, , \, [V_{sr}(Q), iA]^{'}g\rangle & =& \langle V_{sr}(Q)Qf\, ,\, iPg\rangle \, +\, \langle iPf\, ,\,
V_{sr}(Q)Qg\rangle\nonumber\\
\label{eq:commut-v-sr}&&\hspace{.4cm} +\, d\langle f\, ,\, V_{sr}(Q)g\rangle \, ,\\
\langle f\, , \, [V_c(Q), iA]^{'}g\rangle & =& \langle V_c(Q)f\, ,\, \cchi _c(Q)Q\cdot iPg\rangle\, +\,
\langle \cchi _c(Q)Q\cdot iPf\, ,\, V_c(Q)g\rangle\nonumber\\
\label{eq:commut-v-c}&&\hspace{.4cm}+\, d\langle f\, ,\, V_c(Q)g\rangle \, ,\\
\langle f\, , \, [(v\cdot\nabla\tilde{V}_{sr})(Q), iA]^{'}g\rangle&=& \langle \tilde{V}_{sr}(Q)f\, ,\, 
(P\cdot v(Q))(Q.P+2^{-1}d)g\rangle\nonumber\\
\label{eq:commut-v-tilde-sr}&&\, +\, \langle (P\cdot v(Q))(Q.P+2^{-1}d)f\, ,\,
\tilde{V}_{sr}(Q)g\rangle\\
\langle f\, , \, [W_{\alpha \beta}(Q), iA]^{'}g\rangle & =& \langle W_{\alpha \beta}(Q)Qf\, ,\, iPg\rangle \, +\,
\langle iPf\, ,\, W_{\alpha \beta}(Q)Qg\rangle\nonumber\\
\label{eq:commut-w-alpha-beta}&&\hspace{.4cm}+\, d\langle f\, ,\, W_{\alpha \beta}(Q)g\rangle \, .
\end{eqnarray}
Here $\langle V_{sr}(Q)Qf\, ,\, iPg\rangle$ means $\sum_{j=1}^d\langle V_{sr}(Q)Q_jf\, ,\, iP_jg\rangle$. 

Thanks to Assumption~\ref{a:cond-pot}, we see that the forms $[V_{sr}(Q), iA]$, $[V_c(Q), iA]$, 
$[(v\cdot\nabla\tilde{V}_{sr})(Q), iA]^{'}$, and $[V_{lr}(Q), iA]$ are bounded on $\Fc$ and associated to a compact 
operator from $\Fc$ to its dual $\Fc '$, for $\Fc$ given by
$\Hc^1(\R^d )$, $\Hc^2(\R^d )$, $\Hc^2(\R^d )$ again, and $\rL^2(\R^d )$, respectively. In particular, 
\eqref{eq:bound-form-h2} holds true with $H$ replaced by $H-W_{\alpha \beta}(Q)$. This proves that 
$H-W_{\alpha \beta}(Q)\in\Cc^1(A)$. 

\begin{proposition}\label{p:not-c1}
Assume Assumption~\ref{a:cond-pot} with $w\neq 0$ and $|\alpha -1|+\beta <1$. Then $H\not\in\Cc^1(A)$. 
\end{proposition}  
\begin{rem}\label{r:mourre-non}
The Mourre theory with conjugate operator $A$ requires a $C^{1,1}(A)$ regularity for $H$, a regularity that is stronger than the 
$\Cc^1(A)$ regularity (cf. \cite{abg}, Section 7). Thus this Mourre theory cannot be applied to prove our Theorem~\ref{th:main}, 
by Proposition~\ref{p:not-c1}. \\
As mentioned in Remark~\ref{r:W-regu}, Theorem~\ref{th:tal-abg} applies if $|\alpha -1|+\beta >1$. In fact, the proof of this 
theorem relies on the fact that, in that case, $H$ has actually the $C^{1,1}(A)$ regularity. \\
According to \cite{mar}, $H$ would have the $C^{1,1}(A')$ regularity for some other conjugate operator $A'$ 
if $2\alpha +\beta>3$. \\
Concerning the proof of the absence of positive eigenvalue in \cite{fhhh2,fh}, it is assumed in those papers that 
\eqref{eq:bound-form-h2} holds true for $H$ replaced by $V$. Proposition~\ref{p:not-c1} shows that this assumption is not 
satisfied if $|\alpha -1|+\beta <1$. In particular, our Theorem~\ref{th:no-posivite-eigenvalue} is not covered by the results in 
\cite{fhhh2,fh}. \\
If $|\alpha -1|+\beta <1$, the form $[H, A]$ is not bounded from $\Hc^2$ to $\Hc^{-2}$. However, we shall prove in 
Proposition~\ref{p:esti-mourre} that, for appropriate function $\theta$, the form $\theta (H)[H, A]\theta (H)$ does 
extend to a bounded one on $\rL^2(\R^d)$. This will give a meaning to the Mourre estimate and we shall prove its validity. 
Although $H\not\in\Cc^1(A)$, we shall be able to prove the ``virial theorem'' (see Proposition~\ref{p:viriel}). \\
Finally, we note that the proof of Theorem 4.15 in \cite{gj2} (and also the one of our Theorem~\ref{th:tal-Wigner-gj2}) 
uses at the very begining that $H\in\Cc^1(A)$. We did not see how to modify this proof when $H\not\in\Cc^1(A)$. This explains 
why we chose to use the ideas of \cite{gj} to prove Theorem~\ref{th:main} (see Section~\ref{s:lap}). 
\end{rem}
\Pfof{Proposition~\ref{p:not-c1}} Thanks to the considerations preceeding Proposition~\ref{p:not-c1}, we know that 
$H-W_{\alpha \beta}(Q)\in\Cc^1(A)$. Thus, for $w\neq 0$, $H\in\Cc^1(A)$ if and only if the bound \eqref{eq:bound-form-h2} holds 
true with $H$ replaced by $W_{\alpha \beta}(Q)$.\\ 
Let $w\neq 0$ and $(\alpha ; \beta )$ such that $2|\alpha -1|+\beta <1$. Let $\epsilon\in ]2|\alpha -1|; 1-\beta +|\alpha -1|]$. 
We set, for all $x\in\R^d$, 
\begin{eqnarray*}
 &&f(x)\, =\, \bigl(1-\kappa (|x|)\bigr)\cdot |x|^{|\alpha -1|-2^{-1}(d+\epsilon)}\\
 \mbox{and}\hspace{.4cm}&&g(x)\, =\, -\bigl(1-\kappa (|x|)\bigr)\cdot |x|^{1-\alpha-2^{-1}(d+\epsilon)}\cdot
 \cos\bigl(k|x|^\alpha\bigr)\, .
\end{eqnarray*}
Notice that $f\in\Hc^2$, $f\in\Dc(Q\cdot P)=\Dc (A)$, and  $g\in\Hc^2$. 
Furthermore, there exists $f_1\in\rL^2(\R^d)$ such that, for all $x\in\R^d$, 
\[x\cdot\nabla g (x)\ =\ f_1(x)\, +\, k\alpha\bigl(1-\kappa (|x|)\bigr)|x|^{1-2^{-1}(d+\epsilon)}\cdot 
\sin\bigl(k|x|^\alpha\bigr)\, . \]
For $n\in\N^\ast$, let $g_n : \R^d\dans\R$ be defined by $g_n(x)=\kappa (n^{-1}|x|)g(x)$. 
It belongs to $\Hc^2(\R^d)$. By the dominated convergence theorem, the sequence $(g_n)_n$ converges to $g$ in $\Hc^2(\R^d)$.  
Moreover the following limits exist and we have 
\begin{eqnarray*}
 \langle iPf\, ,\, W_{\alpha \beta}(Q)Qg\rangle&=& \lim_{n\to\infty}\langle iPf\, ,\, W_{\alpha \beta}(Q)Qg_n\rangle\\
 \hspace{.4cm}\mbox{and}\hspace{.4cm}\, \langle f\, ,\, W_{\alpha \beta}(Q)g\rangle&=& \lim_{n\to\infty}\langle f\, ,\, 
W_{\alpha \beta}(Q)g_n\rangle \, .
\end{eqnarray*}
By the previous computation, 
\[\langle W_{\alpha \beta}(Q)Qf\, ,\, iPg_n\rangle \ =\ \langle W_{\alpha \beta}(Q)f\, ,\, if_1\rangle\, +\, o(1)\]
\[\, +\, wk\alpha\int_{\R^d}\kappa (n^{-1}|x|)\bigl(1-\kappa (|x|)\bigr)^3|x|^{1-\beta +|\alpha -1|-(d+\epsilon)}\cdot 
\sin^2\bigl(k|x|^\alpha\bigr)\, dx\, ,\]
as $n\to\infty$. By the monotone convergence theorem, the above integrals tend to 
\begin{equation}\label{eq:int-alpha-beta}
\int_{\R^d}\bigl(1-\kappa (|x|)\bigr)^3|x|^{1-\beta +|\alpha -1|-(d+\epsilon)}\cdot 
\sin^2\bigl(k|x|^\alpha\bigr)\, dx\, ,
\end{equation}
as $n\to\infty$. By Lemma~\ref{l:infinite-int}, the integral \eqref{eq:int-alpha-beta} is infinite. If 
\eqref{eq:bound-form-h2} would hold 
true with $H$ replaced by $W_{\alpha \beta}(Q)$, the sequence 
\[
\bigl(\langle f, [W_{\alpha \beta}(Q), iA]g_n\rangle\bigr)_n\]
would converge. Therefore the integral \eqref{eq:int-alpha-beta} 
would be finite, by \eqref{eq:commut-w-alpha-beta}. Contradiction. Thus $H\not\in\Cc^1(A)$. \qed

\begin{figure}
\setlength{\unitlength}{1cm}
\begin{center}
\begin{picture}(8.5,6)(-0.5,-0.5)
\put(0,0){\vector(1,0){10}}
\put(0,0){\vector(0,1){5}}
\put(3,-0.1){\line(0,1){0.2}}
\put(-0.1,3){\line(1,0){0.2}}
\put(-0.1,1.5){\line(1,0){0.2}}
\put(-0.1,-0.5){0}
\put(2.9,-0.5){1}
\put(5.9,-0.5){2}
\put(-0.7,1.4){1/2}
\put(9.8,-0.5){$\alpha$}
\put(-0.4,2.9){1}
\put(-0.4,4.8){$\beta$}
\put(0,0){\line(1,1){3}}
\put(3,3){\line(1,-1){3}}
\color{red}
\put(2.8,1.7){red}
\put(1.5,0.8){red}
\put(4,0.8){red}
\color{blue}
\put(0.2,1){blue}
\put(1.4,4){blue}
\put(5.9,4){blue}
\put(5.9,1){blue}
\end{picture}
\end{center}
\caption{$H\in\Cc^{1,1}(A)$ in the \textcolor{blue}{blue} region; $H\not\in\Cc^1(A)$ in the \textcolor{red}{red} region.}
\label{dessin:3}
\end{figure}

In Fig.~\ref{dessin:3} , we summarised the above results. Note that the results of \cite{fhhh2,fh} on the 
absence of positive eigenvalue apply the blue region. 

Keeping $A$ as conjugate operator, we could try to apply another version of Mourre commutator method, namely the one that 
relies on ``local regularity'' (see \cite{sa}). 

Let us recall this type of regularity. Remember that a bounded operator $T$ belongs to $\Cc^1(A)$ 
if the map $t\donne \exp (itA)T\exp (-itA)$ is strongly $\Cc^1$ (cf. Appendix~\ref{app:s:regu}). We say that such an 
operator $T$ belongs to $\Cc^{1, u}(A)$ if the previous map is norm $\Cc^1$. 
Let $\Ic$ be an open subset of $\R$. We say that $H\in\Cc^1_\Ic(A)$ (resp. $H\in\Cc^{1, u}_\Ic(A)$) if, for any function 
$\varphi\in\Cc^\infty_c(\R; \C)$ with support in $\Ic$, $\varphi (H)\in\Cc^1(A)$ (resp. $\Cc^{1, u}(A)$). 
The Mourre theory with ``local regularity'' requires some $\Cc^{1+0}_\Ic(A)$ regularity, that is stronger than the 
$\Cc^{1, u}_\Ic(A)$, to prove the LAP inside $\Ic$. In our situation, we focus on open, relatively compact interval 
$\Ic\subset ]0; +\infty[$ and denote by $\overline{\Ic}$ the closure of $\Ic$. We first recall a result in \cite{gj2}. 
\begin{proposition}\label{p:not-c1u-I}\cite{gj2}.
Assume Assumption~\ref{a:cond-pot} with $w\neq 0$, $\alpha =\beta =1$, and $\tilde{V}_{sr}=V_c=0$. Then, for any open interval 
$\Ic\subset\overline{\Ic}\subset ]0; +\infty[$, $H\not\in\Cc^{1, u}_\Ic(A)$. 
\end{proposition}  
\begin{rem}\label{r:result-gj2}
Note that, in the framework of Proposition~\ref{p:not-c1u-I}, $H\in\Cc^1(A)$. This implies (cf. \cite{gj2}) that, for any 
open interval $\Ic\subset\overline{\Ic}\subset ]0; +\infty[$, $H\in\Cc^1_\Ic(A)$. But, since the $\Cc^{1+0}_\Ic(A)$ regularity 
is not available, the Mourre theory with conjugate operator $A$, that is developped in \cite{sa}, cannot apply. \\
We believe that Proposition~\ref{p:not-c1u-I} still holds true for nonzero $\tilde{V}_{sr}$ and $V_c$.
\end{rem}
\begin{proposition}\label{p:not-c1-I}
Assume Assumption~\ref{a:cond-pot} with $V_c=\tilde{V}_{sr}=0$, $w\neq 0$, $\alpha =1$, $\beta\in ]1/2; 1[$, and $\rho _{lr}>1/2$. 
Then, for any open interval $\Ic\subset\overline{\Ic}\subset ]0; +\infty[$, $H\not\in\Cc^1_\Ic(A)$. 
\end{proposition}  
\begin{rem}\label{r:mourre-local-non}
By Proposition~\ref{p:not-c1-I}, the Mourre theory with local regularity w.r.t. the conjugate operator $A$ cannot be 
applied to recover Theorem~\ref{th:main} in the region $\Vc\cap\{(1; \beta ); 0<\beta <1\}$. \\
The proof of Proposition~\ref{p:not-c1-I} below is close to the one of Proposition~\ref{p:not-c1u-I} in \cite{gj2}. 
Since $H\not\in\Cc^1(A)$, we need however to be a little bit more careful. 
\end{rem}
\Pfof{Proposition~\ref{p:not-c1-I}} We proceed by contradiction. Assume that, for some open interval interval 
$\Ic\subset\overline{\Ic}\subset ]0; +\infty[$, $H\in\Cc^1_\Ic(A)$. Then, for all $\varphi\in\Cc^\infty_c(\R; \C)$ with support in $\Ic$, 
$\varphi (H)\in\Cc^1(A)$, by definition. Take such a function $\varphi$. Since $H_0\in\Cc^1(A)$, $\varphi (H_0)\in\Cc^1(A)$. 
Therefore, the form $[\varphi (H)-\varphi (H_0), iA]$ extends to a bounded form on $\rL^2(\R^d)$. We shall show that, for some bounded 
operator $B$ and $B'$ on $\rL^2()\R^d$, the form $B[\varphi (H)-\varphi (H_0), iA]B'$ coincides, modulo a bounded 
form on $\rL^2(\R^d)$, with the form associated to a pseudodifferential operator $c^w$ w.r.t.  
the metric $g_0$ (cf. \eqref{eq:metric}), the symbol of which, $c$, is not bounded. By \eqref{eq:caract-pseudo-borne}, $c^w$ is not 
bounded and we arrive at the desired contradiction. \\
Let $f, g$ be functions in the Schwartz space $\Sch (\R^d; \C)$ on $\R^d$. We write 
\begin{eqnarray*}
\langle f\, ,\, Cg\rangle&:=& \langle f\, ,\, [\varphi (H)-\varphi (H_0), iA]g\rangle\\
&=&\bigl\langle \bigl(\varphi (H)^\ast-\varphi (H_0)^\ast\bigr)f\, ,\, iAg\bigr\rangle\, -\, 
\bigl\langle Af\, ,\, i\bigl(\varphi (H)-\varphi (H_0)\bigr)g\bigr\rangle\, .
\end{eqnarray*}
Now, we use \eqref{eq:int} with $k=0$ and the resolvent formula to get 
\begin{eqnarray*}
\langle f\, ,\, Cg\rangle&=&\int_{\C}\partial_{\bar{z}}\varphi^\C(z)\Bigl\{\bigl\langle (\bar{z}-H)^{-1}V(Q)(\bar{z}-H_0)^{-1}f
\, ,\, iAg\bigl\rangle\\
&&\hspace{.4cm}-\, \bigl\langle Af\, ,\, i(z-H)^{-1}V(Q)(z-H_0)^{-1}g\bigr\rangle\Bigr\}dz\wedge d\bar{z}\, .
\end{eqnarray*}
Recall that $V=V_{sr}+W$ with $W=V_{lr}+W_{1\beta}$. 
Using \eqref{eq:bound-resolv-weights}, we can find a bounded operator $B_1$ such that 
\begin{eqnarray*}
\langle f\, ,\, (C-B_1)g\rangle&=&\int_{\C}\partial_{\bar{z}}\varphi^\C(z)\Bigl\{\bigl\langle (\bar{z}-H)^{-1}W(Q)(\bar{z}-H_0)^{-1}f
\, ,\, iAg\bigl\rangle\\
&&\hspace{.4cm}-\, \bigl\langle Af\, ,\, i(z-H)^{-1}W(Q)(z-H_0)^{-1}g\bigr\rangle\Bigr\}dz\wedge d\bar{z}\, .
\end{eqnarray*}
Using again the resolvent formula and \eqref{eq:bound-resolv-weights} and the fact that $2\beta_{lr}>1$, we can find another bounded 
operator $B_2$ such that 
\begin{eqnarray}
\langle f\, ,\, (C-B_2)g\rangle&=&\int_{\C}\partial_{\bar{z}}\varphi^\C(z)\Bigl\{\bigl\langle (\bar{z}-H_0)^{-1}W(Q)(\bar{z}-H_0)^{-1}f
\, ,\, iAg\bigl\rangle\nonumber\\
&&\hspace{.4cm}-\, \bigl\langle Af\, ,\, i(z-H_0)^{-1}W(Q)(z-H_0)^{-1}g\bigr\rangle\Bigr\}dz\wedge d\bar{z}\, .\label{eq:comm-phi-A}
\end{eqnarray}
Since the form $[V_{lr}(Q), iA]$ is bounded from $\Hc^2$ to $\Hc^{-2}$, $H_1:=H_0+V_{lr}(Q)$ has the $\Cc^1(A)$ regularity. Therefore, 
we can redo the above computation with $H$ replaced by $H_1$ to see that the contribution of $V_{lr}$ in \eqref{eq:comm-phi-A} is 
actually bounded. Thus, for some bounded operator $B_3$, 
\begin{eqnarray*}
\langle f\, ,\, (C-B_3)g\rangle&=&\int_{\C}\partial_{\bar{z}}\varphi^\C(z)\Bigl\{\bigl\langle (\bar{z}-H_0)^{-1}W_{1\beta}(Q)(\bar{z}-H_0)^{-1}f
\, ,\, iAg\bigl\rangle\\
&&\hspace{.4cm}-\, \bigl\langle Af\, ,\, i(z-H_0)^{-1}W_{1\beta}(Q)(z-H_0)^{-1}g\bigr\rangle\Bigr\}dz\wedge d\bar{z}\, .
\end{eqnarray*}
Recall that $W_{1\beta}=w(2i)^{-1}(e_+-e_-)$, where $e_\pm=e_\pm^\alpha$ is given by \eqref{eq:def-e_pm-alpha} with $\alpha =1$. 
Let $\cchi _\beta : [0; +\infty[\dans\R$ be a smooth function such that $\cchi _\beta =0$ near $0$ and $\cchi _\beta (t)=t^{-\beta}$ when 
$t$ belongs to the support of $1-\kappa$. Thus, $\langle f\, ,\, (C-B_3)g\rangle$ is 
\begin{eqnarray*}
&=&-\frac{w}{2}\sum_{\sigma\in\{\pm 1\}}\sigma\int_{\C}\partial_{\bar{z}}\varphi^\C(z)\Bigl\{\bigl\langle 
e_\sigma (Q)(\bar{z}-H_0)^{-1}f
\, ,\, \cchi _\beta(|Q|)(z-H_0)^{-1}Ag\bigl\rangle\\
&&\hspace{.4cm}-\, \bigl\langle \cchi _\beta(|Q|)(\bar{z}-H_0)^{-1}Af\, ,\, e_\sigma (Q)(z-H_0)^{-1}g\bigr\rangle\Bigr\}dz\wedge d\bar{z}\, .
\end{eqnarray*}
Now, we use the arguments of the proof of Lemma 5.5 in \cite{gj2} to find a symbol $b\in S(1; g_0)$ such that, for $B'=e^{ik|Q|}$, 
for all $f, g\in\Sch (\R^d; \C)$, $\langle f\, ,\, b^w(C-B_3)B'g\rangle =\langle f\, ,\, c^wg\rangle$, where $c$ is unbounded. 
Actually, there exist $\xi\in\R^d$, $R>0$ and $C>0$ such that $|c(x; \xi )|\geq C|x|^{1-\beta}$, for $|x|\geq R$. \qed

%%%%%%%%%%%%%%%%%%%%%%%%
\section{The Mourre estimate.} 
\label{s:mourre}
\setcounter{equation}{0}
%%%%%%%%%%%%%%%%%%%%%%%%

In this section, we establish a Mourre estimate for the operator $H$ near appropriate positive energies. 
In the spirit of \cite{fh}, we deduce from it spacial decaying, polynomial bounds on the possible 
eigenvectors of $H$ at that energies. Since $H$ does not have a good regularity w.r.t. 
the conjugate operator $A$ (cf. Section~\ref{s:regu}), the abstract setting of Mourre theory does not help much and 
we have to look more precisely at the structure of $H$. The properties derived in Section~\ref{s:oscillations}
play a key role in the result.

Still working under Assumption~\ref{a:cond-pot}, we shall modify, only in the case $\alpha =1$, 
Assumption~\ref{a:cond-alpha-beta-gene} by requiring the following
\begin{assumption}\label{a:cond-alpha-beta} 
Let $\alpha , \beta >0$. Recall that $\beta _{lr}=\min (\beta ; \rho _{lr})$. Unless $|\alpha -1|+\beta >1$, 
we take $\alpha\geq 1$ and we take $\beta$
and $\rho_{lr}$ such that $\beta +\beta _{lr}>1$ or, equivalently, $\beta>1/2$ and $\rho_{lr}>1-\beta$. 
We consider a compact interval $\Jc$ such that $\Jc\subset ]0; +\infty[$, except when $\alpha =1$ and 
$\beta\in ]1/2; 1]$, and, in the latter case, 
we consider a small enough, compact interval $\Jc$ such that $\Jc\subset ]0; k^2/4[$. 
\end{assumption}
\begin{rem}\label{r:smallness}
Assumption~\ref{a:cond-alpha-beta} is identical to Assumption~\ref{a:cond-alpha-beta-gene}, except for the change of 
the name of the interval and for the smallness requirement when $\alpha =1$ and $\beta\in ]1/2; 1]$. We actually need 
to work in a slightly larger interval $\Jc$ than the interval $\Ic$ considered in Theorem~\ref{th:main}. In the case 
$\alpha =1$ and $\beta\in ]1/2; 1]$, the smallness of $\Jc$ (and thus of the above $\Ic$) is the one that matches the 
smallness required in Proposition~\ref{p:oscillations-energy-alpha-leq-1}. It depends only on the distance of the 
middle point of $\Jc$ to $k^2/4$. 
\end{rem}

As pointed out in Section~\ref{s:regu}, the form $[H, A]$ does not extend to a bounded form from $\Hc^2$ to $\Hc^{-2}$ for a 
certain range of the parameters $\alpha$ and $\beta$. Thus, given a function $\theta\in\Cc_c^\infty(\R; \C)$, we do not know 
a priori if the forms $\theta (H)[H, iA]\theta (H)$ and $\theta (H)[H, iA]'\theta (H)$ extend to a bounded one on $\rL^2$. 
Recall that $[H, iA]'$ is defined in \eqref{eq:commutateur}. Nethertheless these two forms are well defined and coincide 
on $\Dc(\langle Q\rangle)$, by Lemma~\ref{l:Q-regu}. By Section~\ref{s:regu} again, we know that the difficulty is 
concentrate in the contribution of the oscillating potential $W_{\alpha\beta}$, namely \eqref{eq:commut-w-alpha-beta}. 
Thanks to the interaction between the oscillations and the kinetic operator, we are able to show the following 
\begin{proposition}\label{p:oscillations-op-compact}
Under Assumptions~\ref{a:cond-pot} and~\ref{a:cond-alpha-beta}, let $\theta\in\Cc_c^\infty(\R; \R)$ with support inside 
$\mathring{\Jc}$, the interior of $\Jc$, the form $\theta (H)[W_{\alpha \beta}(Q), iA]\theta (H)$ 
extends to a bounded form on $\rL^2(\R^d)$ that is associated
to a compact operator.
\end{proposition}  
\begin{rem}\label{r:alpha>1-compact}
In dimension $d=1$ with $\alpha =1$, the result still holds true if the function $\theta$ is supported inside 
$]0; +\infty[\setminus\{k^2/4\}$. 
\end{rem}
Our proof of Proposition~\ref{p:oscillations-op-compact} relies on Propositions~\ref{p:oscillations-energy-alpha-leq-1},
~\ref{p:oscillations-energy-alpha>1}, and on the following
\begin{lemma}\label{l:diff-decroit-Q} Assume Assumptions~\ref{a:cond-pot} and~\ref{a:cond-alpha-beta-gene} satisfied.
Let $\theta\in\Cc_c^\infty(\R; \C)$. Then $\langle Q\rangle^{\beta _{lr}}
(\theta (H)-\theta (H_0))$ and $\langle Q\rangle^{\beta _{lr}}P(\theta (H)-\theta (H_0))$ are bounded on $\rL^2(\R^d)$.
\end{lemma}
\proof See Lemma~\ref{l:diff-decroit-Q-bis}. \qed

\Pfof{Proposition~\ref{p:oscillations-op-compact}} It suffices to study the form $\theta (H)[W_{\alpha \beta}(Q), iA]'\theta (H)$, 
where $[W_{\alpha \beta}(Q), iA]'$ is defined in \eqref{eq:commut-w-alpha-beta}.\\
Consider first the case where $|\alpha -1|+\beta >1$. By Remark~\ref{r:W-regu}, the form $[W_{\alpha \beta}(Q), iA]'$ 
is of one of the types $[V_{lr}(Q), iA]'$, \eqref{eq:commut-v-sr}, and \eqref{eq:commut-v-tilde-sr}. It is thus compact from $\Hc^2$ to $\Hc^{-2}$. Since 
$\langle P\rangle^2\theta (H)$ is bounded, the form $\theta (H)[W_{\alpha \beta}(Q), iA]'\theta (H)$ extends to a bounded 
one on $\rL^2(\R^d)$, that is associated to a compact operator on $\rL^2(\R^d)$.\\
We assume now that $|\alpha -1|+\beta \leq 1$. Since $\beta>0$, the form $\theta (H)W_{\alpha \beta}(Q)\theta (H)$ 
extends to a bounded form associated to a compact operator. We study the form $(f, g)\mapsto \langle P\theta (H)f\, ,\, 
W_{\alpha \beta}(Q)Q\theta (H)g\rangle$, the remainding term being treated in a similar way. 
We write this form as
\begin{eqnarray}
&\hspace{-1.5cm}\theta (H)P\cdot QW_{\alpha \beta}(Q)\theta (H)\ =\ \bigl(\theta (H)-\theta (H_0)\bigr)P\cdot QW_{\alpha \beta}(Q)
\bigl(\theta (H)-\theta (H_0)\bigr)\nonumber\\
&\hspace{.4cm}+\, \bigl(\theta (H)-\theta (H_0)\bigr)P\cdot QW_{\alpha \beta}(Q)\theta (H_0)\nonumber\\
&\hspace{.4cm} +\, \theta (H_0)P\cdot QW_{\alpha \beta}(Q)
\bigl(\theta (H)-\theta (H_0)\bigr)\label{eq:H-H_0}\\
&+\ \theta (H_0)P\cdot QW_{\alpha \beta}(Q)\theta (H_0)\, .\nonumber
\end{eqnarray}
Using Lemma~\ref{l:diff-decroit-Q} and the fact that $\beta +\beta _{lr}-1>0$, we see that
the first three terms on the r.h.s. of \eqref{eq:H-H_0} extends to a compact operator.
So does also the last term, by Proposition~\ref{p:oscillations-energy-alpha-leq-1} with $\epsilon =1-\beta$, if
$\alpha =1$, and by Proposition~\ref{p:oscillations-energy-alpha>1} with $p=1-\beta$, if $\alpha >1$. \qed

Now, we are in position to prove the Mourre estimate.
\begin{proposition}\label{p:esti-mourre}
Under Assumptions~\ref{a:cond-pot} and~\ref{a:cond-alpha-beta}, let $\theta\in\Cc_c^\infty(\R , \R)$ with support inside
the interior $\mathring{\Jc}$ of the interval $\Jc$. Denote by $c>0$ the infimum of $\Jc$.
Then the form $\theta (H)[H, iA]\theta (H)$ extends to a bounded one on $\rL^2(\R^d)$ and there exists 
a compact operator $K$ on $\rL^2(\R^d)$ such that 
\begin{equation}\label{eq:esti-mourre}
\theta (H)[H, iA]\theta (H)\ \geq\ 2c\, \theta (H)^2\, +\, K\, . 
\end{equation}
\end{proposition}  
\proof Let $K_0$ be the operator associated with the form 
\begin{eqnarray*}
&&\theta (H)[V_{sr}(Q), iA]\theta (H)\, +\, \theta (H)[(v\cdot\nabla\tilde{V}_{sr})(Q), iA]^{'}\theta (H)\, +\, 
\theta (H)[V_{lr}(Q), iA]\theta (H)\\
&&\, +\, \theta (H)[V_c(Q), iA]\theta (H)\, +\, \theta (H)[W_{\alpha \beta}(Q), iA]\theta (H)\, .
\end{eqnarray*}
It is compact by Section~\ref{s:regu} and Proposition~\ref{p:oscillations-op-compact}. Thus, as forms, 
\[\theta (H)[H, iA]\theta (H)\ =\ \theta (H)[H_0, iA]\theta (H)\, +\, K_0\, .\]
Since $[H_0, iA]=2H_0$, the form 
\[\bigl(\theta (H)-\theta (H_0)\bigr)[H_0, iA]\theta (H)\, +\, \theta (H_0)[H_0, iA]\bigl(\theta (H)-\theta (H_0)\bigr)\]
is associated to a compact operator $K_1$, by Lemma~\ref{l:diff-decroit-Q}, and
\begin{eqnarray*}
\theta (H)[H, iA]\theta (H)&=& \theta (H_0)[H_0, iA]\theta (H_0)\, +\, K_0\, +\, K_1\\
&\geq &2c\, \theta (H_0)^2\, +\, K_0\, +\, K_1\\
&\geq &2c\, \theta (H)^2\, +\, K_0\, +\, K_1\, +\, K_3\, , 
\end{eqnarray*}
with compact $K_3=2c(\theta (H_0)^2-\theta (H)^2)$. \qed
%

%%%%%%%%%%%%%%%%%%%%%%%%
\section{Polynomial bounds on possible eigenfunctions with positive energy.}
\label{s:poly-bounds}
\setcounter{equation}{0}
%%%%%%%%%%%%%%%%%%%%%%%%

In this section, we shall show a polynomially decaying bound on the possible eigenfunctions of $H$ with positive energy.
Because of the oscillating behaviour of the potential $W_{\alpha \beta}$, the corresponding
result in \cite{fh} does not apply (cf. Section~\ref{s:regu}) but it turns out that one can adapt the arguments from
\cite{fh} to the present situation. We note further that the abstract results in \cite{ca,cgh}
cannot be applied here because of the lack of regularity w.r.t. the generator of dilations (cf. Section~\ref{s:regu}).

\begin{proposition}\label{p:decroissace-poly}
Under Assumptions~\ref{a:cond-pot} and~\ref{a:cond-alpha-beta}, let $E\in\mathring{\Jc}$ and $\psi\in\Dc(H)$ such that $H\psi =E\psi$.
Then, for all $\lambda \geq 0$, $\psi\in\Dc(\langle Q\rangle^\lambda)$ and $\nabla\psi\in\Dc(\langle Q\rangle^\lambda)$. 
\end{proposition}
\begin{corollary}\label{c:regu-A}
Under Assumptions~\ref{a:cond-pot} and~\ref{a:cond-alpha-beta}, for $E\in\mathring{\Jc}$, ${\rm Ker}(H-E)\subset\Dc (A)$.
\end{corollary}
\proof Let $\psi\in {\rm Ker}(H-E)$. By Proposition~\ref{p:decroissace-poly}, $\nabla\psi\in\Dc(\langle Q\rangle)$ thus
$\psi\in\Dc (A)$. \qed

\Pfof{Proposition~\ref{p:decroissace-poly}} We take a function $\theta\in\Cc_c^\infty(\R ; \R)$ with support inside $\mathring{\Jc}$ such that
$\theta (E)=1$. By Proposition~\ref{p:esti-mourre}, the Mourre estimate \eqref{eq:esti-mourre} holds true. \\
Now we follow the beginning of the proof of Theorem 2.1 in \cite{fh}, making appropriate adaptations. 
For $\lambda\geq 0$ and $\epsilon>0$, we consider the function $F : \R^d\to\R$ defined by $F(x)=\lambda \ln
(\langle x\rangle (1+\epsilon\langle x\rangle)^{-1})$. For all $x\in\R^d$, $\nabla F(x)=g(x)x$ with $g(x)=\lambda \langle x\rangle^{-2}
(1+\epsilon\langle x\rangle)^{-1}$. Let $H(F)$ be the operator defined on the domain $\Dc (H(F)):=\Dc (H_0)=\Hc^2(\R^d)$ by
\begin{equation}\label{eq:def-H-F}
 H(F)\ =\ e^{F(Q)}He^{-F(Q)}\ =\ H-|\nabla F|^2+(iP\cdot \nabla F + \nabla F\cdot iP)\, .
\end{equation}
Setting $\psi _F=e^{F(Q)}\psi$, one has $\psi_F\in\Dc (H_0)$,
$H(F)\psi_F=E\psi_F$, and $\langle \psi _F , H\psi_F\rangle = \langle \psi _F , (|\nabla F|^2+E)\psi_F\rangle$.\\
Note that, since $e^F$ does not contain decay in $\langle x\rangle$, we a priori need some argument to give a meaning to
$\langle \psi _F, [H, iA]\psi _F\rangle$ when $\beta <1$, because of the contribution of $W_{\alpha\beta}$ in \eqref{eq:commutateur}.\\
Let $\cchi\in\Cc_c^\infty(\R ; \R)$ with $\cchi =1$ near $0$ and, for $R\geq 1$, let $\cchi _R(t)=\cchi (t/R)$. To replace
Equation (2.9) in \cite{fh}, we claim that
\begin{eqnarray}\label{eq:commutateur-psi-F}
\lim_{R\to +\infty}\langle \cchi_R(\langle Q\rangle)\psi_F\, ,\, [H\, ,\, iA]\cchi_R(\langle Q\rangle)\psi_F\rangle &=&
-4\cdot\bigl\|g(Q)^{1/2}A\psi_F\bigr\|^2\\
&&\, +\,
\bigl\langle \psi_F\, ,\, G(Q)\psi_F\bigr\rangle\, ,\nonumber
\end{eqnarray}
where $G : \R^d\ni x\donne ((x\cdot\nabla)^2g)(x)-(x\cdot\nabla|\nabla F|^2)(x)$. Notice that
$\cchi_R(\langle Q\rangle)\psi_F\in\Dc (\langle Q\rangle\langle P\rangle)$, so the bracket on the l.h.s. of \eqref{eq:commutateur-psi-F}
is well defined. Since, for $x\in\R^d$,  $|g(x)|\leq \lambda \langle x\rangle^{-1}$ and $|G(x)|=O(\langle x\rangle^{-2})$, so is the r.h.s.
By a direct computation, 
\begin{eqnarray}\label{eq:commutateur-partie-reelle-psi-F}
&&2\Re \bigl\langle A\cchi _R(\langle Q\rangle)\psi_F\, ,\, i(H(F)-E)\cchi_R(\langle Q\rangle)\psi_F\bigr\rangle\nonumber\\
&=&
-\bigl\langle \cchi_R(\langle Q\rangle)\psi_F\, ,\, [H\, ,\, iA]\cchi_R(\langle Q\rangle)\psi_F\bigr\rangle\ -\
4\cdot\bigl\|g(Q)^{1/2}A\cchi_R(\langle Q\rangle)\psi_F\bigr\|^2\nonumber\\
&&\, +\, \bigl\langle \cchi_R(\langle Q\rangle)\psi_F\, ,\, G(Q)\cchi_R(\langle Q\rangle)\psi_F\bigr\rangle\, .
\end{eqnarray}
Note that the commutator $[H(F)\, ,\, \cchi_R(Q)]_\circ$ is well-defined since $\cchi_R(Q)$ preserves the domain of $H(F)$.
Furthermore $[H(F)\, ,\, \cchi_R(Q)]_\circ=[H_0(F)\, ,\, \cchi_R(Q)]_\circ$, where $H_0(F)=e^{F(Q)}H_0e^{-F(Q)}$ is a pseudodifferential
operator. Notice that the l.h.s of \eqref{eq:commutateur-partie-reelle-psi-F} is given by
\[2\Re \bigl\langle A\cchi _R(Q)\psi_F\, ,\, i[H(F)\, ,\, \cchi_R(Q)]_\circ\psi_F\bigr\rangle\, .\]
Using an explicit expression for the commutator and the fact that the family of functions  
$x\donne \langle x\rangle\cchi _R'(\langle x\rangle)$ is bounded, uniformly w.r.t. $R$, and converges pointwise to $0$,
as $R\to +\infty$, we apply the
the dominated convergence theorem to see that the l.h.s. of \eqref{eq:commutateur-partie-reelle-psi-F} tends to $0$ and
that the last two terms in \eqref{eq:commutateur-partie-reelle-psi-F} converge to the r.h.s. of \eqref{eq:commutateur-psi-F}.
Thus the limit in \eqref{eq:commutateur-psi-F} exists and \eqref{eq:commutateur-psi-F} holds true. \\
Next we claim that 
\begin{eqnarray}
\lim_{R\to +\infty}\bigl\langle \cchi_R(\langle Q\rangle)\psi_F\, ,\, [H\, ,\, iA]\cchi_R(\langle Q\rangle)\psi_F\bigr\rangle &=&
\bigl\langle \theta (H)\psi_F\, ,\, [H\, ,\, iA]\theta (H)\psi_F\bigr\rangle\nonumber\\
&&+\, \bigl\langle \psi_F\, ,\, (K_1B_{1, \epsilon}+B_{2, \epsilon}K_2)\psi_F\bigr\rangle\label{eq:localisation-psi-F}
\end{eqnarray}
where, on $\rL^2(\R^d)$, $K_1$, $K_2$ are $\epsilon$-independent compact operators and
$B_{1, \epsilon}, B_{2, \epsilon}$ are bounded operators satisfying $\|B_{1, \epsilon}\|+\|B_{2, \epsilon}\|=O(\epsilon^0)$.
Notice that, by Proposition~\ref{p:esti-mourre}, the first term on the r.h.s of \eqref{eq:localisation-psi-F} is well defined
and equal to
\[\lim_{R\to +\infty}\bigl\langle \theta (H)\cchi_R(\langle Q\rangle)\psi_F\, ,\, [H\, ,\, iA]\theta (H)
\cchi_R(\langle Q\rangle)\psi_F\bigr\rangle\, .\]
Writing each $\cchi_R(\langle Q\rangle)\psi_F$ as $\cchi_R(\langle Q\rangle)\psi_F=\theta (H)\cchi_R(Q)+
(1-\theta (H))\cchi_R(\langle Q\rangle)\psi_F$, we split
$\langle \cchi_R(\langle Q\rangle)\psi_F\, ,\, [H\, ,\, iA]\cchi_R(\langle Q\rangle)\psi_F\rangle$ into four terms, one of them
tending to the first term on the r.h.s of \eqref{eq:localisation-psi-F}. We focus on the others.
Since $(1-\theta (H))\psi=0$,
\begin{eqnarray}
\hspace{.7cm}\bigl(1-\theta (H)\bigr)\cchi_R(\langle Q\rangle)\psi_F&=& -\, [\theta (H), \cchi _R(\langle Q\rangle)]_\circ\psi_F\nonumber\\
&&\, -\, \cchi _R(\langle Q\rangle)[\theta (H), e^{F(Q)}]_\circ\psi\, , \label{eq:mal-localise}\\
P\bigl(1-\theta (H)\bigr)\cchi_R(\langle Q\rangle)\psi_F&=& -\, P[\theta (H), \cchi _R(\langle Q\rangle)]_\circ\psi_F\nonumber\\
&&\, -\, [P, \cchi _R(\langle Q\rangle)][\theta (H), e^{F(Q)}]_\circ\psi\nonumber\\
&&-\, \cchi _R(\langle Q\rangle)P[\theta (H), e^{F(Q)}]_\circ\psi\, .
\label{eq:mal-localise-P}
\end{eqnarray}
\begin{lemma}\label{l:commut-fonct-Q-fonct-H}
Recall that $\beta _{lr}=\min (\beta ; \rho _{lr})\leq 1$. For intergers $1\leq i, j\leq d$, let $\tau (P)=1$, or 
$\tau (P)=P_i$, or $\tau (P)=P_iP_j$. 
\begin{itemize}
 \item[(1)] For $\sigma\in [0; 1]$, the operators
\[\langle Q\rangle^{1-\sigma}\tau (P)\bigl[\theta (H), e^{F(Q)}\bigr]_\circ e^{-F(Q)}\langle Q\rangle^\sigma\]
are bounded on $\rL^2(\R^d)$, uniformly w.r.t. $\epsilon\in ]0; 1]$. 
 \item[(2)] For $R\geq 1$, the operators
\[\langle Q\rangle^{1-\beta_{lr}}\tau (P)\bigl[\theta (H), \cchi _R(\langle Q\rangle)\bigr]_\circ\]
are bounded on $\rL^2(\R^d)$ and their norm are $O(R^{-\beta_{lr}})$.
\end{itemize}
\end{lemma}
\proof For the result (2), see the proof of Lemma~\ref{l:commut-fonct-Q-fonct-H-bis}.  \\
Let us prove (1). Making use of Helffer-Sj\"ostrand formula \eqref{eq:int} and of \eqref{eq:bound-resolv-weights}, for
$H'=H$, we can show by induction that, for all $j\in\N^\ast$,
\begin{equation}\label{eq:iterated-commut-bounded}
\langle Q\rangle^{1-\sigma}\cdot\ad_{\langle Q\rangle}^j\bigl(\theta (H)\bigr)\cdot\langle Q\rangle^\sigma
\end{equation}
is bounded on $\rL^2(\R^d)$. Note that the function $e^F$ can be written as $\varphi _\epsilon (\langle \cdot\rangle)$, where $\varphi _\epsilon$
stays in a bounded set in $\Sc^\lambda$, when $\epsilon$ varies in $]0; 1]$. Since $\theta (H)\in\Cc^\infty(\langle Q\rangle)$
(cf. Lemma~\ref{l:Q-regu}), we can apply Propositions~\ref{p:regu} with $B=\theta (H)$ and $k>\lambda +1$. By
\eqref{eq:iterated-commut-bounded}, the first terms are all bounded on $\rL^2(\R^d)$. Let us focus on the last one, that contains an integral.
Exploiting \eqref{eq:dg1} with $\ell =k+1$, \eqref{eq:dg2}, \eqref{eq:majoA}, \eqref{eq:iterated-commut-bounded}, and the fact that
$\varphi _\epsilon (\langle \cdot\rangle)$ is bounded below by $1/2$ for $\epsilon\in ]0; 1]$, we see that the last term is also
bounded on $\rL^2(\R^d)$. \qed

\Pfof{Proposition~\ref{p:decroissace-poly} continued} Using Lemma~\ref{l:commut-fonct-Q-fonct-H} and~\eqref{eq:mal-localise},
we get that 
\begin{eqnarray*}
&&\lim_{R\to +\infty}\bigl\langle \theta (H)\cchi_R(\langle Q\rangle)\psi_F\, ,\, P\cdot QW_{\alpha \beta}(Q)\bigl(1-\theta (H)\bigr)
\cchi_R(\langle Q\rangle)\psi_F\bigr\rangle \\
=&&\ -\, \bigl\langle K\psi_F\, ,\, W_{\alpha \beta}(Q)\langle Q\rangle^{\beta}Q\langle Q\rangle^{-1}\cdot\langle Q\rangle
[\theta (H), e^{F(Q)}]_\circ e^{-F(Q)}\psi_F\bigr\rangle
\end{eqnarray*}
where $K$ is an $\epsilon$-independent vector of compact operators and the bounded operator acting on the right $\psi _F$ is
uniformly bounded w.r.t. $\epsilon$. Similarly, using Lemma~\ref{l:commut-fonct-Q-fonct-H} and \eqref{eq:mal-localise-P},
we see that 
\begin{eqnarray*}
&&\lim_{R\to +\infty}\bigl\langle \theta (H)\cchi_R(\langle Q\rangle)\psi_F\, ,\, W_{\alpha \beta}(Q)Q\cdot P\bigl(1-\theta (H)\bigr)
\cchi_R(\langle Q\rangle)\psi_F\bigr\rangle \\
=&&\ -\, \bigl\langle K'\psi_F\, ,\, W_{\alpha \beta}(Q)\langle Q\rangle^{\beta}Q\langle Q\rangle^{-1}\cdot\langle Q\rangle
P[\theta (H), e^{F(Q)}]_\circ e^{-F(Q)}\psi_F\bigr\rangle
\end{eqnarray*}
with $K'$ compact and an uniformly bounded operator acting on the right $\psi _F$. Using again \eqref{eq:mal-localise} and
\eqref{eq:mal-localise-P}, we also get 
\begin{eqnarray*}
&&\lim_{R\to +\infty}\bigl\langle\bigl(1-\theta (H)\bigr)\cchi_R(\langle Q\rangle)\psi_F\, ,\, W_{\alpha \beta}(Q)Q\cdot
P\bigl(1-\theta (H)\bigr)\cchi_R(\langle Q\rangle)\psi_F\bigr\rangle \\
=&&\ \bigl\langle \langle Q\rangle^{-\beta/2}[\theta (H),e^{F(Q)}]_\circ e^{-F(Q)}\langle Q\rangle^{\beta/2}\langle P\rangle
K''\psi_F\, ,\\
&&\hspace{1.5cm} W_{\alpha \beta}(Q)\langle Q\rangle^{\beta}Q\langle Q\rangle^{-\beta /2}
P[\theta (H), e^{F(Q)}]_\circ e^{-F(Q)}\psi_F\bigr\rangle
\end{eqnarray*}
with compact $K''=\langle P\rangle^{-1}\langle Q\rangle^{-\beta/2}$ and uniformly bounded operators acting on the right $\psi _F$
and on $K''\psi _F$. \\
In a similar way, we can treat the last term in the contribution of $[W_{\alpha \beta}(Q), iA]'$ and the contribution of the forms 
$[H_0, iA]'$, $[V_{lr}(Q), iA]'$, $[V_{sr}(Q), iA]'$, $[V_c(Q), iA]'$, and $[(v\cdot\nabla\tilde{V}_{sr})(Q), iA]^{'}$
(cf. \eqref{eq:commutateur}, \eqref{eq:commut-v-sr}, \eqref{eq:commut-v-c}, \eqref{eq:commut-v-tilde-sr}). This ends 
the proof of \eqref{eq:localisation-psi-F}, yielding, together with \eqref{eq:commutateur-psi-F},
\begin{eqnarray}\label{eq:nouveau-fh}
\hspace{.8cm}\bigl\langle \theta (H)\psi_F\, ,\, [H\, ,\, iA]\theta (H)\psi_F\bigr\rangle&=& -4\cdot\bigl\|g(Q)^{1/2}A\psi_F\bigr\|^2\, +\,
\bigl\langle \psi_F\, ,\, G(Q)\psi_F\bigr\rangle\\
&&\hspace{.8cm}-\, \bigl\langle \psi_F\, ,\, (K_1B_{1, \epsilon}+B_{2, \epsilon}K_2)\psi_F\bigr\rangle\, .\nonumber
\end{eqnarray}
Assume that, for some $\lambda >0$, $\psi\not\in\Dc(\langle Q\rangle^{\lambda})$. 
We define $\Psi _\epsilon =\|\psi _F\|^{-1}\psi _F$. As in \cite{fh}, $(H_0+1)\Psi _\epsilon$ and thus $\Psi _\epsilon$ both go to $0$,
weakly in $\rL^2(\R^d)$, as $\epsilon\to 0$. Therefore $\|K_1\Psi _\epsilon\|+\|K_2\Psi _\epsilon\|\to 0$, as $\epsilon\to 0$.
Since $G(Q)(H_0+1)^{-1}$ is compact, $\|G(Q)\Psi _\epsilon\|\to 0$. Since $(1-\theta (H))\psi=0$,
\[(1-\theta (H))\Psi _\epsilon = \bigl[\theta (H), e^{F(Q)}\bigr]_\circ e^{-F(Q)}\langle Q\rangle \langle Q\rangle^{-1}(H_0+1)^{-1}
(H_0+1)\Psi _\epsilon\, .\]
Since $[\theta (H), e^{F(Q)}]_\circ e^{-F(Q)}\langle Q\rangle$ is uniformly bounded w.r.t. $\epsilon$, by
Lemma~\ref{l:commut-fonct-Q-fonct-H}, and
$\langle Q\rangle^{-1}(H_0+1)^{-1}$ is compact,
the weak convergence to $0$ of $(H_0+1)\Psi _\epsilon$ implies the norm convergence to $0$ of $(1-\theta (H))\Psi _\epsilon$.
Thus $\lim _{\epsilon\to 0}\|\theta (H)\Psi _\epsilon\|=1$. \\
Dividing by $\|\psi _F\|^2$ in \eqref{eq:nouveau-fh} and
then taking the ``$\liminf _{\epsilon\to 0}$'', we get
\[\liminf _{\epsilon\to 0}\bigl\langle \theta (H)\Psi _\epsilon\, ,\, [H\, ,\, iA]\theta (H)\Psi _\epsilon\bigr\rangle\ =\
-4\cdot\liminf _{\epsilon\to 0}\bigl\|g(Q)^{1/2}A\Psi _\epsilon\bigr\|^2\, \leq \, 0\, .\]
Now, we apply the Mourre estimate \eqref{eq:esti-mourre} to $\Psi _\epsilon$, yielding
\[\liminf _{\epsilon\to 0}\bigl\langle \theta (H)\Psi _\epsilon\, ,\, [H\, ,\, iA]\theta (H)\Psi _\epsilon\bigr\rangle\ \geq \
2c\, \liminf _{\epsilon\to 0}\|\theta (H)\Psi _\epsilon\|^2\, +\, 0\, =\, 2c\, >\, 0\]
and a contradiction. Therefore $\psi\in\Dc(\langle Q\rangle^{\lambda})$, for all $\lambda >0$.\\
Take $\lambda >0$. Since $V(Q)$ is $H_0$-bounded with relative bound $0$, we can find, for any $\delta\in]0; 1[$, some $C_\delta >0$
such that, for all $\epsilon>0$,
\[|\langle \psi_F\, ,\, V(Q)\psi _F\rangle |\ \leq \ \delta\langle \psi_F\, ,\, H_0\psi _F\rangle \, +\, C\, \|\psi_F\|^2
\ =\ \, \delta\|\nabla\psi_F\|^2\, +C\, \|\psi_F\|^2\, .\]
Using the equality $\langle \psi _F , H\psi_F\rangle = \langle \psi _F , (|\nabla F|^2(Q)+E)\psi_F\rangle$, we can find $C', C''>0$
such that, for all $\epsilon>0$,
\begin{equation}\label{eq:borne-gradient}
\|\nabla\psi_F\|^2\ \leq \ C'\|\psi_F\|^2\ \leq \, C'\|\langle Q\rangle^\lambda\psi\|^2\ =:\ (C'')^2\, .
\end{equation}
Now, $\nabla\psi _F=(\nabla F)(Q)\psi_F+e^{F(Q)}\nabla\psi$, yielding, for all $\epsilon>0$,
\[\|e^{F(Q)}\nabla\psi\|\ \leq \ C''\, +\, \|\psi_F\|\ \leq \ C''\, +\, \|\langle Q\rangle^\lambda\psi\|\, .\]
This shows that $\nabla\psi$ belongs to $\Dc (\langle Q\rangle^\lambda)$. \qed

%%%%%%%%%%%%%%%%%%%%%%%%
\section{Local finitness of the point spectrum.}
\label{s:finitness}
\setcounter{equation}{0}
%%%%%%%%%%%%%%%%%%%%%%%%

In the usual Mourre theory, one easily deduces from a Mourre estimate on some compact interval $\Jc$
the finitness of the point spectrum in any compact interval $\Ic\subset\mathring{\Jc}$, the interior of $\Jc$, thanks to the
virial Theorem. In the present situation, for some values of the parameters $\alpha$ and $\beta$, we do not have the 
required regularity of $H$ w.r.t. $A$ (cf. Section~\ref{s:regu}) to apply the abstract virial Theorem. But, thanks to 
Corollary~\ref{c:regu-A}, we are able to get it in a trivial way.

\begin{proposition}\label{p:viriel}
Under Assumptions~\ref{a:cond-pot} and~\ref{a:cond-alpha-beta}, let $E\in\mathring{\Jc}$ and $\psi\in\Dc(H)$ such that $H\psi =E\psi$.
Then $\langle \psi , [H, A]\psi\rangle = 0$.
\end{proposition}
\proof Since $\psi\in\Dc(A)$ by Corollary~\ref{c:regu-A}, $\langle \psi , [H, A]\psi\rangle$ is well defined and
\[\langle \psi \, , \, [H, A]\psi\rangle\ =\ \langle H\psi \, ,\, A\psi\rangle \, -\, \langle A\psi \, ,\, H\psi\rangle
\ =\ 0\, ,\]
because $E$ is real and $A$ is self-adjoint. \qed

Now, the Mourre estimate in Proposition~\ref{p:esti-mourre} gives the 
\begin{corollary}\label{c:finitness}
Under Assumptions~\ref{a:cond-pot} and~\ref{a:cond-alpha-beta}, for any compact interval $\Ic\subset\mathring{\Jc}$,
the point spectrum of $H$ inside $\Ic$ is finite (counted with multiplicity).
\end{corollary}
\proof One can follow the usual proof. See \cite{abg} p. 295 or \cite{m}, for instance. \qed

Thanks to Corollaries~\ref{c:regu-A} and~\ref{c:finitness}, we are able to prove the following regularity result.
The precise definition of the mentioned regularity is given in Appendix~\ref{app:s:regu}. 

\begin{corollary}\label{c:pi-C1}
Under Assumptions~\ref{a:cond-pot} and~\ref{a:cond-alpha-beta}, for any $\theta\in\Cc_c^\infty(\R ; \C)$ with support included in
$\mathring{\Jc}$, $\theta (H)\Pi\in\Cc^1(A)$ and $\theta (H)\Pi\in\Cc^\infty(\langle Q\rangle)$.
\end{corollary}
\proof For $\psi\in\Dc(A)$, the projector $\langle\psi , \cdot\rangle\psi$ belongs
to $\Cc^1(A)$ since the form
\[\Dc(A)^2\ni (\varphi_1; \varphi_2)\, \donne\, \bigl\langle\varphi _1, \bigl[\langle\psi , \cdot\rangle\psi\, ,\, A\bigr]\varphi_2\bigr\rangle\ =\
\overline{\langle\psi , \varphi _1\rangle}\langle A\psi , \varphi _2\rangle\, -\, \langle\psi ,  \varphi_2\rangle\,
\langle \varphi _1 , A\psi\rangle\]
extends to a bounded one. By Corollary~\ref{c:finitness},
the point spectrum of $H$ inside the support of $\theta$ is some $\{\lambda _1; \cdots ; \lambda _n\}$ and 
there exist $\psi _1, \cdots , \psi _n\in\Dc(H)$ such that $H\psi _j=\lambda _j\psi_j$, for all $j$.
By Corollary~\ref{c:regu-A}, $\psi_j\in\Dc(A)$, for all $j$. Since
\begin{equation}\label{eq:decomp-theta-h-pi}
 \theta (H)\, \Pi\ =\ \sum_{j=1}^n\, \theta (\lambda _j)\, \langle\psi_j , \cdot\rangle\psi_j\, ,
\end{equation}
$\theta (H)\Pi\in\Cc^1(A)$.\\
Similarly, we show $\theta (H)\Pi\in\Cc^\infty(\langle Q\rangle)$ using \eqref{eq:decomp-theta-h-pi} and
Proposition~\ref{p:decroissace-poly}. \qed

%%%%%%%%%%%%%%%%%%%%%%%%
\section{Exponential bounds on possible eigenfunctions with positive energy.}
\label{s:exp-bounds}
\setcounter{equation}{0}
%%%%%%%%%%%%%%%%%%%%%%%%

In this section, unless $|\alpha -1|+\beta >1$, we impose $\alpha >1$. We 
consider positive energies and show that, a possible eigenfunction of $H$, 
associated to such energies, must satisfy some exponential bound in the $\rL^2$-norm. The result and the proof are almost 
identical to Theorem 2.1 in \cite{fh} and its proof. We only change some argument to take into account the influence of our 
oscillating potential. We try to explain in Remark~\ref{r:cond-alpha>1} below why we do not treat here the case $\alpha =1$. 
However, we have some information at high energy in the case $\alpha =\beta =1$ (see Remark~\ref{r:case-alpha=beta=1}).

\begin{proposition}\label{p:borne-exp}
Under Assumptions~\ref{a:cond-pot} and~\ref{a:cond-alpha-beta-gene} with $\alpha >1$ when $|\alpha -1|+\beta \leq 1$, 
let $E>0$ and $\psi\in\Dc(H)$ such that $H\psi =E\psi$. Let
\[r\ =\ \sup\Bigl\{t^2+E\, ;\, t\in [0; +\infty [\hspace{.4cm}\mbox{and}\hspace{.4cm}e^{t\langle Q\rangle}\psi\in
\rL^2 (\R^d)\Bigr\}\ \geq \ E\, .\]
Then $r=+\infty$.
\end{proposition}
\proof We exactly follow the lines of the last part of Theorem 2.1 in \cite{fh}, except for one important argument
and some details. Just after formula (2.35) in \cite{fh}, the authors use the boundedness of $(H_0+1)^{-1}[H, iA](H_0+1)^{-1}$ 
to show that the l.h.s. of this formula (2.35) is bounded w.r.t. $\lambda$. Here we cannot do so (the previous form is 
actually unbounded, by Section~\ref{s:regu}) but provide another argument (see \eqref{eq:bound-commut}) to get the 
same conclusion. For completeness, we recall the main lines of this last
part of the proof of Theorem 2.1 in \cite{fh}. \\
Assume that the result is false. Then $r$ is finite. By Proposition~\ref{p:esti-mourre},
the Mourre estimate \eqref{eq:esti-mourre} holds true for any $\theta\in\Cc_c^\infty(\R)$ with small enough support
around $r$. Let us take such a function $\theta$ that is also identically $1$ on some open interval $\Ic '$ centered
at $r$. If $r=E$, let $r_0=r=E$, else let $r_0<r$ such that $r_0\in\Ic '$. We set $r_0=t_0^2+E$ with $t_0\geq 0$. We
take $t_1>0$ such that $r_1:=(t_0+t_1)^2+E>r$ and $r_1\in\Ic '$. We may assume that $t_1\leq 1$.\\
For $\lambda \geq 0$, let $F : \R^d\dans\R$ be defined by $F(x)=t_0\langle x\rangle +\lambda
\ln (1+t_1\lambda ^{-1}\langle x\rangle)$. By the definition of $r$, 
we know that $\langle Q\rangle^\lambda e^{t_0\langle Q\rangle}\psi\in \rL^2 (\R^d)$
(if $r=E$ i.e. $t_0=0$, this follows from Proposition~\ref{p:decroissace-poly}). Thus $\psi$ belongs to the
domain of the multiplication operator $e^{F(Q)}$. We define $\psi_F=e^{F(Q)}\psi$ and $\Psi_\lambda =\|\psi_F\|^{-1}\psi_F$.
By the end of the proof of Proposition~\ref{p:decroissace-poly}, we can show that $\nabla\psi _F$ belongs to the
domain of $\langle Q\rangle$. Thus $\psi_F\in \Dc (A)$, therefore the expectation value $\langle \psi _F, [H, iA]\psi _F\rangle$ is
well defined, and a direct computation gives
\begin{equation}\label{eq:quad-comm-psi_F}
 \bigl\langle \psi _F , [H, iA]\psi_F\bigr\rangle \ =\ -4\cdot\bigl\|g(Q)^{1/2}A\psi_F\bigr\|^2\, +\,
\bigl\langle \psi_F\, ,\, G(Q)\psi_F\bigr\rangle\, ,
\end{equation}
where $g$ is defined by $F(x)=g(x)x$ and $G(x)=((Q.P)^2g)(x)-(Q.P|\nabla F|^2)(x)$. Uniformly w.r.t. $\lambda\geq 1$,
$|\nabla F(x)|=O(\langle x\rangle^0)$ and the matrix norm $|(\nabla \otimes\nabla )F(x)|=O(\langle x\rangle^{-1})$. Notice that
$e^{(t_0+t_1)\langle Q\rangle}\psi\not\in \rL^2 (\R^d)$. As in \cite{fh}, we
can show that $\lambda\donne\Psi_\lambda$, $\lambda\donne\nabla\Psi_\lambda$, and $\lambda\donne H_0\Psi_\lambda$ are bounded 
for the $\rL^2 (\R^d)$-norm and tend to $0$ weakly in $\rL^2 (\R^d)$, as $\lambda\to +\infty$. This implies, in particular, that,
for any $\delta>0$,
\begin{equation}\label{eq:norm-limit}
\lim_{\lambda\to +\infty}\bigl\|\langle Q\rangle^{-\delta}\Psi_\lambda\bigr\| \ =\ 0\hspace{.4cm}\mbox{and}\hspace{.4cm}
\lim_{\lambda\to +\infty}\bigl\|\langle Q\rangle^{-\delta}\nabla\Psi_\lambda\bigr\| \ =\ 0\, .
\end{equation}
Since $|G(x)|=O(\langle x\rangle^{-1})+t_1(t_0+t_1)$, uniformly w.r.t. $\lambda\geq 1$,
we derive from \eqref{eq:quad-comm-psi_F} and \eqref{eq:norm-limit} that 
\begin{equation}\label{eq:bound-sup}
\limsup _{\lambda\to +\infty}\, \langle \Psi _\lambda , [H, iA]\Psi_\lambda\rangle \ \leq\ t_1(t_0+t_1)\, .
\end{equation}
Now, we claim that 
\begin{equation}\label{eq:bound-commut}
\sup _{\lambda\geq 1}\, \bigl|\langle \Psi _\lambda , [H, iA]\Psi_\lambda\rangle\bigr| \ <\ +\infty\, .
\end{equation}
Thanks to \eqref{eq:bound-commut}, we can follow the arguments of \cite{fh} to get the desired contradiction
for small enough $t_1$.\\
We are left with the proof of \eqref{eq:bound-commut}. The form $\langle P\rangle^{-2}
[H -W_{\alpha\beta}(Q), iA]\langle P\rangle^{-2}$ extends to a bounded one, by Section~\ref{s:regu}. Since 
the family $(\langle P\rangle^{2}\Psi_\lambda)_{\lambda\geq 1}$ is bounded, so is also 
$(|\langle \Psi _\lambda , [H-W_{\alpha\beta}(Q), iA]\Psi_\lambda\rangle |))_{\lambda\geq 1}$. \\
In the case $|\alpha -1|+\beta >1$, the form $\langle P\rangle^{-2}
[H, iA]\langle P\rangle^{-2}$ also extends to a bounded one, by Section~\ref{s:regu} and Remark~\ref{r:W-regu}. 
Thus we get the bound \eqref{eq:bound-commut}. \\
Now assume that $|\alpha -1|+\beta \leq 1$ and $\alpha >1$. In this case, the form $(f, g)\donne \langle 
W_{\alpha\beta}(Q)f , iAg\rangle$ is not bounded from 
$\Hc^2$ to $\Hc^{-2}$ (cf. Section~\ref{s:regu}). To get the result, we shall use the fact that $\psi$ is localised w.r.t. 
$H$ at energy $E$ and ``move'' this property through the $e^{F(Q)}$ factors appearing in \eqref{eq:bound-commut}. \\
To get the boundedness of 
$(|\langle \Psi _\lambda , [W_{\alpha\beta}(Q), iA]\Psi_\lambda\rangle|)_{\lambda\geq 1}$, it suffices
to show 
\begin{equation}\label{eq:bound-W}
\sup _{\lambda\geq 1}\, \bigl|\langle W_{\alpha\beta}(Q)\Psi _\lambda , Q\cdot P\Psi_\lambda\rangle\bigr| \ <\ +\infty\, .
\end{equation}
Since $V(Q)$ is $H_0$-compact, there exists some $c_0>0$ such that $H\geq -c_0$. For $m>c_0$, $m+H$ is invertible with bounded inverse. 
Recall that $H(F)$ is defined in \eqref{eq:def-H-F}. Let $H_0(F)=e^{F(Q)}H_0e^{-F(Q)}$. 
Since $|\nabla F(x)|=O(\langle x\rangle^0)$, uniformly w.r.t. $\lambda\geq 1$, we can find $m>0$ large enough such that, for all 
$\lambda\geq 1$, $m+H(F)$ and $m+H_0(F)$ are invertible with uniformly bounded inverse. Moreover, we see that $V(Q)(m+H(F))^{-1}$ and 
$V(Q)(m+H_0(F))^{-1}$ are uniformly bounded. \\
For $\lambda\geq 1$, $F$ stays in a bounded set of the symbol class $S(1; g)$ (see Appendix~\ref{app:s:calcul-pseudo} 
for details). Thus, by pseudodifferential calculus, $\langle P\rangle^2(m+H_0(F))^{-1}$ is uniformly bounded. By the resolvent 
formula, so is also $\langle P\rangle^2(m+H(F))^{-1}$. \\
Since $H_0\in\Cc^1(\langle Q\rangle)$ and $H\in\Cc^1(\langle Q\rangle)$ by Lemma~\ref{l:Q-regu}, since $F$ is smooth, 
$H_0(F)\in\Cc^1(\langle Q\rangle)$ and $H(F)\in\Cc^1(\langle Q\rangle)$. Using Propositions~\ref{p:regu} and~\ref{p:est3}, we see that, 
for $\epsilon\in [0; 1]$, $\langle Q\rangle^\epsilon (m+H_0(F))^{-1}\langle Q\rangle ^{-\epsilon}$ and 
$\langle Q\rangle^\epsilon (m+H(F))^{-1}\langle Q\rangle ^{-\epsilon}$ are bounded, uniformly w.r.t. $\lambda\geq 1$.\\
For $\ell\in\N$, we can write $\psi =(m+E)^{\ell}(m+H)^{-\ell}\psi$. By a direct computation, 
\[e^{F(Q)}(m+H)^{-1}e^{-F(Q)}=(m+H(F))^{-1}\, .\] 
Thus, for $\ell_1, \ell_2\in\N$, 
\begin{align}
&\langle W_{\alpha\beta}(Q)\Psi_\lambda , Q\cdot P\Psi_\lambda\rangle\nonumber\\
=\ &(m+E)^{\ell_1+\ell_2}\bigl\langle QW_{\alpha\beta}(Q)\bigl(m+H(F)\bigr)^{-\ell_1}\Psi_\lambda \, ,\,  
P\bigl(m+H(F)\bigr)^{-\ell_2}\Psi_\lambda\bigr\rangle\, .\label{eq:puis-resolv}
\end{align}
In \eqref{eq:puis-resolv}, we write 
\[\bigl(m+H(F)\bigr)^{-\ell_1}\ =\ \Bigl(\bigl(m+H_0(F)\bigr)^{-1}\, -\, \bigl(m+H_0(F)\bigr)^{-1}V(Q)
\bigl(m+H(F)\bigr)^{-1}\Bigr)^{\ell_1}\, ,\]
\[\bigl(m+H(F)\bigr)^{-\ell_2}\ =\ \Bigl(\bigl(m+H_0(F)\bigr)^{-1}\, +\, \bigl(m+H(F)\bigr)^{-1}V(Q)
\bigl(m+H_0(F)\bigr)^{-1}\Bigr)^{\ell_2}\, ,\]
and expand the products. The expansion contains, up to the factor $(m+E)^{\ell_1+\ell_2}$, terms of the form 
\begin{equation}\label{eq:term-form1}
 \bigl\langle QW_{\alpha\beta}(Q)\bigl(m+H_0(F)\bigr)^{-1}V(Q)
\bigl(m+H(F)\bigr)^{-1}B_1\Psi_\lambda \, ,\, B_2\Psi_\lambda\bigr\rangle\, ,
\end{equation}
where $B_1$ and $B_2$ are uniformly bounded operators. By Assumption~\ref{a:cond-alpha-beta-gene}, 
$\langle Q\rangle^{1-\beta -\beta_{lr}}$ is bounded. For $W=V_{sr}$, $W=V_{lr}$, and $W=W_{\alpha\beta}$, 
$\langle Q\rangle^{\beta_{lr}}W(Q)$ is bounded. Since, by the resolvent formula, 
\begin{align*}
&\langle Q\rangle^{\beta_{lr}}V_c(Q)(m+H(F))^{-1}\\
=\, &\langle Q\rangle^{\beta_{lr}}\chi _c(Q)V_c(Q)\langle P\rangle^{-2}\langle P\rangle^2(m+H_0(F))^{-1}\\
&-\, \langle Q\rangle^{\beta_{lr}}\chi _c(Q)V_c(Q)\langle P\rangle^{-2}\langle P\rangle^2(m+H_0(F))^{-1}V(m+H(F))^{-1}\, ,
\end{align*}
the operator $\langle Q\rangle^{\beta_{lr}}V_c(Q)(m+H(F))^{-1}$ is uniformly bounded. Furthermore, 
\begin{align*}
&\langle Q\rangle^{\beta_{lr}}(m+H_0(F))^{-1}(v\cdot\nabla\tilde{V}_{sr})(Q)(m+H(F))^{-1}\\
=\, &\langle Q\rangle^{\beta_{lr}}(m+H_0(F))^{-1}(v(Q)\cdot iP)\langle Q\rangle^{-\beta_{lr}}\cdot
\langle Q\rangle^{\beta_{lr}}\tilde{V}_{sr}(Q)(m+H(F))^{-1}\\
&-\, \langle Q\rangle^{\beta_{lr}}(m+H_0(F))^{-1}\langle Q\rangle^{-\beta_{lr}}\cdot\langle Q\rangle^{\beta_{lr}}
\tilde{V}_{sr}(Q)\cdot (v(Q)\cdot iP)(m+H(F))^{-1}\, ,
\end{align*}
so it is also uniformly bounded. Therefore all the terms of the form \eqref{eq:term-form1} are bounded, uniformly 
w.r.t. $\lambda\geq 1$. Up to the factor $(m+E)^{\ell_1+\ell_2}$, the previous expansion contains also terms of the form 
\begin{equation}\label{eq:term-form2}
 \bigl\langle QW_{\alpha\beta}(Q)B_1'\Psi_\lambda \, ,\, P\bigl(m+H(F)\bigr)^{-1}V(Q)
\bigl(m+H_0(F)\bigr)^{-1}B_2'\Psi_\lambda\bigr\rangle\, ,
\end{equation}
for uniformly bounded operators $B_1'$ and $B_2'$. We note that $\langle Q\rangle^{\beta_{lr}}P\langle Q\rangle ^{-\beta_{lr}}
\langle P\rangle^{-1}$ is bounded and that $\langle Q\rangle^{\beta_{lr}}\langle P\rangle^1(m+H(F))^{-1}\langle Q\rangle^{-\beta_{lr}}$ 
is uniformly bounded, use again the above arguments to conclude that all the terms of the form \eqref{eq:term-form2} are bounded 
functions of $\lambda$. We are left with the term 
\[(m+E)^{\ell_1+\ell_2}\bigl\langle QW_{\alpha\beta}(Q)\bigl(m+H_0(F)\bigr)^{-\ell_1}\Psi_\lambda \, ,\,  
P\bigl(m+H_0(F)\bigr)^{-\ell_2}\Psi_\lambda\bigr\rangle\, .\]
By pseudodifferential calculus, 
\[\langle P\rangle^{2\ell _1}(m+H_0(F))^{-\ell_1}\hspace{.4cm}\mbox{and}\hspace{.4cm}\langle P\rangle^{2\ell _2-1}P(m+H_0(F))^{-\ell_2}\]
are uniformly bounded. Thus, by Proposition~\ref{p:oscillations-energy-alpha>1}, this last term is bounded, 
if we choose $\ell_1$ and $\ell_2$ large enough. This proves \eqref{eq:bound-W} and therefore \eqref{eq:bound-commut}. \qed

\begin{rem}\label{r:cond-alpha>1}
In the second part of the above proof, we used the assumption $\alpha >1$ to get \eqref{eq:bound-W}. Indeed, we managed to move a
''localisation'' $(m+H)^{-\ell}$ through the multiplication operator $e^{F(Q)}$, creating in this way the factors
$\langle P\rangle^{-\ell_1}$ and $\langle P\rangle^{-\ell_2}$. Then we applied Proposition~\ref{p:oscillations-energy-alpha>1} 
that only holds true for
$\alpha >1$ (see Remark~\ref{r:alpha>1}). In the case $\alpha =1$, it is natural to try to move an appropriate localisation
$\theta (H)$ through $e^{F(Q)}$ and then use Proposition~\ref{p:oscillations-energy-alpha-leq-1}. We do not know how to
bound the operator $e^{F(Q)}\theta (H)e^{-F(Q)}$ uniformly w.r.t. $\lambda$, when $\theta$ is smooth and compactly supported. Formally,
$e^{F(Q)}\theta (H)e^{-F(Q)}=\theta (H(F))$ where $H(F)=e^{F(Q)}He^{-F(Q)}$, but the latter is not self-adjoint 
(see \eqref{eq:def-H-F}). 
\end{rem}
\begin{rem}\label{r:case-alpha=beta=1}
In the case $\alpha =\beta =1$, the Mourre estimate is valid at high energy, say on any compact interval included in some $[a; +\infty[$ 
with $a>0$ (cf. the proof of Proposition~\ref{p:esti-mourre}). Take an energy $E>a$ and $\psi\in\Dc(H)$
such that $H\psi =E\psi$. The proof of Theorem 2.1 in \cite{fh} works in this 
situation and yields the conclusion of Proposition~\ref{p:borne-exp}, namely $r=+\infty$. 
\end{rem}
%

%%%%%%%%%%%%%%%%%%%%%%%%
\section{Eigenfunctions cannot satisfy unlimited exponential bounds.}
\label{s:strong-exp-bounds}
\setcounter{equation}{0}
%%%%%%%%%%%%%%%%%%%%%%%%

In this section, we work under Assumption~\ref{a:cond-pot} with $|\alpha -1|+\beta >1$ or with $\beta\geq 1/2$ and 
$|\alpha -1|+\beta \leq 1$, but, in contrast to Section~\ref{s:exp-bounds}, we impose some 
lower bound on the form $[V(Q), iA]$. Again, we study the states 
$\psi\in\Dc(H)$ such that $H\psi =E\psi$, for some $E\in\R$, but also assume that $\psi$ belongs to the domain of the 
multiplication operator $e^{\gamma \langle Q\rangle}$,
for all $\gamma \geq 0$. We shall show that such $\psi$ must be zero. Our proof is inspired by the
corresponding result in \cite{fhhh2} (see also Theorem 4.18 in \cite{cfks}). In fact, when $|\alpha -1|+\beta >1$, 
we just apply \cite{fhhh2}. Our new contribution concerns the case where $\alpha >1$, $1\geq\beta\geq 1/2$, and $\alpha +\beta\leq 2$. 
In that case, that is not covered by the result in \cite{fhhh2}, we still arrive at the same conclusion using an 
appropriate bound on the contribution of the oscillating potential $W_{\alpha\beta}$ to the commutator form $[H, iA]$. 
This provides in particular a proof of 
Theorem~\ref{th:no-posivite-eigenvalue}.

Under Assumption~\ref{a:cond-pot}, we demand, unless $|\alpha -1|+\beta>1$, that $\beta\geq 1/2$. 
We require further, as in \cite{fhhh2}, that the form
$[(V_c+v\cdot\nabla\tilde{V}_{sr})(Q), iA]$ is $H_0$-form-lower-bounded with relative bound less
than $2$. Precisely, we demand that
\begin{eqnarray}\label{eq:commut-v_c}
&&\exists\epsilon _c>0\, ,\ \exists\lambda _c>0\, ;\ \forall\varphi\in\Dc(H)\cap\Dc(A)\, ,\\
&&\langle\varphi , \bigl[(V_c+v\cdot\nabla\tilde{V}_{sr})(Q), iA\bigr]\varphi \rangle
\ \geq \ (\epsilon _c-2)\langle\varphi , H_0\varphi \rangle\, -\, \lambda_c\|\varphi\|^2\, .\nonumber
\end{eqnarray}
We shall need the following known 
\begin{lemma}\label{l:form-bounded}
Under the previous assumptions,
\begin{eqnarray}
&&\forall\delta\in ]0; 1[\, ,\ \exists\mu_\delta >0\, ;\ \forall\varphi\in\Dc(H)\cap\Dc(A)\, ,\nonumber\\
&&\hspace{1cm}\langle\varphi , H_0\varphi \rangle
\ \geq \ \delta\langle\varphi , H\varphi \rangle\, -\, \mu_\delta\|\varphi\|^2\label{eq:H-form-bounded}\, .\\
&&\forall\epsilon >0\, ,\ \exists\lambda_\epsilon >0\, ;\ \forall\varphi\in\Dc(H)\cap\Dc(A)\, ,\nonumber\\
&&\hspace{1cm}\langle\varphi , [H-W_{\alpha\beta}(Q), iA]\varphi \rangle
\ \geq \ (\epsilon _c-\epsilon)\langle\varphi , H_0\varphi \rangle\, -\, \lambda_\epsilon \|\varphi\|^2\, .
\label{eq:commut-form-bounded}
\end{eqnarray}
\end{lemma}
\proof Since $V(Q)$ is $H_0$-compact, it is $H_0$-bounded with relative bound $0$. This implies \eqref{eq:H-form-bounded}
(see \cite{k}). Recall that the form $[V_{sr}(Q)+V_{lr}(Q), iA]$ is compact from $\Hc^1$ to $\Hc^{-1}$ (cf . 
(cf. \eqref{eq:commut-v-sr}, \eqref{eq:commut-v-c}, \eqref{eq:commut-v-tilde-sr}). Thus it is $H_0$ form bounded 
with relative bound $0$. Take $\epsilon>0$. There exists $\mu_\epsilon>0$ such that, for all $\varphi\in\Dc(H)\cap\Dc(A)$, 
\[\bigl|\langle\varphi , [V_{sr}(Q)+V_{lr}(Q), iA]\varphi \rangle\bigr|
\ \leq \ \epsilon \langle\varphi , H_0\varphi \rangle\, +\, \mu_\epsilon\|\varphi\|^2\, .\]
Therefore, for such $\varphi$, the l.h.s. of \eqref{eq:commut-form-bounded} is 
\[
\ \geq \ (2-\epsilon +\epsilon _c-2)\langle\varphi , H_0\varphi \rangle\, -\, (\lambda_c +\mu_\epsilon)\|\varphi\|^2\, ,\]
by \eqref{eq:commut-v_c}. This yieds \eqref{eq:commut-form-bounded} with $\lambda_\epsilon =\lambda_c +\mu_\epsilon$. \qed

As in Section~\ref{s:exp-bounds}, we shall use a conjugaison by an appropriate $e^{F(Q)}$. For $\gamma >0$, let 
$F : \R^d\dans\R$ be the smooth function defined by
$F(x)=\gamma\langle x\rangle$. Setting $g(x)=\gamma \langle x\rangle^{-1}$, $\nabla F(x)=g(x)x$ and
\begin{equation}\label{eq:calcul-explicite-0}
\bigl|\nabla F(x)\bigr|^2\ =\ \gamma^2\bigl(1\, -\, \langle x\rangle^{-2}\bigr)\, .
\end{equation}
A direct computation gives
\begin{eqnarray}\label{eq:calcul-explicite-1}
\bigl((Q.P)^2g\bigr)(x)&=&\gamma \langle x\rangle^{-1}\bigl(1\, -\, \langle x\rangle^{-2}\bigr)
\bigl(1\, -\, 3\langle x\rangle^{-2}\bigr)
\ ,\\
-\bigl((Q.P)(|\nabla F|^2)\bigr)(x)&=&-2\gamma^2\langle x\rangle^{-2}\bigl(1\, -\, \langle x\rangle^{-2}\bigr)\, \leq \, 0\, . 
\label{eq:calcul-explicite-2}
\end{eqnarray}
\begin{proposition}\label{p:pas-de-vp}
Assume Assumption~\ref{a:cond-pot} and \eqref{eq:commut-v_c}. Unless $|\alpha -1|+\beta>1$, take $\beta\geq 1/2$. 
Unless $\alpha +\beta\leq 2$ or $\beta \geq 1/2$, take $|w|$ small enough. 
Let $\psi\in\Dc(H)$ and $E\in\R$ such that $H\psi =E\psi$. Assume further that, for all $\gamma \geq 0$, 
$\psi$ belongs to the domain of the multiplication operator $e^{\gamma \langle Q\rangle}$. Then $\psi =0$.  
\end{proposition}
\begin{rem}\label{r:larger-framework}
Note that Proposition~\ref{p:pas-de-vp} applies under \eqref{eq:commut-v_c} and Assumptions~\ref{a:cond-pot} 
and~\ref{a:cond-alpha-beta-gene}. In particular, the case $\alpha =1$ is allowed. 
\end{rem}

\Pfof{Proposition~\ref{p:pas-de-vp}} First of all, we focus on the cases where a result in \cite{fhhh2} applies. 
Assume that $\beta >1$ or $\alpha<\beta\leq 1$. 
We use Remark~\ref{r:W-regu} to derive, thanks to \eqref{eq:commut-form-bounded}, the following property: for any 
$\epsilon>0$, there exists $\lambda_\epsilon>0$, such that, for all $\varphi\in\Dc(H)\cap\Dc(A)$, 
\begin{equation}\label{eq:commut-form-bounded-V}
\langle\varphi , [V, iA]\varphi \rangle
\ \geq \ (\epsilon _c-\epsilon)\langle\varphi , H_0\varphi \rangle\, -\, \lambda_\epsilon \|\varphi\|^2\, .
\end{equation}
Therefore \cite{fhhh2} applies. \\
Assume now that $\alpha +\beta >2$ and $\beta\leq 1$. Again by Remark~\ref{r:W-regu}, we know that the form
$[W_{\alpha\beta}(Q), iA]$ extends to a bounded one from $\Hc^2$ to $\Hc^{-2}$. Thus, for $|w|$ small enough, 
\eqref{eq:commut-form-bounded-V} still holds true and \cite{fhhh2} applies.\\
Now, we treat the last case: $|\alpha -1|+\beta\leq 1$ and $\beta\geq 1/2$. We always consider $\gamma \geq 1$. By assumption, 
$\psi$ belongs to the domain of the multiplication operator $e^{F(Q)}$. Setting $\psi _F=e^{F(Q)}\psi$, we claim that
\begin{equation}\label{eq:commut-w}
\bigl|\bigl\langle\psi _F , [W_{\alpha\beta}(Q), iA]\psi _F \bigr\rangle\bigr|
\ \leq \ \bigl\|g(Q)^{1/2}A\psi _F\bigr\|^2\, +\, |w|^2\gamma^{-1}\|\psi _F\|^2\, .
\end{equation}
From the definition of the form $[W_{\alpha\beta}(Q), iA]$, we observe that
\begin{eqnarray*}
\bigl|\bigl\langle\psi _F , [W_{\alpha\beta}(Q), iA]\psi _F \bigr\rangle\bigr|&\leq&2|w|\cdot\bigl\|g(Q)^{1/2}A\psi _F\bigr\|\cdot
\bigl\|g(Q)^{-1/2}\langle Q\rangle^{-\beta}\psi _F\bigr\| \\
&\leq&2|w|\cdot\bigl\|g(Q)^{1/2}A\psi _F\bigr\|\cdot\gamma ^{-1/2}\cdot\bigl\|\langle Q\rangle^{1/2-\beta}\psi _F\bigr\|\\
&\leq&2\cdot\bigl\|g(Q)^{1/2}A\psi _F\bigr\|\cdot\gamma ^{-1/2}|w|\cdot\bigl\|\psi _F\bigr\|
\end{eqnarray*}
since we assumed that $\beta \geq 1/2$. Now \eqref{eq:commut-w} follows from the use of the inequality $2ab\leq a^2+b^2$, for $a, b\geq 0$.\\
Now, we essentially follows the argument in the proof of Theorem 4.18 in \cite{cfks} and prove the result by contradiction.
Assume that $\psi\neq 0$. Let $\psi _F=e^{F(Q)}\psi$. 
The formula \eqref{eq:quad-comm-psi_F} is valid with the new function $F$. As in the proof of
Proposition~\ref{p:decroissace-poly}, we also have
\begin{equation}\label{eq:quad-psi_F}
 \langle \psi _F \, ,\,  H\psi _F\rangle \ =\ \bigl\langle \psi _F \, ,\,  \bigl(|\nabla F|^2(Q)+E\bigr)
\psi _F\bigr\rangle\, .
\end{equation}
Combining \eqref{eq:quad-comm-psi_F} and \eqref{eq:commut-w}, we get, for $\gamma\geq 1$,
\begin{eqnarray}\label{eq:commut-sans-w}
\hspace{.2cm}\bigl\langle\psi _F , [H-W_{\alpha\beta}(Q), iA]\psi _F\bigr\rangle
&\leq&-3\cdot\bigl\|g(Q)^{1/2}A\psi _F\bigr\|^2\, +\, |w|^2\gamma^{-1}\|\psi _F\|^2\nonumber\\
&&\, +\, \langle\psi _F  , G(Q)\psi _F\rangle\nonumber\\
\hspace{.2cm}\bigl\langle\psi _F , [H-W_{\alpha\beta}(Q), iA]\psi _F\bigr\rangle
&\leq&\langle\psi _F  , G(Q)\psi _F\rangle\, +\, |w|^2\gamma^{-1}\|\psi _F\|^2\, ,
\end{eqnarray}
where $G(Q)=(Q.P)^2g-(Q.P)(|\nabla F|^2)$. Next we deduce from \eqref{eq:commut-form-bounded} and \eqref{eq:H-form-bounded}
in Lemma~\ref{l:form-bounded}, and \eqref{eq:quad-psi_F},
that, for all $\delta\in ]0; \epsilon _c[$, there exist some $\rho_\delta , \rho_\delta '>0$ such that, for all $\gamma\geq 1$,
\begin{eqnarray}
\bigl\langle\psi _F , [H-W_{\alpha\beta}(Q), iA]\psi _F \bigr\rangle &\geq &
\delta\langle\psi _F , H_0\psi _F\rangle\, -\,
\rho_\delta\|\psi _F\|^2\nonumber\\
&\geq&2^{-1}\delta\bigl(\langle\psi _F , H\psi _F\rangle\, -\,
2\mu_{1/2}\bigr)\, -\, \rho_\delta\|\psi _F\|^2\nonumber\\
&\geq &2^{-1}\delta\langle\psi _F , (H-E)\psi _F\rangle\, -\, \rho_\delta '\|\psi _F\|^2\nonumber\\
&\geq&2^{-1}\delta\bigl\langle\psi _F , |\nabla F|^2(Q)\psi _F\bigr\rangle\, -\,
\rho_\delta '\|\psi _F\|^2\, .\label{eq:minoration-commut-total}
\end{eqnarray}
In view of \eqref{eq:calcul-explicite-0}, we introduce the function $f : [0; +\infty [\to [0; +\infty [$ given by
\begin{equation}\label{eq:function-f}
 f(\gamma )\ =\ \bigl\langle\psi _F \, ,\, \bigl(1\, -\, \langle Q\rangle^{-2}\bigr)\psi _F\bigl\rangle\ =\
\gamma^{-2}\bigl\langle\psi _F , |\nabla F|^2(Q)\psi _F\bigr\rangle\, .
\end{equation}
Since $\psi\neq 0$, we can find $\epsilon >0$ such that $\|\un_{|\cdot |\geq 2\epsilon}(Q)\psi\|>0$.
For all $\gamma \geq 0$,
\[\frac{\bigl\|\un_{|\cdot |\leq \epsilon}(Q)e^{\gamma \langle Q\rangle}\psi\bigr\|^2}{\bigl\|e^{\gamma \langle Q\rangle}\psi\bigr\|^2}
\ \leq \ \frac{e^{2\gamma \langle\epsilon\rangle}\bigl\|\un_{|\cdot |\leq \epsilon}(Q)\psi\bigr\|^2}
{e^{2\gamma \langle 2\epsilon\rangle}\bigl\|\un_{|\cdot |\geq 2\epsilon}(Q)\psi\bigr\|^2}\ \leq \
e^{2\gamma (\langle\epsilon\rangle - \langle 2\epsilon\rangle)}\frac{\|\psi\|^2}{\|\un_{|\cdot |\geq 2\epsilon}(Q)\psi\|^2}\]
and
\begin{eqnarray*}
f(\gamma)&\geq & \bigl(1\, -\, \langle\epsilon\rangle^{-2}\bigr)\bigl\|\un_{|\cdot |\geq \epsilon}(Q)\psi _F\bigr\|^2\\
&\geq &\bigl(1\, -\, \langle\epsilon\rangle^{-2}\bigr)\cdot\Bigl(\|\psi _F\|^2\, -\, \bigl\|\un_{|\cdot |\leq \epsilon}(Q)
e^{\gamma \langle Q\rangle}\psi\bigr\|^2\Bigr)\\
&\geq &\bigl(1\, -\, \langle\epsilon\rangle^{-2}\bigr)\cdot\|\psi _F\|^2\cdot\Bigl(1\, -\, C_\epsilon 
e^{2\gamma (\langle\epsilon\rangle - \langle 2\epsilon\rangle)}\Bigr)\, ,
\end{eqnarray*}
where $C_\epsilon :=\|\psi\|^2\cdot\|\un_{|\cdot |\geq 2\epsilon}(Q)\psi\|^{-2}$. Thus, there exist $C>0$ and $\Gamma\geq 1$ such that, for
$\gamma\geq\Gamma$,
\begin{equation}\label{eq:mino-f}
 f(\gamma )\ \geq \ C\, \|\psi _F\|^2\ \geq \ C\, \|\psi \|^2\ >\ 0\, .
\end{equation}

We derive from \eqref{eq:commut-sans-w} and \eqref{eq:minoration-commut-total},
thanks to \eqref{eq:function-f} and \eqref{eq:calcul-explicite-2},
that, for all $\gamma\geq 1$,
\[2^{-1}\delta\gamma^2f(\gamma )\, -\, \bigl(\rho_\delta '+|w|^2\gamma^{-1}\bigr)\|\psi _F\|^2\ \leq \
\bigl\langle\psi _F  , G(Q)\psi _F \bigr\rangle\ \leq \ \bigl\langle\psi _F  , \bigl((Q.P)^2g\bigr)(Q)
\psi _F \bigr\rangle\, . \]
By \eqref{eq:calcul-explicite-1}, $((Q.P)^2g)(x)\leq \gamma (1-\langle x\rangle^{-2})$, for all $x\in\R^d$, yielding,
for all $\gamma\geq \Gamma$,
\begin{eqnarray*}
 &&2^{-1}\delta\gamma^2f(\gamma )\, -\, \bigl(\rho_\delta '+|w|^2\gamma^{-1}\bigr)\|\psi _F\|^2\ \leq \ \gamma f(\gamma )\\
\mbox{and}&&\bigl(2^{-1}\delta\gamma^2\, -\, \gamma\, -\, (\rho_\delta '+|w|^2\gamma^{-1})C^{-1}\bigr)\cdot f(\gamma )\ \leq \ 0\, ,
\end{eqnarray*}
by \eqref{eq:mino-f}. We get a contradiction for $\gamma$ large enough. \qed

%%%%%%%%%%%%%%%%%%%%%%%%
\section{LAP at suitable energies.}
\label{s:lap}
\setcounter{equation}{0}
%%%%%%%%%%%%%%%%%%%%%%%%

In this section, we prove the limiting absorption principle for $H$ for appropriate energy regions.
As already pointed out in \cite{gj2} and in Section~\ref{s:regu}, one cannot use the usual Mourre theory w.r.t. 
the generator of dilations $A$, since the Hamiltonian is not regular enough w.r.t. $A$. For the same reason, one 
cannot follow the lines in \cite{ge}. As explained in Remark~\ref{r:mourre-non}, we were not able to apply 
the ``weighted Mourre theory'' developed in \cite{gj2}, which is inspired by \cite{ge} and is a kind of 
``localised'' Putnam argument. Instead, we follow the more complicated path introduced in \cite{gj}. 

To prepare our result, we need some notation. For $\delta >0$ and $y\in\R^d$, we set
\begin{equation}\label{eq:f-delta}
g_{\delta}(y)\ = \ \bigl(2\, -\, \langle y\rangle^{-\delta}\bigr)\langle y\rangle^{-1}y\, .
\end{equation}
Let $\cchi\in\Cc_c^\infty(\R)$ with $\cchi (t)=1$ if and only if $|t|\leq 1$ and $\supp\cchi\subset [-2; 2]$.
Let $\tilde\cchi =1-\cchi$. For $R\geq 1$ and $t\in\R$, we set $\cchi _R(t)=\cchi (t/R)$ and $\tilde\cchi _R(t)=\tilde\cchi (t/R)$.
We also set $g_{\delta , R}(y)=\tilde\cchi _R(\langle y\rangle)^2g_{\delta}(y)$. Recall that we set $\beta _{lr}=\min (\rho _{lr}, \beta)$.

First, we show a kind of weighted Mourre estimate at infinity for the position operators $Q$ (meaning for large $|Q|$), which can be
seen as an energy localised (i.e. localised in $H$) Putnam positivity, that is also localised in $|Q|$ at infinity. 
It should be compared with Section 2 in \cite{l1}. 
\begin{proposition}\label{p:weighted-mourre}
Assume Assumption~\ref{a:cond-pot}. Under Assumption~\ref{a:cond-alpha-beta}, take any compact interval $\Ic '\subset\mathring{\Jc}$, the
interior of $\Jc$.
Let $\delta$ be a small enough positive number (depending only on the potential) and $s=(1+\delta )/2$.
There exist $c_1>0$ and $R_1>1$ such that, for $R\geq R_1$,
there exists a bounded, self-adjoint operator $B_R$ such that, for $f\in\rL^2(\R^d)$ with $E_{\Ic '}(H)f=f$, we have the estimate:
\begin{eqnarray}
\bigl\langle f\, ,\, \, [H, iB_R]\, f\bigr\rangle&\geq&c_1\, \bigl\|\tilde\cchi _R(\langle Q\rangle)\langle Q\rangle ^{-s}f\bigr\|^2
\, -\, O\bigl(R^{-\gamma}\bigr)\bigl\|\tilde\cchi _R(\langle Q\rangle)\langle Q\rangle ^{-s}f\bigr\|\nonumber\\
&&\, -\, O\bigl(R^{-\gamma -1}\bigr)\, ,\label{eq:weighted-mourre}
\end{eqnarray}
with $\gamma =1 -\delta >1/2$, if $|\alpha -1|+\beta >1$, else $\gamma =\beta -\delta >1/2$. Here 
\[B_R\ =\ g_{\delta , R}(Q)\cdot P\, +\, P\cdot g_{\delta , R} (Q)\, .\]
The ''O'' terms in the estimate can be chosen independent
of $f$ when $f$ stays in a bounded set for the norm $\|\langle Q\rangle ^{-s}\cdot\|$.   
\end{proposition}
\begin{rem}\label{r:size-delta}
In fact, we can give a precise upper bound on $\delta$ in Proposition~\ref{p:weighted-mourre}. 
We demand that $\delta<\min (\beta ; \rho _{sr}; \rho _{lr}'; 1/2)$. In the case where $\alpha\geq 1$ and 
$\alpha + \beta\leq 2$, we know that $\beta+\beta _{lr}>1$ and $\beta>1/2$, by Assumption~\ref{a:cond-alpha-beta}, 
and we further require that $\delta<\min (\beta+\beta _{lr}-1; \beta -1/2)$.\\
Denoting by $c$ the infimum of $\Jc$, one can take $c_1=\delta c/2$ in \eqref{eq:weighted-mourre}.
\end{rem}
\proof We choose $\delta$ according to Remark~\ref{r:size-delta}.
We take $f$ satisfying $E_{\Ic '} (H)f=f$ and belonging to some fix bounded set for the norm $\|\langle Q\rangle ^{-s}\cdot\|$.
Let $\theta\in\Cc_c^\infty (\R; \R)$ such that $\theta =1$ on $\Ic '$ and $\supp\, \theta\subset\mathring{\Jc}$. 
We have $\theta (H)f=f$. Take $R_1$ large enough such that, for $R\geq R_1$, $\tilde\cchi _RV_c=0$. In particular, 
\begin{eqnarray*}
\langle f\, , \, [V_c(Q)\, ,\,  iB_R]f\rangle & =& 2\langle V_c(Q)f\, ,\, g_{\delta , R}(Q)\cdot iPf\rangle\, +\,
2\langle g_{\delta , R}(Q)\cdot iPf\, ,\, V_c(Q)f\rangle\, =\, 0\, .
\end{eqnarray*}
The other contributions of the potential are given by 
\begin{eqnarray*}
\langle f\, , \, [V_{lr}(Q), iB_R]f\rangle &=&-\langle f\, ,\, (g_{\delta , R}\cdot\nabla V_{lr})(Q)f\rangle\\
\langle f\, , \, [V_{sr}(Q), iB_R]f\rangle & =& 2\Re \langle iPf\, ,\, V_{sr}(Q)g_{\delta , R}(Q)f\rangle\, ,\\
\langle f\, , \, [W_{\alpha \beta}(Q), iB_R]f\rangle & =& 2\Re \langle iPf\, ,\, W_{\alpha \beta}(Q)g_{\delta , R}(Q)f\rangle\, ,
\end{eqnarray*}
and
\begin{eqnarray}
\langle f\, , \, \bigl[(v\cdot\nabla\tilde{V}_{sr})(Q), iB_R\bigr]f\rangle & =& 2\Re \langle iPf\, ,\, 
(v\cdot\nabla\tilde{V}_{sr})(Q)g_{\delta , R}(Q)f\rangle\nonumber\\
&=&2\Re \langle iPf\, ,\, [v(Q)\cdot iP\, ,\, \tilde{V}_{sr}(Q)]g_{\delta , R}(Q)f\rangle\nonumber\\
&=&2\Re \langle (P\cdot v(Q))Pf\, ,\, \tilde{V}_{sr}(Q)g_{\delta , R}(Q)f\rangle\nonumber\\
&&-\, 2\Re \langle \tilde{V}_{sr}(Q)Pf\, ,\, (v(Q)\cdot P)g_{\delta , R}(Q)f\nonumber\rangle\\
\label{eq:comm-tilde-v-sr-R}
&=&2\Re \langle (v(Q)\cdot P)Pf\, ,\, \tilde{V}_{sr}(Q)g_{\delta , R}(Q)f\rangle\\
&&+\, 2\Re \langle (\nabla \cdot v)(Q)Pf\, ,\, i\tilde{V}_{sr}(Q)g_{\delta , R}(Q)f\rangle\nonumber\\
&&-\, 2\Re \langle \tilde{V}_{sr}(Q)Pf\, ,\, g_{\delta , R}(Q)(v(Q)\cdot P)f\rangle\nonumber\\
&&+\, 2\Re \langle \tilde{V}_{sr}(Q)Pf\, ,\, i(v\cdot\nabla g_{\delta , R})(Q)f\rangle\nonumber
\, .
\end{eqnarray}
Note that the term $\langle f\, ,\, (g_{\delta , R}\cdot\nabla V_{lr})(Q)f\rangle$ is 
$O(R^{-\epsilon})\|\tilde\cchi _R(\langle Q\rangle)\langle Q\rangle ^{-s}f\|^2$, for $\epsilon =\rho _{lr}'-\delta >0$.
We shall evaluate the size of the other terms. To this end, we shall repeatedly make use of Lemma~\ref{l:diff-decroit-Q-bis}, of
Lemma~\ref{l:commut-fonct-Q-fonct-H-bis} and of the fact that the term $\|\langle Q\rangle ^{-s}f\|$ stays in a bounded region, for
the considered $f$. Note that those lemmata follow from the regularity
of $H$ w.r.t. $\langle Q\rangle$. \\
Writing 
\begin{eqnarray*}
&&\langle (V_{sr}g_{\delta , R})(Q)f\, ,\, iPf\rangle\\
&=&\langle (V_{sr}g_{\delta , R})(Q)f\, ,\, iP\theta (H)f\rangle\\
&=&\bigl\langle \langle Q\rangle^{s}(V_{sr}g_{\delta})(Q)\tilde\cchi _R(\langle Q\rangle)\langle Q\rangle^{-s}f\, ,\,
\bigl[\tilde\cchi _R(\langle Q\rangle)\, ,\, iP\theta (H)\bigr]\langle Q\rangle^{s}\cdot\langle Q\rangle^{-s}
f\bigr\rangle\\
&&+\, \bigl\langle \langle Q\rangle^{2s}(V_{sr}g_{\delta})(Q)\tilde\cchi _R(\langle Q\rangle)\langle Q\rangle^{-s}f\, ,\,
\langle Q\rangle^{-s}iP\theta (H)\langle Q\rangle^{s}\cdot \tilde\cchi _R(\langle Q\rangle)\langle Q\rangle^{-s}
f\bigr\rangle\, ,
\end{eqnarray*}
the first term is $O(R^{\delta -1-\rho _{sr}})\|\tilde\cchi _R(\langle Q\rangle)\langle Q\rangle ^{-s}f\|$ and the second term
is at most of size $O(R^{\delta -\rho _{sr}})\|\tilde\cchi _R(\langle Q\rangle)\langle Q\rangle ^{-s}f\|^2$, by
Lemma~\ref{l:commut-fonct-Q-fonct-H-bis}. Notice that $O(R^{\delta -1-\rho _{sr}})=O(R^{-\gamma})$ and that 
$\delta -\rho _{sr}<0$.\\
Using that 
\begin{eqnarray}
\tilde\cchi _R(\langle Q\rangle)Pf&=&\tilde\cchi _R(\langle Q\rangle)P\theta (H)f\nonumber\\
\label{eq:comm-P-chi-R}
&=&[\tilde\cchi _R(\langle Q\rangle)\, ,\, P\theta (H)]\langle Q\rangle^{s}\cdot\langle Q\rangle^{-s}f\\
&&+\, \langle Q\rangle^{s}\cdot \langle Q\rangle^{-s}P\theta (H)\langle Q\rangle^{s}\cdot
\tilde\cchi _R(\langle Q\rangle)\langle Q\rangle^{-s}f\, ,\nonumber
\end{eqnarray}
we see that the second term in \eqref{eq:comm-tilde-v-sr-R} is 
\[O(R^{\delta -1-\rho _{sr}})\|\tilde\cchi _R(\langle Q\rangle)\langle Q\rangle ^{-s}f\|
\, +\, O(R^{\delta -\rho _{sr}})\|\tilde\cchi _R(\langle Q\rangle)\langle Q\rangle ^{-s}f\|^2\]
and the fourth term is even better. For the third term, we use \eqref{eq:comm-P-chi-R} twice to see that it is 
\[O(R^{\delta -2-\rho _{sr}})\, +\, 
O(R^{\delta -1-\rho _{sr}})\|\tilde\cchi _R(\langle Q\rangle)\langle Q\rangle ^{-s}f\|
\, +\, O(R^{\delta -\rho _{sr}})\|\tilde\cchi _R(\langle Q\rangle)\langle Q\rangle ^{-s}f\|^2\, .\]
Note that $O(R^{\delta -2-\rho _{sr}})=O(R^{-\gamma -1})$. \\
To evaluate the contribution of $W_{\alpha \beta}$, we use Remark~\ref{r:W-regu}. If $1<\beta$, then we can treat  
this contribution as the one of $V_{sr}$. If $\beta\leq 1$ and $\alpha <\beta$, then 
it is treated as the one of $V_{lr}$. If $\beta\leq 1$ and $\alpha + \beta >2$, we follow the above 
treatment of the contribution of $v\cdot\nabla\tilde{V}_{sr}$. Thus, we are left with the case 
$\alpha\geq 1\geq\beta$ and $\alpha + \beta \leq 2$. By Assymption~\ref{a:cond-alpha-beta}, 
$\beta + \beta_{lr}>1$ and, by Remark~\ref{r:size-delta}, $\beta + \beta_{lr}>1+ \delta$. We write
\begin{eqnarray*}
&&\langle W_{\alpha \beta}g_{\delta , R}(Q)f\, ,\, iPf\rangle  \\
& =& \langle\tilde\cchi _R(\langle Q\rangle)^2W_{\alpha \beta}g_{\delta}(Q)\theta (H)f\, ,\, iP\theta (H) f\rangle\\
&=&\langle \tilde\cchi _{R/2}(\langle Q\rangle)W_{\alpha \beta}g_{\delta}(Q)\theta (H)\tilde\cchi _R(\langle Q\rangle)f\, ,\,
iP\theta (H)\tilde\cchi _R(\langle Q\rangle)f\rangle\\
&&+\, \langle\tilde\cchi _{R/2}(\langle Q\rangle)W_{\alpha \beta}g_{\delta}(Q)\theta (H)\tilde\cchi _R(\langle Q\rangle)f\, ,\,
[\tilde\cchi _R(\langle Q\rangle), iP\theta (H) ]f\rangle\\
&&+\, \langle\tilde\cchi _{R/2}(\langle Q\rangle)W_{\alpha \beta}g_{\delta}(Q)[\tilde\cchi _R(\langle Q\rangle), \theta (H)]f\, ,\,
iP\theta (H) \tilde\cchi _R(\langle Q\rangle)f\rangle\\
&&+\, \langle\tilde\cchi _{R/2}(\langle Q\rangle)W_{\alpha \beta}g_{\delta}(Q)[\tilde\cchi _R(\langle Q\rangle), \theta (H)]f\, ,\,
[\tilde\cchi _R(\langle Q\rangle), iP\theta (H) ]f\rangle\, .
\end{eqnarray*}
The second and third terms are
$O(R^{\delta -\beta})\|\tilde\cchi _R(\langle Q\rangle)\langle Q\rangle ^{-s}f\|$ and the last term is $O(R^{\delta -1-\beta})$,
by Lemma~\ref{l:commut-fonct-Q-fonct-H-bis}. \\
We now focus on the first term. We write 
\begin{eqnarray}
&&\hspace{.8cm}\langle \tilde\cchi _{R/2}(\langle Q\rangle)W_{\alpha \beta}g_{\delta}(Q)\theta (H)\tilde\cchi _R(\langle Q\rangle)f\, ,\,
iP\theta (H)\tilde\cchi _R(\langle Q\rangle)f\rangle\label{eq:effect-oscillations}\\
&=&\langle \tilde\cchi _{R/2}(\langle Q\rangle)W_{\alpha \beta}g_{\delta}(Q)\theta (H_0)\tilde\cchi _R(\langle Q\rangle)f\, ,\,
iP\theta (H_0)\tilde\cchi _R(\langle Q\rangle)f\rangle\nonumber\\
&+&\langle \tilde\cchi _{R/2}(\langle Q\rangle)W_{\alpha \beta}g_{\delta}(Q)\bigl(\theta (H)-\theta (H_0)\bigr)
\tilde\cchi _R(\langle Q\rangle)f\, ,\, iP\theta (H_0)\tilde\cchi _R(\langle Q\rangle)f\rangle\nonumber\\
&+&\langle \tilde\cchi _{R/2}(\langle Q\rangle)W_{\alpha \beta}g_{\delta}(Q)\theta (H_0)\tilde\cchi _R(\langle Q\rangle)f\, ,\,
iP\bigl(\theta (H)-\theta (H_0)\bigr)\tilde\cchi _R(\langle Q\rangle)f\rangle\nonumber\\
\hspace{-2cm}&+&\langle \tilde\cchi _{R/2}(\langle Q\rangle)W_{\alpha \beta}g_{\delta}(Q)\bigl(\theta (H)-\theta (H_0)\bigr)
\tilde\cchi _R(\langle Q\rangle)f\, ,\, \nonumber\\
&&\hspace{6cm}iP\bigl(\theta (H)-\theta (H_0)\bigr)\tilde\cchi _R(\langle Q\rangle)f\rangle\, .\nonumber
\end{eqnarray}
By Lemma~\ref{l:diff-decroit-Q-bis}, the second and third terms on the r.h.s. of
\eqref{eq:effect-oscillations} are at most of size 
$O(R^{\delta +1-\beta -\beta_{lr}})\|\tilde\cchi _R(\langle Q\rangle)\langle Q\rangle ^{-s}f\|^2$, whereas the fourth one is seen to be 
$O(R^{\delta +1-\beta -2\beta_{lr}})\|\tilde\cchi _R(\langle Q\rangle)\langle Q\rangle ^{-s}f\|^2$. We write the first one as
\begin{eqnarray*}
&&\langle \tilde\cchi _{R/2}(\langle Q\rangle)W_{\alpha \beta}g_{\delta}(Q)\theta (H_0)\tilde\cchi _R(\langle Q\rangle)f\, ,\,
iP\theta (H_0)\tilde\cchi _R(\langle Q\rangle)f\rangle\\
&=&\langle W_{\alpha \beta}\bigl[\tilde\cchi _{R/2}(\langle Q\rangle)g_{\delta}(Q),\theta (H_0)\bigr]\tilde\cchi _R(\langle Q\rangle)f\, ,\,
iP\theta (H_0)\tilde\cchi _R(\langle Q\rangle)f\rangle\\
&&+\, \langle W_{\alpha \beta}\theta (H_0)g_{\delta}(Q)\tilde\cchi _R(\langle Q\rangle)f\, ,\,
iP\theta (H_0)\tilde\cchi _R(\langle Q\rangle)f\rangle\, .
\end{eqnarray*}
By the above arguments, the first term on the r.h.s is
$O(R^{\delta -\beta})\|\tilde\cchi _R(\langle Q\rangle)\langle Q\rangle ^{-s}f\|^2$. So is also the last term by
Propositions~\ref{p:oscillations-energy-alpha-leq-1} and~\ref{p:oscillations-energy-alpha>1}. 

We are left with the contribution of $H_0$ in the l.h.s. of \eqref{eq:weighted-mourre}. A direct computation gives
$[H_0, iB_R]=P^T\cdot\Gc _{\delta , R}\cdot P-h_{\delta , R}$ where the entries of the $d\times d$-matrix valued function
$\Gc _{\delta , R}$ on $\R^d$ are given by
\[\partial _{k}\bigl(\tilde\cchi _R(\langle \cdot\rangle )^2(g_\delta)_j\bigr)(y)\, .\]
Here $g_\delta (y)=((g_\delta)_1(y), \cdots , (g_\delta)_d(y))^T$ and $T$ denotes
the transposition. The real valued function $h_{\delta , R}$ on $\R^d$ is given by
\[h_{\delta , R}(y)\ =\ \sum _{1\leq j, k\leq d}\partial^3 _{kkj}\bigl(\tilde\cchi _R(\langle \cdot\rangle )^2(g_\delta)_j\bigr)(y)\, .\]
The contribution of $h_{\delta , R}$ to \eqref{eq:weighted-mourre} is seen to be $O(R^{\delta -2})=O(R^{-\gamma -1})$. Since
\[\partial _{k}\bigl(\tilde\cchi _R(\langle \cdot\rangle )^2(g_\delta)_j\bigr)(y)\ =\ \tilde\cchi _R(\langle y\rangle )^2
\partial _{k}\bigl((g_\delta)_j\bigr)(y)\, +\, 2(2-\langle y\rangle^{-\delta})\tilde\cchi _R(\langle y\rangle )
\tilde\cchi _R'(\langle y\rangle )\frac{y_jy_k}{\langle y\rangle^2}\, ,\]
$2(2-\langle \cdot\rangle^{-\delta})\tilde\cchi _R(\langle \cdot\rangle )\tilde\cchi _R'(\langle \cdot\rangle )\geq 0$, and the matrix
$(y_jy_k\langle y\rangle^{-2})_{1\leq j, k\leq d}$ is nonnegative,
\[\bigl\langle f\, ,\, \, P^T\cdot\Gc _{\delta , R}\cdot P\, f\bigr\rangle\ \geq \
\bigl\langle f\, ,\, \, P^T\cdot\tilde\cchi _R(\langle Q\rangle )^2\Gc _{\delta }(Q)\cdot P\, f\bigr\rangle\]
where the entries of the $d\times d$-matrix valued function
$\Gc _{\delta}$ on $\R^d$ are given by $\partial _{k}((g_\delta)_j)(y)$. For $y\in\R^d$, $\Gc _{\delta }(y)$ is the sum of two
nonnegative matrices, namely
\begin{eqnarray*}
\Gc _{\delta }(y)&=&\frac{\bigl(2-\langle y\rangle^{-\delta}\bigr)}{\langle y\rangle}\Bigl(\delta _{jk}\, -\,
\frac{y_jy_k}{\langle y\rangle^2}\Bigr)_{1\leq j, k\leq d}\, +\, \frac{\delta}{\langle y\rangle^{1+\delta}}\Bigl(
\frac{y_jy_k}{\langle y\rangle^2}\Bigr)_{1\leq j, k\leq d}\\
&\geq&\frac{\delta}{\langle y\rangle^{1+\delta}}\Bigl(\delta _{jk}\, -\,
\frac{y_jy_k}{\langle y\rangle^2}\Bigr)_{1\leq j, k\leq d}\, +\, \frac{\delta}{\langle y\rangle^{1+\delta}}\Bigl(
\frac{y_jy_k}{\langle y\rangle^2}\Bigr)_{1\leq j, k\leq d}\\
&\geq&\frac{\delta}{\langle y\rangle^{1+\delta}}I_d\, ,
\end{eqnarray*}
where $I_d$ is the $d\times d$ identity matrix. This yields 
\[\bigl\langle f\, ,\, \, P^T\cdot\Gc _{\delta , R}\cdot P\, f\bigr\rangle\ \geq \
\delta\bigl\langle f\, ,\, \, P^T\cdot\tilde\cchi _R(\langle Q\rangle )^2\langle Q\rangle^{-2s}P\, f\bigr\rangle\, .\]
We write
\begin{eqnarray*}
&&\bigl\langle f\, ,\, \, P^T\cdot\tilde\cchi _R(\langle Q\rangle )^2\langle Q\rangle^{-2s}P\, f\bigr\rangle\\
&=&\bigl\langle\theta (H)f\, ,\, \, P^T\cdot\tilde\cchi _R(\langle Q\rangle )^2\langle Q\rangle^{-2s}P\, \theta (H)f\bigr\rangle\\
&=&\bigl\langle f\, ,\, \, \bigl[\theta (H)P^T, \tilde\cchi _R(\langle Q\rangle )\langle Q\rangle^{-s}\bigr]\cdot
\bigl[\tilde\cchi _R(\langle Q\rangle )\langle Q\rangle^{-s}, P\theta (H)\bigr]\, f\bigr\rangle\\
&&+\, \bigl\langle f\, ,\, \, \bigl[\theta (H)P^T, \tilde\cchi _R(\langle Q\rangle )\langle Q\rangle^{-s}\bigr]\cdot
P\theta (H)\tilde\cchi _R(\langle Q\rangle )\langle Q\rangle^{-s}\, f\bigr\rangle\\
&&+\, \bigl\langle \tilde\cchi _R(\langle Q\rangle )\langle Q\rangle^{-s}f\, ,\, \, \theta (H)P^T\cdot
\bigl[\tilde\cchi _R(\langle Q\rangle )\langle Q\rangle^{-s}, P\theta (H)\bigr]\, f\bigr\rangle\\
&&+\, \bigl\langle \tilde\cchi _R(\langle Q\rangle )\langle Q\rangle^{-s}f\, ,\, \, \theta (H)H_0\theta (H)\,
\tilde\cchi _R(\langle Q\rangle )\langle Q\rangle^{-s}f\bigr\rangle\, .
\end{eqnarray*}
By Lemma~\ref{l:commut-fonct-Q-fonct-H-bis}, the first term is $O(R^{-2})=O(R^{-\gamma -1})$, the second and third ones
are $O(R^{-1})\|\tilde\cchi _R(\langle Q\rangle)\langle Q\rangle ^{-s}f\|$, thus also
$O(R^{-\gamma})\|\tilde\cchi _R(\langle Q\rangle)\langle Q\rangle ^{-s}f\|$. Writing $H_0=H-V$ in the last term and using
the fact that $\theta (H)V\langle Q\rangle^{\beta _{lr}}$ is bounded, this last term is
\[\geq\ c\|\theta (H)\tilde\cchi _R(\langle Q\rangle)\langle Q\rangle ^{-s}f\|^2\, -\,
O\bigl(R^{-\beta _{lr}}\bigr)\|\tilde\cchi _R(\langle Q\rangle)\langle Q\rangle ^{-s}f\|^2\, ,\]
where $c$ is the infimum of $\Jc$. Now, we write
\begin{eqnarray*}
&&\|\theta (H)\tilde\cchi _R(\langle Q\rangle)\langle Q\rangle ^{-s}f\|^2\\
&=&\bigl\langle \bigl[\theta (H), \tilde\cchi _R(\langle Q\rangle)\langle Q\rangle ^{-s}\bigr]f\, ,\,
\bigl[\theta (H), \tilde\cchi _R(\langle Q\rangle)\langle Q\rangle ^{-s}\bigr]f\bigr\rangle\\
&&+\, \bigl\langle \bigl[\theta (H), \tilde\cchi _R(\langle Q\rangle)\langle Q\rangle ^{-s}\bigr]f\, ,\,
\tilde\cchi _R(\langle Q\rangle)\langle Q\rangle ^{-s}f\bigr\rangle\\
&&+\, \bigl\langle\tilde\cchi _R(\langle Q\rangle)\langle Q\rangle ^{-s}f\, ,\,
\bigl[\theta (H), \tilde\cchi _R(\langle Q\rangle)\langle Q\rangle ^{-s}\bigr]f\bigr\rangle\\
&&+\, \|\tilde\cchi _R(\langle Q\rangle)\langle Q\rangle ^{-s}f\|^2\, .
\end{eqnarray*}
By Lemma~\ref{l:commut-fonct-Q-fonct-H-bis} again, the first term is $O(R^{-2})=O(R^{-\gamma -1})$, the second and third ones
are $O(R^{-1})\|\tilde\cchi _R(\langle Q\rangle)\langle Q\rangle ^{-s}f\|=
O(R^{-\gamma})\|\tilde\cchi _R(\langle Q\rangle)\langle Q\rangle ^{-s}f\|$.
Gathering all the previous estimates and taking $R_1$ large enough, we get \eqref{eq:weighted-mourre} with $c_1=\delta c/2$. \qed

Now we are in position to prove our first main result, namely Theorem~\ref{th:main}. To this end, we use the characterization of
the LAP in terms of so called ``special sequences'', that was introduced in \cite{gj}. 

\Pfof{Theorem~\ref{th:main}} Without loss of generality, the length of $\Ic$ may be assumed small enough. 
In particular,
we can find a compact interval $\Jc$ satisfying Assumption~\ref{a:cond-alpha-beta} such that $\Ic\subset\mathring{\Jc}$. Since
the validity of \eqref{eq:tal-Q} for some $s>1/2$ implies the validity of \eqref{eq:tal-Q} for any $s'\geq s$, we may choose
$s>1/2$ as close to $1/2$ as we want. Let $\theta\in\Cc^\infty _c(\R;\R)$ such that $\theta =1$ on $\Ic$ and
$\supp\, \theta\subset\mathring{\Jc}$. 
By Proposition~\ref{p:esti-mourre}, \eqref{eq:esti-mourre} holds true. Multiplying each term on both sides by $\cchi (H)\Pi^\perp$, with
$\cchi\theta=\cchi$, and shriking the size of the support of $\cchi$ so much that $\|K\cchi (H)\Pi^\perp\|\leq c$, we get
\eqref{eq:esti-mourre} with $2c$ replaced by $c$, $\theta (H)$ replaced by $\cchi (H)\Pi^\perp$, and $K=0$. This can be done with the
requirement that $\cchi =1$ on a small compact interval. Therefore we may assume that, for $\Ic$, $\Jc$, and $\theta$, as above, we
have the following strict, projected Mourre estimate 
\begin{equation}\label{eq:esti-mourre-projetee}
\Pi^\perp\theta (H)[H, iA]\theta (H)\Pi^\perp\ \geq\ c\, \theta (H)^2\Pi^\perp\, . 
\end{equation}
Recall that $\theta (H)\in\Cc^\infty(\langle Q\rangle)$ and $\theta (H)\Pi\in\Cc^\infty(\langle Q\rangle)$, by Lemma~\ref{l:Q-regu}
and by Corollary~\ref{c:pi-C1}, respectively. Thus $\theta (H)\Pi^\perp\in\Cc^\infty(\langle Q\rangle)$
and we can apply Proposition~3.2 in \cite{gj2}. Therefore the LAP \eqref{eq:tal-Q} is equivalent to the
following statement: \\
Take a sequence $(f_n, z_n)_{n\in\N}$ such that, for all $n$, $z_n\in\C$, $\Re z_n\in\Ic$, $\Im z_n\neq 0$, $f_n\in\Dc (H)$,
$\Pi^\perp f_n=f_n$, $\theta (H)f_n=f_n$, and $(H-z_n)f_n\in\Dc(\langle Q\rangle^s)$. Assume further that $\Im z_n\to 0$,
$\|\langle Q\rangle^s(H-z_n)f_n\|\to 0$, and that $(\|\langle Q\rangle^{-s}f_n\|)_{n\in\N}$ converges to some real number $\eta$.
Then $\eta =0$. \\
We shall prove this statement. Let us consider such a sequence $(f_n, z_n)_{n\in\N}$. Take $R\geq 1$. Notice that  
$\cchi _R(\langle Q\rangle)f_n$ actually belongs to $\Dc(H)\cap\Dc(\langle Q\rangle)$. Note also that the operator
$A^\perp:=\Pi^\perp\theta (H)A\theta (H)\Pi^\perp$ is well-defined on $\Dc(\langle Q\rangle)$, since $P\theta (H)$ is bounded and 
preserves, together with $\theta (H)$ and $\Pi^\perp$, the set $\Dc(\langle Q\rangle)$. Since $H$ commutes with $\theta (H)\Pi^\perp$,
we derive from \eqref{eq:esti-mourre-projetee} applied to 
$\cchi _R(\langle Q\rangle)f_n$ that
\begin{equation}\label{eq:esti-mourre-projetee-f_n}
\bigl\langle \cchi _R(\langle Q\rangle)f_n\, ,\, [H, iA^\perp ]\cchi _R(\langle Q\rangle)f_n\bigr\rangle\ \geq\ c\,
\bigl\|\theta (H)\Pi^\perp\cchi _R(\langle Q\rangle)f_n\bigr\|^2\, .
\end{equation}
Since $\theta (H)\Pi^\perp$ is smooth w.r.t. $\langle Q\rangle$, 
\begin{eqnarray*}
\theta (H)\Pi^\perp\cchi _R(\langle Q\rangle)f_n&=& \cchi _R(\langle Q\rangle)f_n\, +\, \bigl[\theta (H)\Pi^\perp ,
\cchi _R(\langle Q\rangle)\bigr]\langle Q\rangle^s\cdot\langle Q\rangle^{-s}f_n\\
&=&\cchi _R(\langle Q\rangle)f_n\, +\, O\bigl(R^{s-1}\bigr)\, ,
\end{eqnarray*}
thanks to Lemma~\ref{l:commut-fonct-Q-fonct-H-bis}. The above $O(R^{s-1})$ and the following ''$O$'' are all independent of $n$.
Inserting this information in \eqref{eq:esti-mourre-projetee-f_n}, we get 
\begin{eqnarray}
\bigl\langle \cchi _R(\langle Q\rangle)f_n\, ,\, [H, iA^\perp ]\cchi _R(\langle Q\rangle)f_n\bigr\rangle &\geq& c\,
\bigl\|\cchi _R(\langle Q\rangle)f_n\bigr\|^2 \nonumber\\
&&\, +\, O\bigl(R^{s-1}\bigr)\bigl\|\cchi _R(\langle Q\rangle)f_n\bigr\|\, +\,
O\bigl(R^{2s-2}\bigr)\, . \label{eq:esti-mourre-projetee-f_n-bis}
\end{eqnarray}
Now, we need information on the $f_n$ for ''large $\langle Q\rangle$''. Let $\Ic '$ a compact interval such that $\supp\, \theta
\subset\Ic '\subset\mathring{\Jc}$. Since $f_n=\theta (H)f_n$ and $E_{\Ic '}\theta =\theta$, $E_{\Ic '}f_n=E_{\Ic '}(H)\theta (H)f_n=
\theta (H)f_n=f_n$. Furthermore, the sequence $(\|\langle Q\rangle ^{-s}f_n\|)_n$ is bounded since it converges to $\eta$, by assumption.
Therefore we can apply Proposition~\ref{p:weighted-mourre} to $f=f_n$ (choosing $s$ close 
enough to $1/2$, requiring in particular that $s<\gamma$), yielding \eqref{eq:weighted-mourre} with $f$ replaced by $f_n$ and
with $n$-independent ''$O's$''. As in \cite{gj} (cf. Corollary 3.2), we deduce from this that, for $R\geq R_1$,
\begin{equation}\label{eq:esti-f_n-large-Q}
 \limsup _n\bigl\|\tilde\cchi _R(\langle Q\rangle)\langle Q\rangle ^{-s}f_n\bigr\|\ =\ O\bigl(R^{-\gamma}\bigr)\, +\,
O\bigl(R^{-(\gamma +1)/2}\bigr)\ =\ O\bigl(R^{-\gamma}\bigr)\, . 
\end{equation}
We rewrite the l.h.s of \eqref{eq:esti-mourre-projetee-f_n-bis} as
\begin{eqnarray*}
&&\bigl\langle \cchi _R(\langle Q\rangle)f_n\, ,\, [H, iA^\perp ]\cchi _R(\langle Q\rangle)f_n\bigr\rangle\\
&=&\bigl\langle f_n\, ,\, \bigl[H, i\cchi _R(\langle Q\rangle)A^\perp \cchi _R(\langle Q\rangle)\bigr]f_n\bigr\rangle\, +\,
2\Re \bigl\langle \bigl[H, \cchi _R(\langle Q\rangle)\bigr]f_n\, ,\, iA^\perp \cchi _R(\langle Q\rangle)f_n\bigr\rangle\, .
\end{eqnarray*}
Since, as form,
\[\bigl[H, \cchi _R(\langle Q\rangle)\bigr]\, =\, \bigl[H_0, \cchi _R(\langle Q\rangle)\bigr]_\circ\ =\ -2\nabla\cdot
\Bigl(\nabla \bigl(\cchi _R(\langle \cdot\rangle)\bigr)\Bigr)(Q)\, +\,
\Bigl(\Delta \bigl(\cchi _R(\langle \cdot\rangle)\bigr)\Bigr)(Q)\, ,\]
and since $\langle Q\rangle^{-1}\nabla A^\perp$ is bounded, we obtain, using \eqref{eq:esti-f_n-large-Q},
\[2\Re \bigl\langle \bigl[H, \cchi _R(\langle Q\rangle)\bigr]f_n\, ,\, iA^\perp \cchi _R(\langle Q\rangle)f_n\bigr\rangle\ =\
O\bigl(R^{s-\gamma}\bigr)\bigl\|\cchi _R(\langle Q\rangle)f_n\bigr\|\, .\]
Therefore \eqref{eq:esti-mourre-projetee-f_n-bis} yields
\begin{eqnarray}
\bigl\langle f_n\, ,\, \bigl[H, i\cchi _R(\langle Q\rangle)A^\perp \cchi _R(\langle Q\rangle)\bigr]f_n\bigr\rangle
&\geq&c\,
\bigl\|\cchi _R(\langle Q\rangle)f_n\bigr\|^2\nonumber\\
&&\, +\, O\bigl(R^{s-\gamma}\bigr)\bigl\|\cchi _R(\langle Q\rangle)f_n\bigr\|\, +\,
O\bigl(R^{2s-2}\bigr)\, . \label{eq:esti-mourre-projetee-f_n-ter}
\end{eqnarray}
Expanding the commutator as in \cite{gj} (cf. Proposition 2.15), we see that
\begin{equation}\label{eq:almost-virial}
\lim _n\bigl\langle f_n\, ,\, \bigl[H, i\cchi _R(\langle Q\rangle)A^\perp \cchi _R(\langle Q\rangle)\bigr]f_n\bigr\rangle\ = \ 0\, .
\end{equation}
Using \eqref{eq:almost-virial} in \eqref{eq:esti-mourre-projetee-f_n-ter}, we deduce that
\[\limsup _n\bigl\|\cchi _R(\langle Q\rangle)f_n\bigr\|\ =\ O\bigl(R^{s-\gamma}\bigr)\, , \]
with $s-\gamma <0$. It follows from this and \eqref{eq:esti-f_n-large-Q} that $\eta =0$. \qed

%%%%%%%%%%%%%%%%%%%%%%%%
\section{Symbol-like long range potentials.} 
\label{s:symbolic-long-range}
\setcounter{equation}{0}
%%%%%%%%%%%%%%%%%%%%%%%%

This section is devoted to the

\Pfof{Theorem~\ref{th:results-symbol}} Let $H_1$ be the self-adjoint operator $H_0+V_{lr}(Q)$ on $\Dc (H_0)$.
Thanks to the assumption on $V_{lr}$, $H_1$ is actually the Weyl quantization $p^w$ of the symbol
$p\in S(\langle \xi\rangle ^2, g)$ defined by $p(x; \xi )=|\xi |^2+V_{lr}(x)$ (see Appendix~\ref{app:s:calcul-pseudo} for
details). Now we redo the proofs of Theorems~\ref{th:main} and~\ref{th:no-posivite-eigenvalue}, replacing $H_0$ by $H_1$ at some appropriate
places. More precisely, we perform this replacement exactly when the original proofs use the ``decay'' in $\langle Q\rangle$ of
$\theta (H)-\theta (H_0)$. \\
First, we claim that the last statement in Proposition~\ref{p:oscillations-energy-alpha-leq-1} is valid if $H_0$ is replaced 
by $H_1$. Indeed we can follow the proof of Lemma 4.3 in \cite{gj2} and arrive at (4.7) with 
$\theta (|\xi |^2)\theta (|\xi \mp k\hat{x}|^2)$ replaced by $\theta (|\xi |^2+V_{lr}(x))\theta (|\xi \mp k\hat{x}|^2+V_{lr}(x))$.
Since the latter also vanishes for small enough support of $\theta$, we conclude as in \cite{gj2}. \\
For any $\ell\geq 0$ and any $\theta\in\Cc_c^\infty \R; \C)$, $\langle P\rangle^\ell \theta (H_1)$ is bounded by pseudodifferential
calculus (cf. Appendix~\ref{app:s:calcul-pseudo}). Therefore, the last statement in Proposition~\ref{p:oscillations-energy-alpha>1}
holds true with $H_0$ is replaced by $H_1$.\\
We can check that the result in Lemma~\ref{l:diff-decroit-Q} holds true with $H_0$ replaced by $H_1$. 
Thus, performing the same replacement in \eqref{eq:H-H_0}, we get the result of Proposition~\ref{p:oscillations-op-compact}.
We derive the Mourre estimate of Proposition~\ref{p:esti-mourre} by the same proof. Also with the same proofs, we get the results
of Proposition~\ref{p:decroissace-poly}, Corollary~\ref{c:regu-A}, Proposition~\ref{p:viriel}, Corollary~\ref{c:finitness}, and 
Corollary~\ref{c:pi-C1}.\\
In the proof of Proposition~\ref{p:borne-exp}, we modify the argument leading to the bound \eqref{eq:bound-W}. Again, we replace
$H_0$ by $H_1$. We notice that $H_1(F)=e^{F(Q)}H_1e^{-F(Q)}$ is also a pseudodifferential operator in $\Cc^1(\langle Q\rangle)$
such that, for $\epsilon\in [0; 1]$, the operator $\langle Q\rangle^\epsilon (m+H_1(F))^{-1}\langle Q\rangle ^{-\epsilon}$ is
bounded, uniformly w.r.t. $\lambda\geq 1$. Then, we can follow the end of the proof of Proposition~\ref{p:borne-exp} with $\beta _{lr}$
replaced by $\beta$, $H_0(F)$ by $H_1(F)$, and $V$ by $V-V_{lr}$. \\
Next, we redo the proof of Proposition~\ref{p:pas-de-vp} without change. In the proof of Proposition~\ref{p:weighted-mourre},
we only change the treatment of \eqref{eq:effect-oscillations} in the following way. We can check that the results in
Lemma~\ref{l:diff-decroit-Q-bis} are valid with $H_0$ replaced by $H_1$ and $\beta _{lr}$ by $\beta$. Concerning the first term
on the r.h.s of \eqref{eq:effect-oscillations}, we only need to point out that $\langle P\rangle ^\ell\theta (H_1)$ is bounded
for any $\ell$, by pseudodifferential calculus. We thus obtain the result of Proposition~\ref{p:weighted-mourre}. Finally, we
recover the result of Theorem~\ref{th:main} by the same proof. \qed

\appendix 
\renewcommand{\theequation}{\thesection .\arabic{equation}}

\section{Standard pseudodifferential calculus. }
\label{app:s:calcul-pseudo}
\setcounter{equation}{0}

In this appendix, we briefly review some basic facts about pseudodifferential calculus.
We refer to \cite{h3}[Chapters 18.1, 18.4, 18.5, and 18.6] for a traditional study
of the subject but also to \cite{be,bo,bo2,bc,l} for a modern and powerful version.

Denote by $\Sch (M)$ the Schwartz space on the space $M$ and by $\Fc$ the Fourier transform
on $\R^d$ given by
\[\Fc u (\xi) :=  (2\pi )^{-d}\int_{\R^{d}}e^{-ix\cdot\xi}u(x)\, dx\, ,\]
for $\xi\in\R^{d}$ and $u\in\Sch (\R^d)$. For test functions $u, v\in\Sch (\R^d)$,
let $\Omega (u,v)$ and $\Omega '(u,v)$ be the functions in $\Sch (\R^{2d})$ defined by
\begin{align*}
\Omega (u,v)(x,\xi )&:=\overline{v}(x)\Fc u(\xi )e^{ix\cdot\xi}\, ,\\
\Omega '(u,v)(x,\xi )&:=(2\pi
)^{-d}\int_{\R^{d}}u(x-y/2)\overline{v}(x+y/2)e^{-iy\cdot\xi}\, dy\, ,
\end{align*}
respectively. Given a distribution $b\in\Sch '(T^\ast\R^d)$, the formal quantities
\begin{eqnarray*}
(2\pi )^{-d}\int_{\R^{3d}}e^{i(x-y)\cdot\xi}b(x,\xi)v(x)u(y)\, dxdyd\xi\, , \\
(2\pi )^{-d}\int_{\R^{3d}}e^{i(x-y)\cdot\xi}b((x+y)/2,\xi)u(x)u(y)\, dxdyd\xi\,
\end{eqnarray*}
are defined by the duality brackets $\langle b,\Omega (u,v)\rangle$ and
$\langle b,\Omega '(u,v)\rangle$, respectively. They define continuous operators
from $\Sch (\R^d)$ to $\Sch '(\R^d)$ that we denote by $\op b(x,D_x)$
and $b^w(x,D_x)$ respectively. Sometimes we simply write $\op b$ and
$b^w$, respectively.\\
Choosing on the phase space $T^\ast\R^d$ a metric $g$ and a weight function $m$ with
appropriate properties (cf., admissible metric and weight in \cite{l}), let
$S(m, g)$ be the space of smooth functions on $T^\ast\R^d$ such that, for all $k\in \N$,
there exists $c_k>0$ so that, for all $\ph =(x,\xi )\in T^\ast\R^d$, all $(t_1,\cdots ,t_k)
\in (T^\ast\R^d)^k$,
\begin{equation}\label{eq:bound-symbol}
|a^{(k)}(\ph )\cdot (t_1,\cdots ,t_k)|\leq c_k m(x^*)g_{x^*}(t_1)^{1/2}\cdots
g_{x^*}(t_k)^{1/2}\, .
\end{equation}
Here, $a^{(k)}$ denotes the $k$-th derivative of the function $a$.
We equip the vector space $S(m, g)$ with the semi-norms $\|\cdot\|_{\ell ,S(m, g)}$
defined by $\max_{0\leq k\leq\ell}c_k$, where the $c_k$ are
the best constants in \eqref{eq:bound-symbol}. $S(m, g)$ is a Fr\'echet space. The space
of operators $\op b(x, D_x)$ (resp.\  $b^w(x,D_x)$) when $b\in S(m, g)$ has nice properties
(cf., \cite{h3,l}). Defining $\ph =(x, \xi)\in T^\ast\R^d$, we stick here to the following metrics
\begin{equation}\label{eq:metric}
 (g_0)_{\ph}:=\ \frac{dx^2}{\langle x\rangle ^2}+\frac{d\xi ^2}{\langle \xi\rangle ^2}
\hspace{.5cm}\mbox{and}\hspace{.5cm} (g_\alpha )_{\ph}:=\ \frac{dx^2}{\langle x\rangle ^{2(1-\alpha )}}+\frac{d\xi ^2}{\langle \xi\rangle ^2}
\, ,
\end{equation}
for $0<\alpha <1$, and to weights of the form, for $p, q \in \R$,
\begin{align}\label{eq:weight}
m(\ph ):=\langle x\rangle ^p\langle \xi\rangle ^q.
\end{align}
The gain of the calculus associated to each metric in \eqref{eq:metric}
is given respectively by
\begin{align}\label{eq:gain}
h_0(\ph ):=\langle x\rangle ^{-1}\langle \xi\rangle ^{-1}\hspace{.5cm}\mbox{and}\hspace{.5cm}
h_\alpha (\ph )=\langle x\rangle^{1-\alpha}\langle \xi\rangle ^{-1}.
\end{align}
Take weights $m_1$, $m_2$ as in \eqref{eq:weight}, let $g$ be $g_0$ or $g_\alpha$, and denote
by $h$ the gain of $\tilde g$. For any $a\in S(m_1, g)$
and $b\in S(m_2, g)$, there are a symbol $a\# _rb\in S(m_1m_2, g)$ and a symbol
$a\# b\in S(m_1m_2, g)$ such that $\op a\op b=\op (a\# _rb)$ and
$a^wb^w=(a\# b)^w$. The maps $(a,b)\donne a\# _rb$ and $(a,b)\donne a\# b$
are continuous and so are also $(a,b)\donne a\# _rb-ab\in
S(m_1m_2h, g)$ and
$(a,b)\donne a\# b-ab\in S(m_1m_2h, g)$. If $a\in
S(m_1, g)$, there exists
$c\in S(m_1, g)$ such that $a^w=\op c$. The maps $a\donne c$ and
$a\donne c-a
\in S(m_1m_2h, g)$ are continuous. If $a\in S(\langle\xi\rangle^m, g)$ for $m\in\N$, $a^w$ and  $\op a$ are bounded from $\Hc^m(\R^d)$ to
$\rL^2(\R^d)$ and the corresponding operator norms are controlled above by some appropriate semi-norm of $a$ in $S(\langle\xi\rangle^m, g)$. 
In particular, they are bounded on $\rL^2(\R^d)$, if $a\in S(1, g)$. Futhermore, for $a\in S(m, g)$, 
\begin{equation}\label{eq:caract-pseudo-borne}
\op a\mbox{ is bounded}\ssi a^w\mbox{ is bounded}\ssi a\in S(1, g)\, .
\end{equation}
For $a\in S(1, g)$, 
\begin{equation}\label{eq:caract-pseudo-compact}
\op a\mbox{ is compact}\ssi a^w\mbox{ is compact}\ssi \lim _{|\ph|\to
  \infty}a(\ph)=0\, .
\end{equation}
Finally, we recall the following result on some smooth functional calculus for pseudodifferential operators associated to
some admissible metric $g$. This result is essentially contained in \cite{bo} (see \cite{gj2,l}, for details).
We also use it for $g=g_0$ or $g=g_\alpha$.\\
For $\rho\in\R$, we denote by $\Sc^\rho$ the set of smooth functions $\varphi :\R\dans\C$ such that, for all
$k\in\N$, $\sup _{t\in\R}\langle t\rangle ^{k-\rho}|\partial _t^k\varphi (t)|<\infty$.
If we take a real symbol $a\in S(m,g)$, then the operator $a^w$ is self-adjoint on the
domain $\Dc (a^w)=\{u\in\rL^2(\R^d_x);a^wu\in\rL^2(\R^d_x)\}$. In particular, the operator
$\varphi (a^w)$ is well defined by the functional calculus if $\varphi$ is a borelean
function on $\R$. We assume that $m\geq 1$. A real symbol $a\in S(m,g)$ is said elliptic
if $(i-a)^{-1}$ belongs to $S(m^{-1},g)$. Recall that $h$ denotes the gain of the symbolic calculus in $S(m,g)$.
\begin{theorem}\label{t:func}
Let $m\geq 1$ and $a\in S(m, g)$ be real and elliptic. Take a function $\varphi\in \Sc ^\rho$. Then
$\varphi (a)\in S(m^\rho , g)$ and there exists $b\in S(hm^\rho , g)$ such that
\begin{equation}\label{eq:fonct-pseudo}
\varphi\big(a^w(x,D)\big)= \big(\varphi(a)\big)^w(x,D) + b^w(x,D).
\end{equation}
\end{theorem}

\section{Regularity w.r.t. an operator. }
\label{app:s:regu}
\setcounter{equation}{0}

For sake of completness, we recall here important facts on the regularity w.r.t. to a self-adjoint operator.
Further details can be found in \cite{abg,dg,gj2,ggm,gg}.

Let $\Hr$ be a complex Hilbert space. 
The scalar product $\langle \cdot ,\cdot \rangle$ in $\Hr$ is right linear and $\|\cdot\|$
denotes the corresponding norm and also the norm in $\Bc (\Hr )$, the space of bounded
operators on $\Hr$. Let $M$ be a self-adjoint operator in $\Hr$. Let $T$ be a closed operator in $\Hr$. The
form $[T,M]$ is defined on $(\Dc(M)\cap\Dc(T))\times(\Dc(M)\cap\Dc(T))$ by
\begin{equation}\label{eq:sens-forme}
 \langle f\, ,\, [T,M]g\rangle \ :=\ \langle T^\ast f\, ,\, Mg\rangle\, -\,
\langle Mf\, ,\, Tg\rangle \, .
\end{equation}
If $T$ is a bounded operator on $\Hr$ and $k\in \N$, we say that $T\in \Cc^k(M)$
if, for all $f\in \Hr$, the map $\R\ni t\mapsto e^{itM}Te^{-itM}f\in \Hr$ has the usual
$\Cc^k$ regularity. The following characterization is available.
\begin{proposition}\label{p:caract-C1} \cite[p.\ 250]{abg}.
Let $T\in \Bc(\Hr)$. Are equivalent:
\begin{enumerate}
 \item $T\in \Cc^1(M)$.
 \item The form $[T,M]$ defined on $\Dc(M)\times \Dc(M)$ extends to
a bounded form on $\Hr\times\Hr$ associated to a bounded operator denoted by  $\ad_M^1(T):=[T,M]_\circ$.
 \item $T$ preserves $\Dc (M)$ and
the operator $TM-MT$, defined on $\Dc(M)$, extends to a bounded operator on $\Hr$.
\end{enumerate}
\end{proposition}
It immediately follows that $T\in \Cc^k(M)$ if and only if the iterated
commutators $\ad_M^p(T):= [\ad_M^{p-1}(T),M]_\circ$ are bounded for $p\leq k$.\\
It turns out that $T\in \Cc^k(M)$ if and only if, for a $z$ outside
$\sigma (T)$, the spectrum of $T$, $(T-z)^{-1}\in \Cc^k(M)$. Now, let $N$ be
a self-adjoint operator in $\Hr$. It  is
natural to say that $N\in \Cc^k(M)$ if $(N-z)^{-1}\in \Cc^k(M)$ for
some  $z\not\in\sigma (N)$. In that case, $(N-z)^{-1}\in \Cc^k(M)$,
for all $z\not\in\R$. Lemma~6.2.9 and Theorem~6.2.10 in \cite{abg} gives the
following  characterization of this regularity:
\begin{theorem}\cite[p.\ 251]{abg}.\label{th:abg}
Let $M$ and $N$ be two self-adjoint operators in the Hilbert space
$\Hr$. For $z\notin \sigma(N)$, set $R(z):=(N-z)^{-1}$. The following
points are equivalent:
\begin{enumerate}
\item $N\in\Cc^1(M)$.
\item For one (then for all) $z\notin \sigma(N)$, there is a finite
$c$ such that
\begin{align}\label{e:C1a}
|\langle M f, R(z) f\rangle - \langle R(\bar{z}) f, Mf\rangle| \leq c
\|f\|^2, \mbox{ for all $f\in\Dc(M)$}.
\end{align}
\item
\begin{enumerate}
\item [a.]There is a finite $c$ such that for all $f\in \Dc(M)\cap\Dc(N)$:
\begin{equation}\label{e:C1b}
|\langle Mf, N f\rangle- \langle N f, Mf\rangle|\leq \,
 c\big(\|N f\|^2+\|f\|^2\big).
\end{equation}
\item [b.] The set
$\{f\in\Dc(M);\,  R(z)f\in\Dc(M)\, \mbox{\rm and}\, R(\bar{z})f\in\Dc(M)
\}$ is a core for $M$, for some (then for all) $z\notin \sigma(N)$.
\end{enumerate}
\end{enumerate}
\end{theorem}
Note that the condition (3.b) could be uneasy to check, see
\cite{gg}. We mention \cite{GoleniaMoroianu}[Lemma A.2]
to overcome this subtlety. Note that \eqref{e:C1a} yields that the
commutator $[M, R(z)]$ extends to a bounded operator, in the form
sense. We shall denote the extension by $[M, R(z)]_\circ$. In the same
way, from \eqref{e:C1b}, the commutator $[N, M]$ extends to a unique
element of $\Bc\big(\Dc(N), \Dc(N)^*\big)$ denoted by $[N,
  M]_\circ$. Moreover, if $N\in \Cc^1(M)$ and $z\notin \sigma(N)$,
\begin{eqnarray}\label{eq:comm-resolv}
\big[M, (N-z)^{-1}\big]_\circ =\quad  \underbrace{(N-z)^{-1}}_{\Hr
  \leftarrow \Dc(N)^*}\quad  \underbrace{[N, M]_\circ}_{\Dc(N)^*\leftarrow
  \Dc(N)} \quad \underbrace{(N-z)^{-1}}_{\Dc(N)\leftarrow \Hr}.
\end{eqnarray}
Here we used the Riesz lemma to identify $\Hr$ with its anti-dual
$\Hr^*$. It turns out that an easier characterization is available if
the domain of $N$ is conserved under the action of the unitary group
generated by $M$.
\begin{theorem}\cite[p.\ 258]{abg}.\label{th:abg2}
Let $M$ and $N$ be two self-adjoint operators in the Hilbert space
$\Hr$ such that $e^{itM}\Dc (N)\subset\Dc (N)$, for all $t\in\R$.
Then $N\in\Cc^1(M)$ if and only if \eqref{e:C1b} holds true.
\end{theorem}
%

%Finally, the regulatity w.r.t. $N$ is preserved by an appropriate functional calculus as seen in
%the following theorem. Recall that, for $\rho\in\R$, $\Sc^\rho$ is the set of functions $\varphi\in\Cc^\infty(\R; \C)$
%such that, for all $k\in\N$, $\sup _{t\in\R}\langle t\rangle ^{k-\rho}|\partial _t^k\varphi (t)|<\infty$.
%
%\begin{theorem}\cite[p.\ 244]{abg}.\label{th:abg3} \textcolor{red}{Mauvaise ref !}
%Let $M$ and $N$ be two self-adjoint operators in the Hilbert space
%$\Hr$ and $k\in\N^\ast$ such that $N\in\Cc^k(M)$. Then, for any $\rho\leq 0$ and any $\varphi\in\Sc^\rho$; 
%$\varphi (N)\in\Cc^k(M)$. This applies in particular for $\varphi=\theta\in\Cc_c^\infty (\R; \C)$.
%\end{theorem}
%

\section{Commutator expansions. }
\label{app:s:commut-expansions}
\setcounter{equation}{0}

In this appendix, we recall known results on functional calculus and on commutator expansions. Details can be found in \cite{dg,gj, gj2,mo}. 
We then apply these results to get several facts used in the main part of the text. We make use of pseudodifferential calculus
(cf. Appendix~\ref{app:s:calcul-pseudo}) and of the regularity w.r.t. an operator, recalled in
Appendix~\ref{app:s:regu}.

As in Appendix~\ref{app:s:calcul-pseudo}, we consider, for $\rho\in\R$, the set $\Sc^\rho$ of functions $\varphi\in\Cc^\infty(\R; \C)$
such that
\begin{eqnarray}\label{eq:def-S-rho}
\forall k\in\N, \quad C_k(\varphi) :=\sup _{t\in\R}\, \langle t\rangle^{-\rho+k}|
\partial _t^k\varphi (t)|<\infty .
\end{eqnarray}
Equipped with the semi-norms defined by (\ref{eq:def-S-rho}), $\Sc^\rho$ is
a Fr\'echet space. We recall the following result from \cite{dg} on almost analytic extension.
\begin{proposition}\label{p:dg}\cite{dg}.
Let $\varphi\in\Sc^\rho$ with $\rho\in\R$. There is a smooth function
$\varphi^\C:\C \rightarrow \C$, called an
\emph{almost analytic extension} of $\varphi$, such that, for all
$l\in \N$,
\begin{align}
\label{eq:dg1} \varphi^\C|_{\R}=\varphi,\quad &\big|\partial_{\bar{z}}\varphi^\C(z)
\big|\leq c_1 \langle \re(z)\rangle^{\rho-1 -l} |\im(z)|^l\, ,\\
\label{eq:dg2}
& \supp\, \varphi^\C \subset\{x+iy; |y|\leq c_2 \langle
  x\rangle\},\\
\label{eq:dg3} & \varphi^\C(x+iy)= 0, \mbox{ if }
  x\not\in\supp\,\varphi,
\end{align}
for constants $c_1$, $c_2$ only depending on the semi-norms
\eqref{eq:def-S-rho} of $\varphi$ in $\Sc^\rho$.
\end{proposition}
%

%\section{Commutator expansions. } \label{dev-commut}
%\setcounter{equation}{0}

Next we recall Helffer-Sj\"{o}strand's functional calculus
(cf., \cite{hs,dg}). As in Appendix~\ref{app:s:regu}, we consider a self-adjoint operator $M$ acting
in some complex Hilbert space $\Hr$. For $\rho< 0$, $k\in\N$, and $\varphi\in \Sc^{\rho}$, the bounded
operators $(\partial^k\varphi )(M)$ can be recovered by
\begin{eqnarray}\label{eq:int}
(\partial^k\varphi )(M) = \frac{i(k!)}{2\pi}\int_\C\partial_{\bar{z}}\varphi^\C(z)(z-M)^{-1-k}
dz\wedge d\bar{z},
\end{eqnarray}
where the integral exists in the norm topology, by \eqref{eq:dg1}
with $l=1$. For $\rho\geq 0$, we rely on the following approximation:
\begin{proposition}\label{p:ext-hs}\cite{gj}.
Let $\rho \geq 0$ and $\varphi\in \Sc^{\rho}$. Let $\cchi \in\Cc
_c^\infty(\R)$ with $\cchi =1$ near $0$ and $0\leq\cchi \leq 1$, and,
for $R>0$, let $\cchi _R(t)=\cchi (t/R)$. For $f\in
\Dc(\langle M\rangle^{\rho})$ and $k\in\N$, there exists
\begin{eqnarray}\label{eq:int-cv}
(\partial^k\varphi )(M)f =\lim_{R\to +\infty}\ \frac{i}{2\pi}\int_\C\partial_{\bar{z}}
(\varphi \cchi_R)^\C(z)
(z-M)^{-1-k}f\, dz\wedge d\bar{z}.
\end{eqnarray}
The r.h.s.\ converges for the norm in $\Hr$. It is independent of the
choise of $\cchi$.
\end{proposition}

Notice that, for some $c>0$ and $s\in [0;1]$, there exists
some $C>0$ such that, for all $z=x+iy\in\{a+ib\mid 0<|b|\leq c\langle a\rangle \}$
(like in \eqref{eq:dg2}),
\begin{eqnarray}\label{eq:majoA}
\big\| \langle M\rangle^s (M-z)^{-1}\big\|\leq C \langle x \rangle^{s}\cdot |y|^{-1}.
\end{eqnarray}

Observing that the self-adjointness assumption on $B$ is useless, we pick from \cite{gj} the
following result in two parts.
\begin{proposition}\label{p:regu}\cite{bg,dg,gj,mo}.
Let $k\in \N^\ast$, $\rho< k$, $\varphi\in \Sc^{\rho}$, and $B$ be a bounded
operator on $\Hr$ such that $B\in\Cc^k(M)$. As forms on
$\Dc(\langle M\rangle^{k-1})\times \Dc(\langle M\rangle^{k-1})$,
\begin{align}\label{eq:egalite}
[\varphi(M), B] &= \sum_{j=1}^{k-1} \frac{1}{j!}
(\partial^j\varphi )(M)\ad_M^j(B)\\
\label{eq:reste22}
&\quad + \,
\frac{i}{2\pi}\int_\C\partial_{\bar{z}}\varphi^\C(z)(z-M)^{-k}
\ad_M^k(B)(z-M)^{-1} dz\wedge d\bar{z}.
\end{align}
In particular, if $\rho\leq 1$, then $B\in\Cc^1(\varphi(M))$.
\end{proposition}

The rest of the previous expansion is estimated in
\begin{proposition}\label{p:est3} \cite{bg,gj,mo}.
Let $B$ be a bounded operator on $\Hr$ such that $B\in\Cc^k(M)$. Let $\varphi\in\Sc^\rho$, with $\rho< k$. Let
$I_k(\varphi)$ be the rest of the development of order $k$ \eqref{eq:egalite} of $[\varphi(M), B]$,
namely \eqref{eq:reste22}. Let
$s, s'\geq 0$ such that $s'<1$, $s<k$, and $\rho+s+s'<k$. Then, for $\varphi$ staying in a
bounded subset of $\Sc^\rho$, $\langle M\rangle^{s}
I_k(\varphi)\langle M\rangle^{s'}$ is bounded and there exists a $M$ and $\varphi$
independent constant $C>0$ such that $\|\langle M\rangle^{s}
I_k(\varphi)\langle M\rangle^{s'}\|\leq C\|\ad_M^k(B)\|$.
\end{proposition}

Now, we show a serie of results needed in the main text. 
Most of them are more or less known. We provide proofs for completeness.

\Pfof{Lemma~\ref{l:Q-regu}} The assumptions~\ref{a:cond-pot} and~\ref{a:cond-alpha-beta-gene} are not required for the proof of $(1)$.
We note that $(1+H_0)^{-1}=a^w$ and $\langle Q\rangle=b^w$, where $a(x, \xi)=(1+|\xi|^2)^{-1}$ and $b(x, \xi )=\langle x\rangle$,
Since $a\in S(\langle\xi\rangle^{-2}, g_0)$ and $b\in S(\langle x\rangle , g_0)$, where the metric $g_0$ defined in \eqref{eq:metric},
the form $[(1+H_0)^{-1}, \langle Q\rangle]$ is associated to $c^w$ with $c\in S(h\langle\xi\rangle^{-2}\langle x\rangle, g_0)=
S(\langle\xi\rangle^{-3}, g_0)$, by pseudodifferential calculus. Since $S(\langle\xi\rangle^{-3}, g_0)\subset
S(1, g_0)$, the form $[(1+H_0)^{-1}, \langle Q\rangle]$ extends to bounded one on $\rL^2(\R^d)$. Similarly, we can show that
the iterated commutators $\ad_{\langle Q\rangle}^p((1+H_0)^{-1})$ all extend to bounded operator on $\rL^2(\R^d)$. By the
comment just after Proposition~\ref{p:caract-C1}, $(1+H_0)^{-1}\in\Cc^\infty (\langle Q\rangle)$ and $H_0\in\Cc^\infty (\langle Q\rangle)$,
by definition. Since $\langle P\rangle=d^w$ with $d(x, \xi)=(1+|\xi|^2)^{1/2}$, we can follow the same lines to prove that
$\langle P\rangle^{-1}\in\Cc^\infty (\langle Q\rangle)$ and thus $\langle P\rangle\in\Cc^\infty (\langle Q\rangle)$. 
Similarly, $P_i, P_iP_j, \langle P\rangle^2\in\Cc^\infty (\langle Q\rangle)$. Since
the form $[\langle P\rangle, \langle Q\rangle]$ is associated to bounded pseudodifferential operator, we see that 
$\Dc(\langle Q\rangle\langle P\rangle)=\Dc(\langle P\rangle\langle Q\rangle)$.\\
By a direct computation, we see that the group $e^{it\langle Q\rangle}$ (for $t\in\R$) preserves the Sobolev space $\Hc^2(\R^d)$,
which is the domain of $H$. Furthermore the form $[H, \langle Q\rangle]$ co\"incide on $\Dc (H)\cap\Dc (\langle Q\rangle)$ with
$[H_0, \langle Q\rangle]$. The latter is associated, by pseudodifferential calculus, to a pseudodifferential operator that is
bounded from $\Hc^1(\R^d)$ to $\rL^2(\R^d)$. By Theorem~\ref{th:abg2}, $H\in\Cc^1 (\langle Q\rangle)$ and, for $z\in\C\setminus\R$, 
\begin{equation}\label{eq:commut-resolv}
\bigl[(z-H)^{-1}, \langle Q\rangle\bigr]_\circ\ =\ (z-H)^{-1}\, \bigl[H, \langle Q\rangle\bigr]_\circ (z-H)^{-1}\, .
\end{equation}
On $\Dc (\langle Q\rangle)\times\Dc (\langle Q\rangle)$, we can write the form $[[(z-H)^{-1}, \langle Q\rangle]_\circ ]$ as
\begin{align*}
&\bigl[(z-H)^{-1}, \langle Q\rangle\bigr]\, \bigl[H, \langle Q\rangle\bigr]_\circ (z-H)^{-1}\, +\, (z-H)^{-1}
\bigl[H, \langle Q\rangle\bigr]_\circ\, \bigl[(z-H)^{-1}, \langle Q\rangle\bigr]\\
&+\ (z-H)^{-1}\, \bigl[\bigl[H, \langle Q\rangle\bigr]_\circ\, ,\, \langle Q\rangle\bigr]\, (z-H)^{-1}\, .
\end{align*}
Since $[[H, \langle Q\rangle ]_0, \langle Q\rangle]=[[H_0, \langle Q\rangle ]_0, \langle Q\rangle]$ is associated to a
bounded pseudodifferential operator, $H\in\Cc^2 (\langle Q\rangle)$ by Proposition~\ref{p:caract-C1}.
Now we conclude the proof of (2) by induction, making use of \eqref{eq:commut-resolv} and the fact that the form
$\ad_{\langle Q\rangle}^p(H)=\ad_{\langle Q\rangle}^p(H_0)$ extends to a bounded one, if $p\geq 2$.\\
Let $N=H$ or $H_0$. For $z\in\C\setminus\R$, we have \eqref{eq:commut-resolv} with $H$ replaced by $N$, thanks to (1) and (2).
Using the resolvent equality for the difference $(z-N)^{-1}-(i-N)^{-1}$, we see that
\begin{equation}\label{eq:bound-commut-resolv}
 \bigl\|\bigl[(z-N)^{-1}, \langle Q\rangle\bigr]_\circ\bigr\| \ \leq\ C\Bigl(1\, +\, \frac{\langle \Re z\rangle}{|\Im z|}\Bigr)\, .
\end{equation}
where $C$ only depends on the operator norm of  $[N, \langle Q\rangle]_\circ$.
Now we use \eqref{eq:int} with $\varphi =\theta$ to express the form $[\theta (H), \langle Q\rangle]$ and see that it 
extends to a bounded one, thanks to \eqref{eq:bound-commut-resolv}. This shows that $\theta (N)\in\Cc^1(\langle Q\rangle)$. 
In a similar way, we can show by induction that $\theta (N)\in\Cc^\infty (\langle Q\rangle)$. The above arguments actually 
show that $P_i[\theta (N), \langle Q\rangle]_\circ$ is a bounded operator on $\rL^2(\R^d)$. So is also
$[P_i\theta (N), \langle Q\rangle]_\circ$ and, since $P_i\theta (N)$ is bounded, $P_i\theta (N)\in\Cc^1(\langle Q\rangle)$. 
Again we can derive by induction that $P_i\theta (N)\in\Cc^\infty(\langle Q\rangle)$. Similarly we can verify that 
$P_iP_j\theta (N)\in\Cc^\infty(\langle Q\rangle)$. \\
Note that $\theta (H)\Dc (\langle Q\rangle)\subset\Dc (H)=\Dc (H_0)$. Let $z\in\C\setminus\R$. By (2), $(z-H)^{-1}$ preserves
$\Dc (\langle Q\rangle)$ and, on $\Dc (\langle Q\rangle)$,
\[\langle Q\rangle\, (z-H)^{-1}\ =\ (z-H)^{-1}\, \langle Q\rangle\, +\, \bigl[\langle Q\rangle, (z-H)^{-1}\bigr]_\circ\, .\]
Thus $\langle Q\rangle\, (z-H)^{-1}\langle Q\rangle^{-1}$ is bounded and
\[\langle Q\rangle\, (z-H)^{-1}\langle Q\rangle^{-1}\ =\ (z-H)^{-1}\, +\, \bigl[\langle Q\rangle, (z-H)^{-1}\bigr]_\circ\,
\langle Q\rangle^{-1}\, .\]
By \eqref{eq:commut-resolv}, we see that $\langle P\rangle\langle Q\rangle\, (z-H)^{-1}\langle Q\rangle^{-1}$
\[=\ \langle P\rangle(z-H)^{-1}\, +\, \langle P\rangle (z-H)^{-1}\, 
\bigl[\langle Q\rangle, H\bigr]_\circ\, (z-H)^{-1}\langle Q\rangle^{-1}\]
is bounded and, for some $z$-independent $C'>0$,
\[\bigl\|\langle P\rangle\langle Q\rangle\, (z-H)^{-1}\langle Q\rangle^{-1}\bigr\| \ \leq\ \frac{C'}{|\Im z|}\Bigl(1\, +\,
\frac{\langle \Re z\rangle}{|\Im z|}\Bigr)\, .\]
Therefore, $\langle P\rangle\langle Q\rangle\, \theta (H)\langle Q\rangle^{-1}$ is bounded, by \eqref{eq:int} with $k=0$.
This implies that $\theta (H)\Dc (\langle Q\rangle)\subset\Dc (\langle P\rangle\langle Q\rangle)$. \qed

\begin{lemma}\label{l:diff-decroit-Q-bis} Assume Assumptions~\ref{a:cond-pot} and~\ref{a:cond-alpha-beta-gene}. 
For integers $1\leq i, j\leq d$, let the operator $\tau (P)$ be either $1$, or $P_i$, or $P_iP_j$. Then, for any 
$\theta\in\Cc_c^\infty(\R; \C)$ and any $\sigma\geq 0$, $\langle Q\rangle^{\beta _{lr} -\sigma}\tau (P)
(\theta (H)-\theta (H_0))\langle Q\rangle^\sigma$, $\langle Q\rangle^{-\sigma}\tau (P)\theta (H)\langle Q\rangle^\sigma$, 
and $\langle Q\rangle^{-\sigma}\tau (P)\theta (H_0)\langle Q\rangle^\sigma$
are bounded on $\rL^2(\R^d)$.
\end{lemma}
\proof We first note that, for $\delta\in [-1; 1]$,  the form $[H, \langle Q\rangle ^\delta]=[H_0, \langle Q\rangle ^\delta]$ 
extends to a bounded one from $\Hc^1(\R^d )$ to $\rL^2(\R^d )$. Thus, as in the previous proof (the one of Lemma~\ref{l:Q-regu}), 
for $H'=H$ and $H'=H_0$, there exists $C>0$, such that, $z\in\C\setminus\R$, 
\begin{equation}\label{eq:bound-resolv-weights}
 \bigl\|\langle P\rangle^2\langle Q\rangle^{-\delta}\, (z-H')^{-1}\langle Q\rangle^\delta\bigr\| \ \leq\ \frac{C}{|\Im z|}\Bigl(1\, +\,
\frac{\langle \Re z\rangle}{|\Im z|}\Bigr)\, .
\end{equation}
Since, for $\delta\in [0; 1]$, we can write
\begin{eqnarray*}
 \langle Q\rangle^{-1-\delta}(z-H')^{-1}\langle Q\rangle^{1+\delta}&=&\langle Q\rangle^{-\delta}(z-H')^{-1}\langle Q\rangle^{\delta}\\
&&+\, \langle Q\rangle^{-1-\delta}(z-H')^{-1}[H', \langle Q\rangle ]_\circ(z-H')^{-1}\langle Q\rangle^{\delta}
\end{eqnarray*}
with $[H', \langle Q\rangle ]_\circ =[H_0, \langle Q\rangle]_\circ$, \eqref{eq:bound-resolv-weights} implies \eqref{eq:bound-resolv-weights}
with $\delta$ replaced by $\delta +1$. By induction, we get \eqref{eq:bound-resolv-weights} for all $\delta\geq 0$. For $\delta\in [-1; 0]$,
we can similarly show \eqref{eq:bound-resolv-weights} with $\delta$ replaced by $\delta -1$ and then, by induction,
\eqref{eq:bound-resolv-weights} for all $\delta\leq 0$.\\
For $z\in\C\setminus\R$, 
\begin{align*}
 &\langle Q\rangle^{\beta _{lr}-\sigma}V_c(Q)(z-H_0)^{-1}\langle Q\rangle^\sigma\\ 
&=\, \langle Q\rangle^{\beta _{lr}}\cchi _c(Q)\cdot V_c(Q)\langle P\rangle^{-2}\cdot
\langle P\rangle^2\langle Q\rangle^{-\sigma}(z-H_0)^{-1}\langle Q\rangle^\sigma
\end{align*}
and, for $W=W_{\alpha\beta}+V_{lr}+V_{sr}$, 
\[\langle Q\rangle^{\beta _{lr}-\sigma}W(Q)(z-H_0)^{-1}\langle Q\rangle^\sigma\\ 
\ =\ \langle Q\rangle^{\beta _{lr}}W(Q)\ \langle Q\rangle^{-\sigma}(z-H_0)^{-1}\langle Q\rangle^\sigma\, ,\]
and, using $iP\cdot v(Q)=(\nabla\cdot v)(Q)+v(Q)\cdot iP$, 
\begin{align*}
 &\langle Q\rangle^{\beta _{lr}-\sigma}(z-H)^{-1}\bigl(v\cdot\nabla\tilde{V}_{sr}\bigr)(Q)(z-H_0)^{-1}\langle Q\rangle^\sigma\\ 
&=\, \langle Q\rangle^{\beta _{lr}-\sigma}(z-H)^{-1}\bigl(v(Q)\cdot iP\bigr)\tilde{V}_{sr}(Q)(z-H_0)^{-1}\langle Q\rangle^\sigma\\
&\hspace{.4cm}-\, \langle Q\rangle^{\beta _{lr}-\sigma}(z-H)^{-1}\tilde{V}_{sr}(Q)\bigl(v(Q)\cdot iP\bigr)(z-H_0)^{-1}\langle Q\rangle^\sigma\\
&=\langle Q\rangle^{\beta _{lr}-\sigma}(z-H)^{-1}iP\langle Q\rangle^{\sigma -\beta _{lr}}\cdot v(Q)
\langle Q\rangle^{\beta _{lr}}\tilde{V}_{sr}(Q)\langle Q\rangle^{-\sigma}(z-H_0)^{-1}\langle Q\rangle^\sigma\\
&\hspace{.4cm}-\, \langle Q\rangle^{\beta _{lr}-\sigma}(z-H)^{-1}\langle Q\rangle^{\sigma -\beta _{lr}}(\nabla\cdot v)(Q)
\langle Q\rangle^{\beta _{lr}}\tilde{V}_{sr}(Q)\langle Q\rangle^{-\sigma}(z-H_0)^{-1}\langle Q\rangle^\sigma\\
&\hspace{.4cm}-\, \langle Q\rangle^{\beta _{lr}-\sigma}(z-H)^{-1}\langle Q\rangle^{\sigma -\beta _{lr}}
\langle Q\rangle^{\beta _{lr}}\tilde{V}_{sr}(Q)v(Q)\cdot\langle Q\rangle^{-\sigma}iP(z-H_0)^{-1}\langle Q\rangle^\sigma\, . 
\end{align*}
By \eqref{eq:bound-resolv-weights} for $H'=H$ and $\delta =\sigma -\beta _{lr}$, \eqref{eq:bound-resolv-weights} 
for $H'=H_0$ and $\delta =\sigma$, and by the resolvent formula, we see that the operator 
\[\langle P\rangle^2\langle Q\rangle^{\beta _{lr}-\sigma}\Bigl((z-H)^{-1}\, -\, (z-H_0)^{-1}\Bigr)\langle Q\rangle^\sigma\]
is bounded and its norm is dominated by some $z$-independent $C'$ times the r.h.s. of \eqref{eq:bound-resolv-weights} squared. 
Now, we use \eqref{eq:int} with $k=0$ to get the boundedness of $\langle P\rangle^2\langle Q\rangle^{\beta _{lr}-\sigma}
(\theta (H)-\theta (H_0))\langle Q\rangle^\sigma$. This shows the desired result for the first considered operator. \\
The result for the last two operators follows from \eqref{eq:bound-resolv-weights} and \eqref{eq:int} with $k=0$. \qed

\begin{lemma}\label{l:commut-fonct-Q-fonct-H-bis}
Assume Assumptions~\ref{a:cond-pot} and~\ref{a:cond-alpha-beta-gene} satisfied.
Let $\theta\in\Cc_c^\infty(\R; \C)$. Let $\cchi\in\Cc_c^\infty(\R ; \R)$ with $\cchi =1$ near $0$ and, for $R\geq 1$, 
let $\cchi _R(t)=\cchi (t/R)$ and $\tilde\cchi _R(t)=1-\cchi _R(t)$. Let $\tau (P)$ be either $1$, or $P_i$, or 
$P_iP_j$, for $1\leq i, j\leq d$. 
\begin{itemize}
 \item[(1)] For $\sigma\in [0; 1[$ and $\epsilon \geq 0$, the operators 
\[\langle Q\rangle^{\sigma -\epsilon}[\tau (P)\theta (H), \tilde{\cchi }_R(\langle Q\rangle)]_\circ\langle Q\rangle^\sigma
\hspace{.4cm}\mbox{and}\hspace{.4cm}\langle Q\rangle^{\sigma -\epsilon}\tau (P)[\theta (H), 
\tilde{\cchi }_R(\langle Q\rangle)]_\circ\langle Q\rangle^\sigma\]
are bounded on $\rL^2(\R^d)$ and their norm are $O(R^{2\sigma -1-\epsilon})$.
 \item[(2)] The operators
\[\langle Q\rangle^{1-\beta_{lr}}\bigl[\tau (P)\theta (H), \cchi _R(\langle Q\rangle)\bigr]_\circ
\hspace{.4cm}\mbox{and}\hspace{.4cm}\langle Q\rangle^{1-\beta_{lr}}\tau (P)\bigl[\theta (H), \cchi _R(\langle Q\rangle)\bigr]_\circ\]
are bounded on $\rL^2(\R^d)$ and their norm are $O(R^{-\beta_{lr}})$.
\end{itemize}
\end{lemma}
\proof We only prove (1). The proof of (2) is similar since $[\theta (H), \cchi _R(\langle Q\rangle)]=
-[\theta (H), \tilde\cchi _R(\langle Q\rangle)]$ and $[\tau (P)\theta (H), \cchi _R(\langle Q\rangle)]=
-[\tau (P)\theta (H), \tilde\cchi _R(\langle Q\rangle)]$. \\
Note that, on $\Dc (\langle Q\rangle^\sigma)$, 
\begin{align*}
\langle Q\rangle^{\sigma -\epsilon}\bigl[\tau (P)\theta (H), \cchi _R(\langle Q\rangle)\bigr]_\circ\langle Q\rangle^\sigma
&=\langle Q\rangle^{\sigma -\epsilon}\bigl[\tau (P), \cchi _R(\langle Q\rangle)\bigr]_\circ\theta (H)\langle Q\rangle^\sigma\\
&\hspace{.4cm}+\, \langle Q\rangle^{\sigma -\epsilon}\tau (P)\bigl[\theta (H), \cchi _R(\langle Q\rangle)\bigr]_\circ\langle Q\rangle^\sigma\, ,
\end{align*}
where $[\theta (H), \cchi _R(\langle Q\rangle)\bigr]_\circ$ is explicit and satisfies 
\[\bigl\|\langle Q\rangle^{\sigma -\epsilon}[\theta (H), \cchi _R(\langle Q\rangle)\bigr]_\circ\langle Q\rangle^\sigma\bigr\|\ =\ 
O\bigl(R^{2\sigma -1-\epsilon}\bigr)\, .\]
Thus, it suffices to study the second operator in (1). \\
The form $[H, \cchi _R(\langle Q\rangle)]=[H_0, \cchi _R(\langle Q\rangle)]$ extends to a bounded one from 
$\Hc^1(\R^d )$ to $\rL^2(\R^d )$. Furthermore,
\[[H, \tilde\cchi _R(\langle Q\rangle)]_\circ\ =\ [H_0, \tilde\cchi _R(\langle Q\rangle)]_\circ\ =\
-\cchi _R'(\langle Q\rangle)\langle Q\rangle^{-1} Q\cdot P\, +\, B_R\, ,\]
with bounded $B_R$ such that $\|B_R\|=O(R^{-2})$. Using the proofs of Lemma~\ref{l:Q-regu} and of Lemma~\ref{l:diff-decroit-Q-bis},
we get, for $z\in\C\setminus\R$, the operator 
\[\langle Q\rangle^{\sigma -\epsilon}[(z-H)^{-1}, \tilde{\cchi }_R(\langle Q\rangle)]_\circ\langle Q\rangle^\sigma
\ =\ -\langle Q\rangle^{\sigma -\epsilon}(z-H)^{-1}[H, \tilde{\cchi }_R(\langle Q\rangle)]_\circ(z-H)^{-1}\langle Q\rangle^\sigma\]
is bounded and, there exist $C>0$ such that, for all $z\in\C\setminus\R$ and all $R\geq 1$, 
\[\bigl\|\langle Q\rangle^{\sigma -\epsilon}[(z-H)^{-1}, \tilde{\cchi }_R(\langle Q\rangle)]_\circ\langle Q\rangle^\sigma\bigr\|\ \leq \
R^{2\sigma -1-\epsilon}\frac{C}{|\Im z|^2}\Bigl(1\, +\,
\frac{\langle \Re z\rangle}{|\Im z|}\Bigr)^2\, .\]
Using \eqref{eq:int} with $k=0$, we get the boundedness of $\langle Q\rangle^{\sigma -\epsilon}
[\theta (H), \tilde{\cchi }_R(\langle Q\rangle)]_\circ\langle Q\rangle^\sigma$ and the desired upper bound on its norm.
Similarly, we can treat the operators $\langle Q\rangle^{\sigma -\epsilon}
P_i[\theta (H), \tilde{\cchi }_R(\langle Q\rangle)]_\circ\langle Q\rangle^\sigma$ and $\langle Q\rangle^{\sigma -\epsilon}
P_iP_j[\theta (H), \tilde{\cchi }_R(\langle Q\rangle)]_\circ\langle Q\rangle^\sigma$.\qed

\begin{lemma}\label{l:infinite-int}
Let $(\alpha ; \beta )$ such that $|\alpha -1|+\beta <1$. Let $\epsilon\in ]2|\alpha -1|; 1-\beta +|\alpha -1|]$. 
Then the integral \eqref{eq:int-alpha-beta} is infinite. 
\end{lemma}
\proof Denote by $I$ this integral \eqref{eq:int-alpha-beta}. Note that its integrand is nonnegative. Using spherical coordinates, 
\[I\ =\ c_d\int_0^{+\infty}\bigl(1-\kappa (r)\bigr)^3r^{1-\beta +|\alpha -1|-(d+\epsilon)+d-1}\sin^2\bigl(kr^\alpha\bigr)\, dr\]
where $c_d>0$ is the mesure of the unit sphere in $\R^d$. For $n\in\N$, let 
\[a_n\ =\ \frac{1}{k}\Bigl(\frac{\pi}{2}\, -\, \frac{\pi}{4}\, +\, 2n\pi \Bigr)\hspace{.4cm}\mbox{and}\hspace{.4cm}
b_n\ =\ a_n\, +\, \frac{\pi}{2k}\, .\]
For $r\in [a_n^{1/\alpha}; b_n^{1/\alpha}]$, $\sin^2\bigl(kr^\alpha\bigr)\geq 1/2$. Let $N$ be a large enough integer 
such that, for $n\geq N$, $a_n^{1/\alpha}$ lies outside the support of $\kappa (|\cdot|)$. Thus, 
\[\frac{2I}{c_d}\ \geq\ \sum_{n=N}^\infty\int_{a_n^{1/\alpha}}^{b_n^{1/\alpha}}r^{-\beta +|\alpha -1|-\epsilon}\, dr\, .\]
The general term in the above serie is bounded below by $c\cdot n^{\alpha^{-1}(1-\beta +|\alpha -1|-\epsilon )-1}$, 
for some $c>0$. By assumption, $1-\beta +|\alpha -1|-\epsilon \geq 0$, therefore the serie diverges, showing that $I$ 
is infinite. \qed

\section{Strongly oscillating term.}
\label{app:s:oscillation}
\setcounter{equation}{0}

In this section, we focus on the case $\alpha >1$ and prove the key result on oscillations, namely
Proposition~\ref{p:oscillations-energy-alpha>1}. To this end, we recall the following well-known result.

\begin{lemma}\label{l:schur} Schur's lemma. \\
Let $(n; m)\in(\N^\ast)^2$. Let $K : \R^n\times\R^m\dans\C$ be a measurable function such that, there exists $C>0$
such that
\[\sup _{x\in\R^n}\, \int_{\R^m}\, |K(x; y)|\, dy\ \leq\ C\hspace{.4cm}\mbox{and}\hspace{.4cm}
\sup _{y\in\R^m}\, \int_{\R^n}\, |K(x; y)|\, dx\ \leq\ C\, .\]
Then the operator $A : \rL^2(\R^m)\dans\rL^2(\R^n)$, that maps $f\in\rL^2(\R^m)$ to the function
\[x\, \donne \, \int_{\R^m}\, K(x; y)\cdot f(y)\, dy\, ,\]
is well-defined, bounded and its operator norm is bounded above by $C$. 
\end{lemma}

\Pfof{Proposition~\ref{p:oscillations-energy-alpha>1}}
Recall that, by \eqref{eq:def-e_pm-alpha}, denoting $1-\kappa$ by $\cchi$, 
\[
e_\pm^\alpha (Q)\ =\ \Bigl(1-\kappa \bigl(|Q|\bigr)\Bigr)e^{\pm ik|Q|^\alpha}\ =\ \cchi\bigl(|Q|\bigr)e^{\pm ik|Q|^\alpha}\, ,\]
where $\kappa\in \Cc_c^\infty(\R; \R)$ is identically $1$ near $0$. Note that, for $\epsilon , \delta>0$,
$\langle Q\rangle^{-\epsilon}\langle P\rangle^{-\delta}$ is compact on $\rL^2(\R^d; \C)$. By pseudodifferential calculus
(or commutator expansions, cf. \cite{gj}), $\langle Q\rangle^{-\epsilon}\langle P\rangle^{-\ell}\langle Q\rangle^{\epsilon}$ is bounded on
$\rL^2(\R^d; \C)$ for any $\ell\geq 0$. Thus, the desired result follows from the boundedness on $\rL^2(\R^d; \C)$ for all $p\geq 0$ of
%the operator 
$\langle P\rangle^{-\ell_1}\langle Q\rangle^pe_\pm^\alpha (Q)\langle P\rangle^{-\ell_2}$, for appropriate $\ell_1$ and $\ell _2$.
Given $p$, we seek for $\ell _1, \ell _2\geq 0$ and $C>0$ such that, for all function $f\in\Sch (\R^d; \C)$, the Schwartz space on $\R^d$,
\[\bigl\|\langle P\rangle^{-\ell_1}\langle Q\rangle^pe_\pm^\alpha (Q)\langle P\rangle^{-\ell_2}f\bigr\|^2\ =\
\bigl\langle\langle P\rangle^{-\ell_2}f\, ,\, \langle Q\rangle^pe_\mp^\alpha (Q)\langle P\rangle^{-2\ell_1}
\langle Q\rangle^pe_\pm^\alpha (Q)
\langle P\rangle^{-\ell_2}f\bigr\rangle\]
is bounded above by $C\|f\|^2$. \\
Given $f\in\Sch (\R^d; \C)$, we set $g=\langle P\rangle^{-\ell_2}f\in\Sch (\R^d; \C)$ and write
\begin{eqnarray}
f_1(x)& := &\bigl(\langle Q\rangle^pe_\mp^\alpha (Q)\langle P\rangle^{-2\ell_1}\langle Q\rangle^pe_\pm^\alpha (Q)g\bigr)(x)\nonumber\\
&=&\bigl(2\pi \bigr)^{-d}\int_{\R^{2d}}e^{i\varphi _{\alpha , \pm} (x; y; \xi )}\langle x\rangle^p\cchi (x)\langle\xi\rangle^{-2\ell_1}
\langle y\rangle^{p}\cchi (y)g(y)\, dy\, d\xi\, ,\label{eq:def-f1}
\end{eqnarray}
where $\varphi _{\alpha , \pm}(x; y; \xi )=(x-y)\cdot\xi\mp k(|x|^\alpha -|y|^\alpha )$ and the integral converges absolutely,
if $\ell_1>d/2$. Take $\delta\in ]0; 1/2[$ and $\tau\in\Cc^\infty_c (\R)$ such
that $\tau (t)=1$ if $|t|\leq  1-2\delta$ and $\tau (t)=0$ if $|t|\geq 1-\delta$. On the support of $(x; y)\donne
\cchi (x)\cchi (y)\tau (|x-y|\cdot |x|^{-1})$, $|x-y|\leq (1-\delta )|x|$. In particular, on this support,
$0$ does not belong the segment $[x;y]$ and, for all $t\in [0;1]$,
\begin{equation}\label{eq:segment}
(2-\delta )|x|\ \geq\ |tx+(1-t)y|\ \geq \ |x|-(1-t)|y-x|\ \geq \ \delta |x|\, .
\end{equation}
We write $f_1(x)=f_2(x)+f_3(x)$ where $f_2$ (resp. $f_3$) is given by \eqref{eq:def-f1} with $g(y)$ replaced by
$(1-\tau (|x-y|\cdot |x|^{-1}))g(y)$ (resp. $\tau (|x-y|\cdot |x|^{-1})g(y)$). On the support of the function
$(x; y)\donne\cchi (x)\cchi (y)(1-\tau (|x-y|\cdot |x|^{-1}))$, $|x-y|\geq (1-2\delta )|x| >0$ and
$|x-y|\geq C_\delta |y|$, for some $(x; y)$-independent, positive constant $C_\delta$. Since
\[\bigl(L_{x, y, D_\xi}-1\bigr)e^{i(x-y)\cdot \xi\mp ik(|x|^\alpha -|y|^\alpha)}\ =\ 0\hspace{.4cm}\mbox{for}\hspace{.4cm}
L_{x, y, D_\xi}\ =\ |x-y|^{-2}(x-y)\cdot D_\xi\, ,\]
we get, by integration by parts, that, for all
$n\in\N$,
\begin{align*}
f_2(x)&=(2\pi )^{-d}\int e^{i\varphi _{\alpha , \pm} (x; y; \xi )}\langle x\rangle^p\cchi (x)\langle y\rangle^{p}\cchi (y)
g(y)\bigl(1-\tau (|x-y|
\langle x\rangle^{-1})\bigr)\\
&\hspace{4.3cm}\cdot \ \bigl(L_{x, y, D_\xi}^\ast\bigr)^n\bigl(\langle\xi\rangle^{-2\ell_1}\bigr)\, dyd\xi\, .
\end{align*}
Choosing $n$ large enough, we can apply Lemma~\ref{l:schur} to show that the map $f\donne f_2$ is bounded on $\rL^2(\R^d)$.\\
On the support of the function $(x; y)\donne\cchi (x)\cchi (y)\tau (|x-y|\cdot |x|^{-1})$, we can write
$\varphi _{\alpha , \pm}(x; y; \xi )=(x-y)\cdot (\xi\mp kw_\alpha (x; y))$
where
\[w_\alpha (x; y)\ =\ \alpha \int_0^1\, \bigl|tx\, +\, (1-t)y\bigr|^{\alpha -2}\bigl(tx\, +\, (1-t)y\bigr)\, dt\, .\]
Setting, for $j\in\{0; 1\}$,
\[\lambda _j\ =\ \int_0^1\, \bigl|tx\, +\, (1-t)y\bigr|^{\alpha -2}\, t^j\, dt\, ,\]
$\lambda _0\geq\lambda _1>0$ and $\alpha ^{-1}w_\alpha (x; y)=\lambda _1x+(\lambda _0-\lambda _1)y=
\lambda _0((\lambda _1/\lambda _0)x+(1-\lambda _1/\lambda_0)y)$. By \eqref{eq:segment},
\[\lambda _0\ \geq \ \lambda _1\ \geq \ 2^{-1}\bigl(\delta |x|\bigr)^{\alpha -2}\]
and $|w_\alpha (x; y)|\geq \alpha \lambda _0\delta |x|$. Furthermore $|w_\alpha (x; y)|\leq \alpha ((2-\delta)|x|)^{\alpha -1}$, thus
\begin{align}
 &2^{-1}\delta ^{\alpha -1}\ \leq \ \alpha ^{-1}|x|^{1-\alpha }\cdot|w_\alpha (x; y)|\ \leq \ (2-\delta )^{\alpha -1}\, ,
\label{eq:encadrement-w-alpha-x}\\
&2^{-1}\delta ^{\alpha -1}(2-\delta )^{1-\alpha}\ \leq \ \alpha ^{-1}|y|^{1-\alpha }\cdot|w_\alpha (x; y)|\ \leq \
\delta ^{1-\alpha }(2-\delta )^{\alpha -1}\, .\label{eq:encadrement-w-alpha-y}
\end{align}
In the integral defining $f_3$, we make the change of variables $\xi\donne \eta=\xi \mp kw_\alpha (x; y)$ and obtain
\begin{align}
f_3(x)&=(2\pi )^{-d}\int e^{i(x-y)\cdot\eta}\langle x\rangle^p\cchi (x)\langle y\rangle^{p}\cchi (y)
g(y)\tau (|x-y|\cdot |x|^{-1})\nonumber\\
&\hspace{4.3cm}\cdot \ \bigl\langle\eta\pm kw_\alpha (x; y)\bigr\rangle^{-2\ell _1}\, dyd\eta\, .\label{eq:f3}
\end{align}
We write $f_3(x)=f_4(x)+f_5(x)$ where $f_4$ (resp. $f_5$) is given by \eqref{eq:f3} with $g(y)$ replaced by
$\tau (|\eta|\cdot |kw_\alpha (x; y)|^{-1})g(y)$ (resp. $(1-\tau (|\eta|\cdot |kw_\alpha (x; y)|^{-1}))g(y)$).
On the support of the integrand of $f_4$, $|\eta |\leq (1-\delta )|kw_\alpha (x; y)|$ which implies that $|\eta\pm kw_\alpha (x; y)|\geq 
\delta |kw_\alpha (x; y)|$. Take $\ell _1>(\alpha -1)^{-1}(p+d)$. By \eqref{eq:encadrement-w-alpha-x}, 
\eqref{eq:encadrement-w-alpha-y}, and Lemma~\ref{l:schur}, the map $f\donne f_4$ is bounded on $\rL^2(\R^d)$.\\
On the support of the integrand of $f_5$, $|\eta |\geq (1-2\delta )|kw_\alpha (x; y)|>0$. Since
\[M_{\eta , D_x}e^{i(x-y)\cdot\eta}\ =\ e^{i(x-y)\cdot\eta}\ =\ -M_{\eta , D_y}e^{i(x-y)\cdot\eta}\hspace{.4cm}\mbox{for}\hspace{.4cm}
M_{\eta , D_z}\ =\ |\eta |^{-2}\eta\cdot D_z\, ,\]
we get, by integration by parts, that, for all
$n\in\N$,
\begin{align*}
\langle g\, ,\, f_4\rangle&=(2\pi )^{-d}\int e^{i(x-y)\cdot\eta}\bigr(-M_{\eta , D_x}^\ast M_{\eta , D_y}^\ast\bigr)^n
\Bigl[\langle x\rangle^p\cchi (x)\overline{g(x)}\langle y\rangle^{p}\cchi (y)g(y)\\
&\hspace{2.3cm}\cdot \ \tau (|x-y|\cdot |x|^{-1})\bigl(1-\tau (|\eta|\cdot |kw_\alpha (x; y)|^{-1})\bigr)
\Bigr]\, dxdyd\eta\, . 
\end{align*}
Choosing the integer $n$ such that $n(\alpha -1)>p+d$, using \eqref{eq:encadrement-w-alpha-x} and \eqref{eq:encadrement-w-alpha-y}, 
we can apply Lemma~\ref{l:schur} to get some $f$-independent constant $C_0>0$ such that 
\[\bigl|\langle g \, ,\,  f_4\rangle\bigr|\ \leq \ C_0\, \sup_{0\leq |\gamma |\leq n}\bigl(\|g\|^2\, +\, \|P^\gamma g\|^2\bigr)\, .\]
Now the r.h.s. is bounded above by $C\|f\|^2$ if $\ell _2$ is greater than $1$ plus the integer part of $(\alpha -1)^{-1}(p+d)$. \qed

%%%%%%%%%%%%%%%%%%%%%%%%%%%%%%%%%%%%%%%%%%%%%%%%%%%%%%%%%%%%%%%%%%%%%%  
  
%  
\end{document}